%% file: BoundStates_arXiv_v2.tex
\documentclass[10pt,a4paper]{article}

\pdfoutput=1
\usepackage{jheppub}

\usepackage[usenames,dvipsnames]{xcolor}
\usepackage{enumerate,framed}
\usepackage[hang,flushmargin]{footmisc} 
\usepackage{amsfonts,amsmath,amssymb,amsthm}
\usepackage{graphicx,comment,multirow}
\numberwithin{equation}{section}

\input{TIKZ_prelim}

\input{mymacros}

\def\M{\mathcal{M}}
\def\O{\mathcal{O}}
\renewcommand\({\left(}
\renewcommand\){\right)}
\renewcommand\[{\left[}
\renewcommand\]{\right]}
\def\<{\langle}
\def\>{\rangle}

\renewcommand{\vec}[1]{{\bf #1}}

\def\vrel{v_{\rm rel}}
\def\ann{_{\rm ann}}
\def\dec{_{\rm dec}}
\def\BSF{_{_{\rm BSF}}}
\def\BSFgr{ _{_{\rm BSF}}^{\{100\}} }
\def\uni{_{\rm uni}}

\def\PropV{
\tikz[baseline=-0.5ex,line width=1.5 pt]{\draw[violet] (0,0) -- (0.8,0);
\draw[draw=none,fill=violet] (0.4,0) circle (0.12cm);}~} 

\def\PropP{
\tikz[baseline=-0.5ex,line width=1.5 pt]{\draw[purple] (0,0) -- (0.8,0);
\draw[draw=none,fill=purple] (0.4,0) circle (0.12cm);}~} 

\def\PropMed{
\tikz[baseline=-0.5ex,line width=1.5 pt]{\draw[vector] (0,0) -- (1,0);
\draw[draw=none,fill=teal] (0.5,0) circle (0.12cm);}~}

\def\PropAnn{
\tikz[baseline=-0.5ex,line width=1.5 pt]{\draw[scalarnoarrow] (0,0) -- (0.8,0);
\draw[fill=orange!40,draw=none] (0.37,0) circle (0.12cm);}~}

\title{Dark-matter bound states from Feynman diagrams}

\author[1]{Kalliopi Petraki,}
\author[1]{Marieke Postma}
\author[1,2]{and Michael Wiechers}
\affiliation[1]{Nikhef, Science Park 105, 1098 XG Amsterdam, The Netherlands}
\affiliation[2]{GRAPPA Institute, University of Amsterdam, Science Park 904, 1090 GL Amsterdam, The Netherlands}

\emailAdd{kpetraki@nikhef.nl}
\emailAdd{mpostma@nikhef.nl}
\emailAdd{mwiecher@nikhef.nl}

\date{\today}

\abstract{
If dark matter couples directly to a light force mediator, then it may form bound states in the early universe and in the non-relativistic environment of haloes today. In this work, we establish a field-theoretic framework for the computation of bound-state formation cross-sections, de-excitation and decay rates, in theories with long-range interactions. Using this formalism, we carry out specific computations for scalar particles interacting either via a light scalar or vector mediator. At low relative velocities of the interacting particles, the formation of bound states is enhanced by the Sommerfeld effect. For particle-antiparticle pairs, we show that bound-state formation can be faster than annihilation into radiation in the regime where the Sommerfeld effect is important. The field-theoretic formalism outlined here can be generalised to compute bound-state formation cross-sections in a variety of theories, including theories featuring non-Abelian (albeit non-confining) interactions, such as the electroweak interactions.
}

\arxivnumber{1505.00109}

\begin{document}
\maketitle

\section{Introduction \label{Sec:Intro}}

Dark matter (DM) with \emph{long-range} interactions, mediated by a light or massless force carrier, appears in diverse theories motivated on various grounds. Let us mention a few important examples. Dark matter with sizable self-interactions, mediated by a light particle, can explain the observed galactic structure better than collisionless DM~\cite{Spergel:1999mh,Kusenko:2001vu, Feng:2008mu, Loeb:2010gj, Weinberg:2013aya, Peter:2012jh, Rocha:2012jg, Vogelsberger:2014pda, Zavala:2012us, Vogelsberger:2012ku}. Asymmetric DM~\cite{Davoudiasl:2012uw, Petraki:2013wwa, Zurek:2013wia, Boucenna:2013wba} -- motivated in part by the similarity of the dark and the ordinary matter abundances -- resides, in most implementations, in a hidden sector that often includes long-range interactions~\cite{Foot:2003jt, Kaplan:2009de, Petraki:2011mv, vonHarling:2012yn, Cline:2012is, Fargion:2005ep}.  Dark matter which dissipates energy via its coupling to a light force mediator may provide a dynamical explanation for some of the observed scaling relations governing haloes~\cite{Foot:2013nea,Foot:2013lxa,Foot:2013uxa,Foot:2014uba}, and other features~\cite{Foot:2015sia,Fan:2013tia}. Inside haloes, the inelastic scatterings of either symmetric or asymmetric DM with long-range interactions can produce radiative signals~\cite{ArkaniHamed:2008qn, Pospelov:2008jd, Pearce:2013ola, Pearce:2015zca, Frandsen:2014lfa, Boddy:2014qxa, Cline:2014eaa, Detmold:2014qqa, MarchRussell:2008tu}, which can potentially account for anomalous excesses observed in the radiation backgrounds~\cite{ArkaniHamed:2008qn,Pospelov:2008jd, Pearce:2013ola, Pearce:2015zca, Frandsen:2014lfa, Boddy:2014qxa}.  Moreover, long-range DM-nucleon scattering implies a different interpretation of the direct-detection data than the commonly assumed short-range scattering~\cite{Foot:2003iv,Fornengo:2011sz,Foot:2012cs,Foot:2013msa}.  Notably, the long-range character of DM interactions is relevant not only for theories involving hidden sectors; even the electroweak interactions of the Standard Model manifest as long-range if DM is heavier than a few TeV~\cite{Hisano:2002fk,Hisano:2003ec,Cirelli:2007xd}. Clearly, long-range interactions play a central role in the venture to identify  DM. In order to extract accurate predictions for the DM phenomenology, it is then essential to fully understand their implications.

Long-range interactions typically imply the existence of \emph{bound states}. The formation of DM bound states in the early universe and/or in the dense environment of haloes today affects the phenomenology of DM in many important ways. In the early universe, symmetric DM may form unstable bound states, whose decay contributes to the DM annihilation rate and affects the DM relic abundance~\cite{vonHarling:2014kha}. Asymmetric DM may form stable bound states~\cite{Foot:2003jt, Kaplan:2009de, Kaplan:2011yj, Behbahani:2010xa, Petraki:2011mv, vonHarling:2012yn, CyrRacine:2012fz, Cline:2012is, Cline:2013pca, Petraki:2014uza, Boddy:2014yra, Krnjaic:2014xza, Wise:2014jva, Wise:2014ola, Detmold:2014qqa, Fargion:2005ep}, which affect all manifestations of DM today, including the DM self-scattering in haloes~\cite{CyrRacine:2012fz, Cline:2012is, Cline:2013pca, Petraki:2014uza}, the expected indirect-detection~\cite{Pearce:2013ola, Frandsen:2014lfa, Cline:2014eaa, Boddy:2014qxa, Detmold:2014qqa, Pearce:2015zca} and direct-detection~\cite{Laha:2013gva} signals, as well as the kinetic decoupling of DM from dark radiation~\cite{Cyr-Racine:2013fsa}. Inside haloes, DM bound states -- whether they are stable or unstable -- may form radiatively and yield detectable signatures~\cite{Pospelov:2008jd, MarchRussell:2008tu, Pearce:2013ola, Pearce:2015zca}. Radiative signals may also be produced in transitions between the bound-state energy levels~\cite{Frandsen:2014lfa, Cline:2014eaa, Boddy:2014qxa}, or other related inelastic processes~\cite{Detmold:2014qqa}. Moreover, the formation of DM bound states may be detectable at the LHC~\cite{Shepherd:2009sa}. These rich phenomenological implications strongly suggest that it is critical to accurately account for the formation of DM bound states, in order to constrain the DM properties and eventually detect DM. This work is a step toward this goal.

There are, of course, two different classes of bound states: Those arising due to non-confining interactions, such as the atomic bound states in QED, and those arising due to confining interactions, such as the hadrons in QCD. 
Particles charged under a confining force always combine into hadronic states, roughly once the kinetic energy in their center-of-mass frame (or the temperature of their plasma) drops below the confinement scale. On the other hand, in the case of non-confining interactions, the efficiency of bound-state formation (BSF) depends on the corresponding cross-sections and the details of the thermodynamic environment. Here we shall consider bound states due to non-confining interactions, and calculate the cross-sections for their formation in the non-relativistic regime, which is relevant for cosmology and DM indirect detection signals.\footnote{Another class of bound states -- non-topological solitons -- has also been considered in the context of DM~\cite{Kusenko:1997zq,  Kusenko:1997si, Kasuya:1999wu, Kusenko:2001vu,Postma:2001ea, Kusenko:2004yw}.}

The formation of DM bound states in the early universe and inside the non-relativistic environment of haloes differs from BSF in colliders in some important ways. In high-energy colliders, the initial-state particles are highly relativistic.\footnote{See, however, Ref.~\cite{Brodsky:2009gx}.} However, BSF is more efficient when the relative velocity of the interacting particles is lower than the expectation value of the relative velocity of the particles inside the bound state. Equivalently, this is when the kinetic energy in the center-of-mass (CM) frame is lower than the binding energy; clearly, this lies within the non-relativistic regime. In this regime, the long-range interaction distorts the wavefunctions of the incoming particles, which cannot be approximated by momentum eigenstates (plane waves). This is the well-known \emph{Sommerfeld effect}~\cite{Sommerfeld:1931}. If the interaction is attractive, the Sommerfeld effect enhances the cross-section for any process the two particles may participate in, including the formation of bound states. It follows that, cosmologically, DM bound states form most efficiently after the temperature of the dark plasma drops below the binding energy. While the formation of bound states in the early universe eventually freezes out due to the cosmological expansion, bound states may again form efficiently in today's dense and non-relativistic haloes. In either case, to accurately estimate the formation of DM bound states, we must account for the Sommerfeld effect, which is a non-perturbative phenomenon, as is of course the very existence of bound states.

\begin{table}[t!]
\centering

\renewcommand{\arraystretch}{2.1}
\begin{tabular}{|l|c|c|} 
\hline
& Scalar mediator 
& Vector mediator
\\[-3mm]   
& $\a = g_1 g_2 / (16\p), \quad g_1g_2 >0$
& $\a = -c_1 c_2 g^2 / (4\p), \quad c_1c_2 <0$
\\ \hline\hline
\multirow{5}{*}{$\s\BSFgr\vrel \,/\, \s_c$}
& degenerate species: $g_1=g_2, \h_1=\h_2=1/2$
& \multirow{5}{*}{$\!\!\left[\dfrac{(\h_2 c_1 -\h_1 c_2)^2}{-c_1 c_2} \right] 
\! \dfrac{2^7 \z^4  e^{-4\z{\rm arccot}\z}}{3(1+\z^2)^2} \!\!$}
\\[2.5mm] 
& $\dfrac{2^6 \a^2}{15} \, \dfrac{\z^2 (3+2\z^2)}{(1+\z^2)^2} \: e^{-4\z {\rm arccot} \z} $  
&  
\\[2.5mm] \cline{2-2}
& non-degenerate species
& 
\\[2.5mm]
& $ \left[\dfrac{(g_1\h_2-g_2\h_1)^2}{16\p\a}\right] \, \dfrac{2^6 \z^4 \: e^{-4\z {\rm arccot}\z}}{3 (1+\z^2)^2}$
& 
\\[2.5mm] \hline 
  $\s\BSF^{\{210\}} \vrel \,/\, \s_c$
& $\!\!\left[\dfrac{(g_1\h_2+g_2\h_1)^2}{16\p\a}\right] 
   \!\!\dfrac{2^7 \z^6 (28+23\z^2)}{15(4+\z^2)^4} \, e^{-4\z {\rm arccot}(\z/2)}\!\!$
& 
\\[2.5mm] \hline 
  $\s\ann\vrel \, / \, \s_c$ 
& $1$ 
& $\dfrac{1}{2}$ 
\\[2.5mm] \hline
  $\G_{\{100\} \to \vf\vf}$ 
& $\mu \a^5$ 
& $\dfrac{\mu \a^5}{2}$ 
\\[2.5mm] \hline 
  $\G_{\{210\} \to \vf\vf}$ 
& $\dfrac{\mu \a^7 }{2^7 3}$ 
& 
\\[2.5mm] \hline
  $\G_{\{210\} \to \{100\}+\vf\!\!\!}$ 
& $\dfrac{2^7\mu\a^5}{3^7} \left[\dfrac{(g_1 \h_2+g_2\h_1)^2}{16\p\a}\right]$ 
& 
\\[2.5mm]  \hline
\end{tabular}
\caption{Summary of the bound-state-formation and annihilation cross-sections, the decay and de-excitation rates, computed in Sec.~\ref{Sec:Interactions}, for scalar particles interacting via a light scalar or vector mediator. The annihilation cross-sections and the rates of decay into $\varphi$-mediators refer to unbound and bound particle-antiparticle pairs respectively; all other formulae apply to any pair of particles. The cross-sections are normalised to $\sigma_c \equiv (\pi \alpha^2 / \mu^2) \times 2\pi\zeta/(1-e^{-2\pi\zeta})$, where $\alpha$ is the fine-structure constant entering the Coulomb potential, $\mu = m_1 m_2/(m_1+m_2)$ is the reduced mass of the interacting species, and $\zeta = \alpha/\vrel$, with $\vrel$ being the relative velocity of the incoming particles. Also, $\eta_{1,2} = m_{1,2}/(m_1+m_2)$, and $g_{1,2}$, $c_{1,2}$ are the couplings of the interacting particles to the force mediators, as described by the Lagrangian densities of Sec.~\ref{Sec:Interactions}. In our computations, we have neglected the mediator mass. Unitarity suggests that the range of validity of the above computations is $\alpha \lesssim 0.5$. }
\label{tab:CrossSections & Rates}
\end{table}

In this paper, we establish a field-theoretic framework for the calculation of BSF cross-sections and other level-transition rates, as well as the decay rates of unstable bound states. Then, we carry out computations for specific interactions. We organise our work as follows.

We begin, in Sec.~\ref{Sec:Wavefunctions}, with reviewing how to determine the wavefunctions of the two-particle states and the bound states in the presence of a long-range interaction. We derive the Bethe-Salpeter equation for the wavefunctions, and reduce it to the Schr\"{o}dinger equation using the instantaneous approximation in the non-relativistic regime. In Sec.~\ref{Sec:TransAmpl}, we determine the amplitudes for transitions between energy levels with the emission of a force mediator; this includes the radiative capture to a bound state. In the fully relativistic regime, we express the amplitudes for such processes in terms of the Bethe-Salpeter wavefunctions describing the initial and final states, and a perturbative interaction. We then employ the instantaneous and the non-relativistic approximations, to express the transition amplitudes in terms of the Schr\"{o}dinger wavefunctions. In Sec.~\ref{Sec:AnnDec}, we repeat the analysis for the decay of unstable bound states into radiation, as well as for the (co-)annihilation of unbound pairs of particles into radiation -- two closely related processes.  While we lay out our formalism in terms of scalar particles, it is straightforward to extend it to include fermionic species. 

\begin{figure}[t!]
\centering
\includegraphics[width=0.48\linewidth]{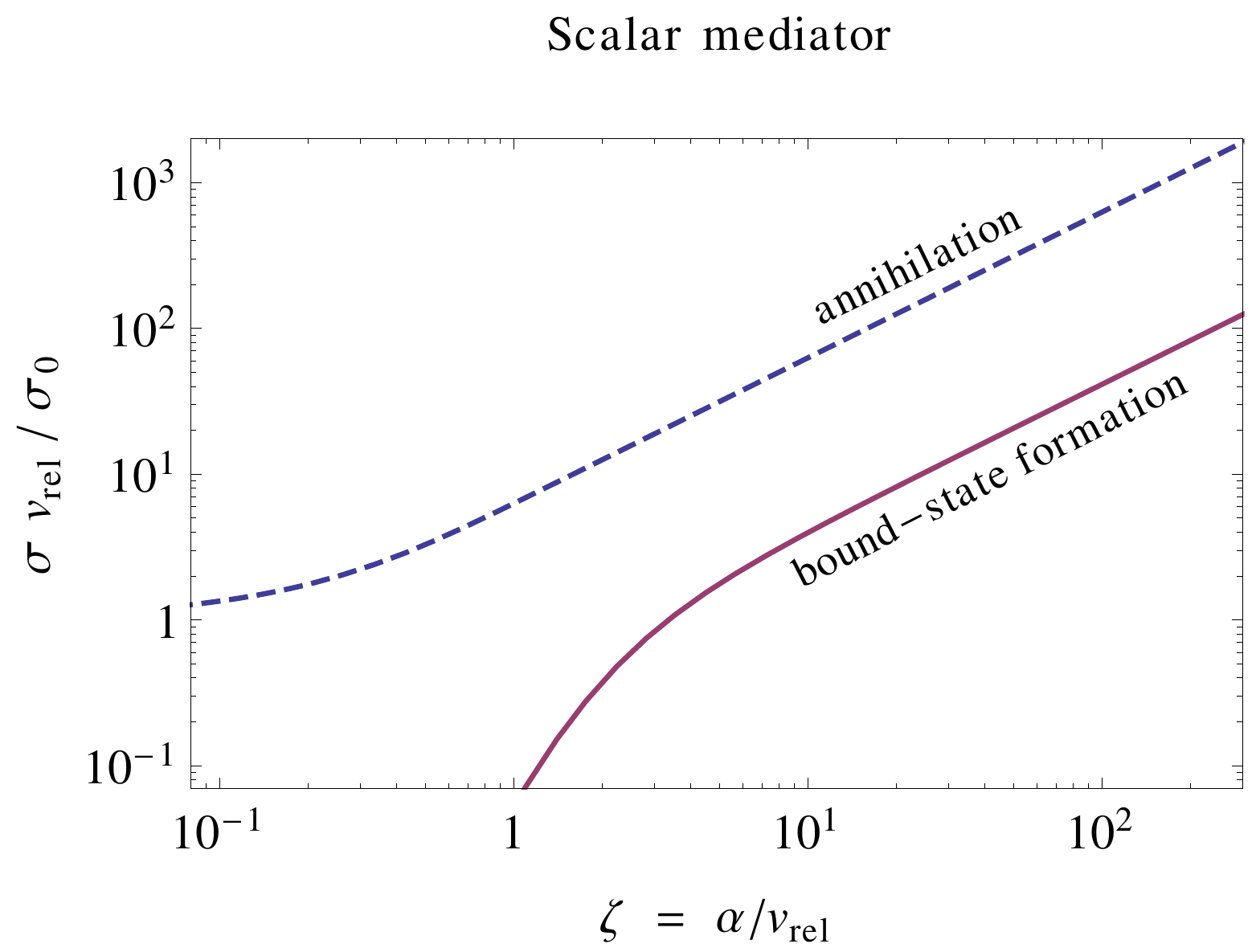}~~~
\includegraphics[width=0.48\linewidth]{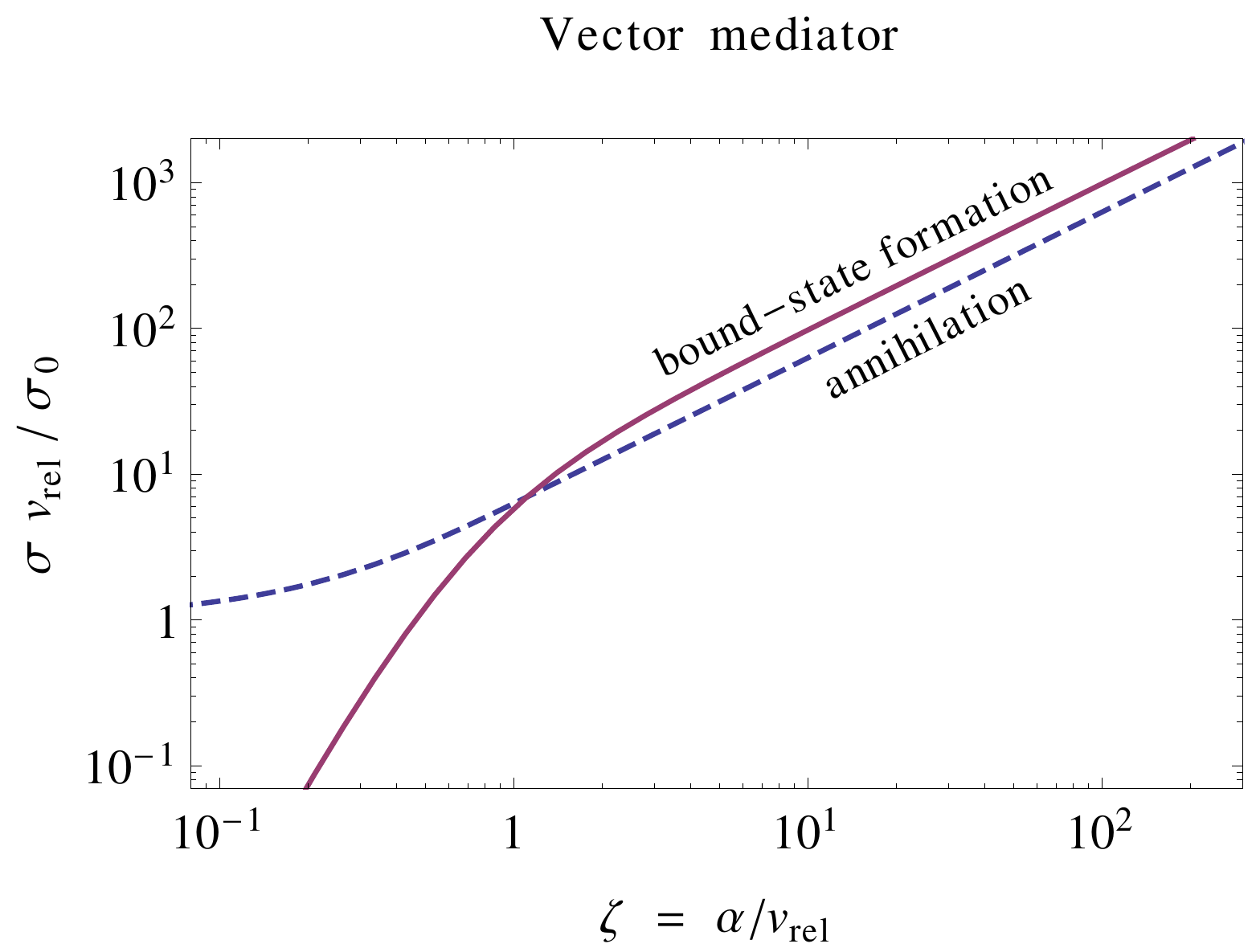}
\caption{Comparison of the cross-sections for annihilation (dashed blue lines) and the dominant radiative capture process to a bound state (solid purple lines), for a scalar particle-antiparticle pair. 
The cross-sections times relative velocity, $\sigma v_{\rm rel}$, are normalised to the perturbative value for $s$-wave annihilation, $\sigma_0$. 
\emph{Left}: For a scalar mediator, $\sigma_0 = \p\a^2/\mu^2$. The leading capture process is to the excited state $\{210\}$ and consists dominantly of the $J=2$ partial wave. At $\zeta \gg 1$, the ratio of the bound-state formation and annihilation cross-sections is $\sigma_{_{\rm BSF}}/\sigma_{\rm ann} \simeq 0.066$. 
\emph{Right}: For a vector mediator, $\sigma_0 = \pi\alpha^2/(2\mu^2)$. The leading capture process is to the ground state $\{100\}$ and consists dominantly of the $J=0$ and $J=2$ partial waves. At $\zeta \gg 1$, $\sigma_{_{\rm BSF}}/\sigma_{\rm ann} \simeq 1.56$. 
}
\label{fig:AnnVsBSF}
\end{figure}

We continue by applying our formalism to specific interactions. In Sec.~\ref{Sec:Interactions}, we consider scalar particles interacting either via a scalar or a vector mediator, and calculate the cross-sections for the dominant radiative capture to a bound state. We estimate the range of validity of our computations, using the unitarity bound on the inelastic cross-section. For bound states made of particle-antiparticle pairs or pairs of self-conjugate particles, we compare BSF with annihilation, and show that BSF can be the dominant inelastic process in the regime where the Sommerfeld effect is important; we sketch this comparison in Fig.~\ref{fig:AnnVsBSF}. In addition, we calculate the decay rates of particle-antiparticle bound states into force mediators. We  cast our results in terms of a minimal parametrisation, which makes their potential implications more transparent, and summarise them in table~\ref{tab:CrossSections & Rates}. We conclude with a discussion of the phenomenological implications of DM bound-state formation in Sec.~\ref{Sec:Conc}, and present many of the detailed calculations in the appendices.

The field-theoretic formalism developed in this work has several advantages in comparison to previous quantum mechanical calculations~\cite{BetheSalpeter_QM, AkhiezerMerenkov_sigmaHydrogen, MarchRussell:2008tu, Pearce:2013ola, Wise:2014jva}. It can accommodate the possibility of DM coupled to non-Abelian interactions. Such interactions can convert the incoming particles into different species which may subsequently form bound states; importantly, DM coupled to the electroweak interactions of the Standard Model belongs to this category. Moreover, the field-theoretic approach allows for a systematic inclusion of higher-order corrections, both in the interaction strength and in the momentum transfer between the interacting degrees of freedom.

We emphasise that BSF is distinct from processes such as the direct annihilation into mediators or elastic scattering, in which particles coupled to a long-range interaction may participate. While all these processes are influenced by the Sommerfeld effect, the final-state particles are obviously different. Field-theoretic treatments of the annihilation processes, analogous to the formalism presented here for BSF and discrete level transitions, have been presented in Refs.~\cite{Iengo:2009ni,Cassel:2009wt}.

\section{Bound-state and two-particle state wavefunctions \label{Sec:Wavefunctions}}

We shall consider two scalar particle species $\x_1$ and $\x_2$, interacting via a light or massless force mediator $\vf$. Specific interaction Lagrangians will be introduced in Sec.~\ref{Sec:Interactions} (c.f. \eqs{eq:Langr SSS r}, \eqref{eq:Langr SSS c} and \eqref{eq:Langr SSV}). In this section, we aim to determine the wavefunctions which describe two-particle states and bound states with the quantum numbers of $\x_1$ and $\x_2$. Our presentation draws largely from the pedagogical discussions of Refs.~\cite{Itzykson:1980rh, Silagadze:1998ri}.

Let us first introduce some notation. We denote the masses of $\x_1, \: \x_2$ and $\vf$ by $m_1, \: m_2$ and $m_\vf$ respectively. We define the total and the reduced $\x_1, \, \x_2$ masses, 
\begin{align}
m &\equiv m_1 + m_2 \, , \\
\m &\equiv \frac{m_1 \, m_2}{m_1+m_2} \: .
\end{align}
Obviously, $\mu\leqslant m/4$. In the following, $|{\cal B}_{\vec Q, n}\>$ stands for a $\x_1 - \x_2$ bound state, of total momentum $\vec Q$ and energy  $\w_{\vec Q, n} = \sqrt{\vec Q^2 +M_n^2}$, where $n$ denotes collectively all the discrete quantum numbers characterizing the bound state, and $M_n<m$ is the bound-state mass. $|{\cal U}_{\vec Q, \vec q} \>$ stands for a  $\x_1 - \x_2$ unbound two-particle state, with total momentum $\vec Q$, expectation value of relative velocity $\vec \vrel = \vec q/\mu$, and energy $\w_{\vec Q, \vec q} \geqslant m$. (As is common in scattering theory, we shall often refer to the unbound states as scattering states.) Moreover, $|\vf_{\vec Q}\>$ stands for a $\vf$ particle state with momentum $\vec Q$ and energy $\w_\vf = \sqrt{\vec Q^2 + m_\vf^2}$.

\subsection{The Bethe-Salpeter wavefunctions  \label{sec:BSW}}

We shall now introduce the Bethe-Salpeter wavefunctions that will appear in the fully relativistic version of the transition amplitudes of sections~\ref{Sec:TransAmpl} and \ref{Sec:AnnDec}. The Bethe-Salpeter wavefunctions are related to the more familiar Schr\"{o}dinger wavefunctions, which we will use in evaluating the amplitudes of interest in the non-relativistic regime.

We are interested in the following wavefunctions:
\begin{align}
\Psi_{\vec Q, n}(x_1, x_2) 
&\equiv  \langle \W|T \x_1 (x_1) \x_2(x_2) |{\cal B}_{\vec Q, n} \rangle \: , 
\label{eq:BSW bound}
\\
\Psi_{\vec Q, n}^\star(x_1, x_2) 
&\equiv  \langle {\cal B}_{\vec Q, n} |T \x_1^\dagger (x_1) \x_2^\dagger (x_2) | \W \rangle 
\label{eq:BSW* bound}
\end{align}
and
\begin{align}
\F_{\vec Q, \vec q}(x_1, x_2) 
&\equiv \langle \W|T \x_1 (x_1) \x_2(x_2) |{\cal U}_{\vec Q, \vec q} \rangle  \: , 
\label{eq:BSW free}
\\
\F_{\vec Q, \vec q}^\star(x_1, x_2) 
&\equiv \langle {\cal U}_{\vec Q, \vec q}| T \x_1^\dagger (x_1) \x_2^\dagger (x_2) |\W \rangle  \: ,
\label{eq:BSW* free}
\end{align}
where $T$ is the time-ordering operator, and $|\W\rangle$ is the vacuum of the interacting theory. 
If $\x_1, \x_2$ are a particle-antiparticle pair, in the above definitions we replace $\x_1 \to \x, \: \x_2 \to \x^\dagger$.
Note that ${}^\star$ does not denote complex conjugation, for which we shall use the symbol ${}^*$, as usual. In fact
\begin{align}
\Psi_{\vec Q, n}^\star(x_1, x_2) = \<\W|\bar T \x_1(x_1) \x_2(x_2)|{\cal B}_{\vec Q, n}\>^* \: , 
\label{eq:Psi star=Tbar cc}
\\
\F_{\vec Q, \vec q}^\star(x_1, x_2) = \<\W|\bar T \x_1(x_1) \x_2(x_2)|{\cal U}_{\vec Q, \vec q}\>^* \: ,
\label{eq:Phi star=Tbar cc}
\end{align}
where $\bar T$ is the anti-time-ordering operator.

We define the coordinate transformation and its inverse
\begin{align}
x \equiv x_1 - x_2 \, ,    &\qquad  X \equiv \h_1 x_1 + \h_2 x_2 \, ,
\label{eq:x,X}
\\
x_1 \equiv X + \h_2 x \, , &\qquad x_2 \equiv X - \h_1 x \, ,
\label{eq:x1,x2}
\end{align}
where $\h_1 + \h_2 =1$ for the Jacobian to be 1, and we choose specifically
\beq
\h_{1,2} = \frac{m_{1,2}}{m_1+m_2} \: .
\label{eq:eta}
\eeq
In the non-relativistic regime, this choice will enable us to separate the motion of the CM from the relative motion.
Using the 4-momentum operator $\hat P$, we obtain 
\begin{align}
\x_1 (x_1) &= \exp(i \hat P  X) \x_1 (\h_2 x) \exp(-i \hat P X) \: , \\
\x_2 (x_2) &= \exp(i \hat P  X) \x_2 (-\h_1 x) \exp(-i \hat P X) \: .
\end{align}
Then, the wavefunction of \eqs{eq:BSW bound} becomes
\begin{align}
&\Ps_{\vec Q, n}(x_1, x_2)  = 
\th(x^0 ) \< \W | \x_1(x_1) \x_2 (x_2) | {\cal B}_{\vec Q, n} \> 
+ \th(-x^0 ) \< \W |\x_2(x_2) \x_1 (x_1)| {\cal B}_{\vec Q, n} \>
\nn \\
&= \th(x^0 ) \< \W | e^{i \hat P X} \x_1(\h_2 x) \x_2 (-\h_1 x) e^{-i \hat P X} | {\cal B}_{\vec Q, n} \>
+ \th(-x^0 ) \< \W | e^{i \hat P X} \x_2(-\h_1 x) \x_1 (\h_2 x) e^{-i \hat P X}| {\cal B}_{\vec Q, n} \>
\nn \\
&= e^{-i Q X} \< \W | T \x_1(\h_2 x) \x_2 (-\h_1 x) | {\cal B}_{\vec Q, n} \>
\nn \\
&\equiv e^{-i Q X} \Psi_{\vec Q,n} (x) \: ,
\end{align}
where it is understood that $Q^0 = \w_{\vec Q, n}$, and we defined
\beq
\Psi_{\vec Q, n} (x) \equiv \< \W | T \x_1(\h_2 x) \x_2 (-\h_1 x) | {\cal B}_{\vec Q, n} \>  \: .
\eeq
This is the first step in separating the motion of the CM from the relative motion. For notational simplicity, we are using the same symbol for $\Psi_{\vec Q, n} (x_1,x_2)$ and $\Psi_{\vec Q, n} (x)$.
We also define the Fourier transforms
\begin{align}
\Psi_{\vec Q,n} (x) &\equiv \int \frac{d^4 p}{(2\p)^4} \, \tilde \Psi_{\vec Q,n} (p) \, e^{- i p x} 
\:, \qquad
\tilde \Psi_{\vec Q,n} (p) \equiv \int d^4 x \, \Psi_{\vec Q,n} (x) \, e^{i p x}
\: .
\end{align}
We repeat the above for the amplitudes of \eqs{eq:BSW* bound} -- \eqref{eq:BSW* free}, and summarise the definitions in appendix~\ref{App:BSW def}.

\subsection{The 4-point Green's function and Dyson-Schwinger equation \label{sec:DE}}

\begin{figure}[t]
\centering
\begin{tikzpicture}[line width=1.5 pt, scale=1.5]
\begin{scope}
\node at (-.15,1){$x_1$};
\node at (-.15,0){$x_2$};
\draw[violet] (0,1)--(1,1);
\draw[purple] (0,0)--(1,0);
\node at (1.15,1){$y_1$};
\node at (1.15,0){$y_2$};
\draw[fill=lightgray,draw=none] (0.2,-0.1) rectangle (0.8,1.1);
\node at (0.5,0.5){$G^{(4)}$};
\node at (1.5,0.5){$=$};
\end{scope}
\begin{scope}[shift={(2.05,0)}]
\node at (-.15,1){$x_1$};
\node at (-.15,0){$x_2$};
\draw[violet] (0,1)--(.7,1);
\draw[purple] (0,0)--(.7,0);
\node at (.85,1){$y_1$};
\node at (.85,0){$y_2$};
\draw[draw=none,fill=violet] (.35,1) circle (0.08cm);
\draw[draw=none,fill=purple] (.35,0) circle (0.08cm);
\node at (1.2,0.5){$+$};
\end{scope}
\begin{scope}[shift={(3.75,0)}]
\node at (-.15,1){$x_1$};
\node at (-.15,0){$x_2$};
\draw[violet] (0,1)--(1.2,1);
\draw[purple] (0,0)--(1.2,0);
\node at (1.35,1){$y_1$};
\node at (1.35,0){$y_2$};
\draw[fill=white] (.6,.5) ellipse (0.3cm and 0.6cm);
\draw[draw=none,fill=violet] (.2,1) circle (0.08cm);
\draw[draw=none,fill=purple] (.2,0) circle (0.08cm);
\draw[draw=none,fill=violet]  (1,1) circle (0.08cm);
\draw[draw=none,fill=purple]  (1,0) circle (0.08cm);
\node at (0.6,0.5){$W$};
\node at (1.7,0.5){$+$};
\end{scope}
\begin{scope}[shift={(5.95,0)}]
\node at (-.15,1){$x_1$};
\node at (-.15,0){$x_2$};
\draw[violet] (0,1)--(2.2,1);
\draw[purple] (0,0)--(2.2,0);
\node at (2.35,1){$y_1$};
\node at (2.35,0){$y_2$};
\draw[fill=white] (.6,.5) ellipse (0.3cm and 0.6cm);
\node at (.6,.5){$W$};
\draw[fill=white] (1.6,.5) ellipse (0.3cm and 0.6cm);
\node at (1.6,.5){$W$};
\draw[draw=none,fill=violet] (0.2,1) circle (0.08cm);
\draw[draw=none,fill=purple] (0.2,0) circle (0.08cm);
\draw[draw=none,fill=violet] (1.1,1) circle (0.08cm);
\draw[draw=none,fill=purple] (1.1,0) circle (0.08cm);
\draw[draw=none,fill=violet]   (2,1) circle (0.08cm);
\draw[draw=none,fill=purple]   (2,0) circle (0.08cm);
\node at (2.9,0.5){$+ \: \cdots$};
\end{scope}
%
\node at (1.5,-1.5){$=$};
\begin{scope}[shift={(2.05,0)}]
\node at (-.15,-1){$x_1$};
\node at (-.15,-2){$x_2$};
\draw[violet] (0,-1)--(.7,-1);
\draw[purple] (0,-2)--(.7,-2);
\node at (.85,-1){$y_1$};
\node at (.85,-2){$y_2$};
\draw[draw=none,fill=violet] (.35,-1) circle (0.08cm);
\draw[draw=none,fill=purple] (.35,-2) circle (0.08cm);
\node at (1.2,-1.5){$+$};
\end{scope}
\begin{scope}[shift={(3.75,0)}]
\node at (-.15,-1){$x_1$};
\node at (-.15,-2){$x_2$};
\draw[violet] (0,-1)--(1.9,-1);
\draw[purple] (0,-2)--(1.9,-2);
\node at (2.05,-1){$y_1$};
\node at (2.05,-2){$y_2$};
\draw[draw=none,fill=violet] (.2,-1) circle (0.08cm);
\draw[draw=none,fill=purple] (.2,-2) circle (0.08cm);
\draw[fill=white] (.6,-1.5) ellipse (0.3cm and 0.6cm);
\node at (.6,-1.5){$W$};
\draw[fill=lightgray,draw=none] (1.1,-2.1) rectangle (1.7,-0.9);
\node at (1.4,-1.5){$G^{(4)}$};
\end{scope}
\end{tikzpicture}
\caption{Diagrammatic representation of the Dyson-Schwinger equation~\eqref{eq:DE symb} for the 4-point function $G^{(4)}(x_1,x_2,y_1,y_2)$. \protect\PropV and \protect\PropP stand for the $\x_1$ and $\x_2$ full propagators, respectively.}
\label{fig:Dyson-Schwinger}
\end{figure}
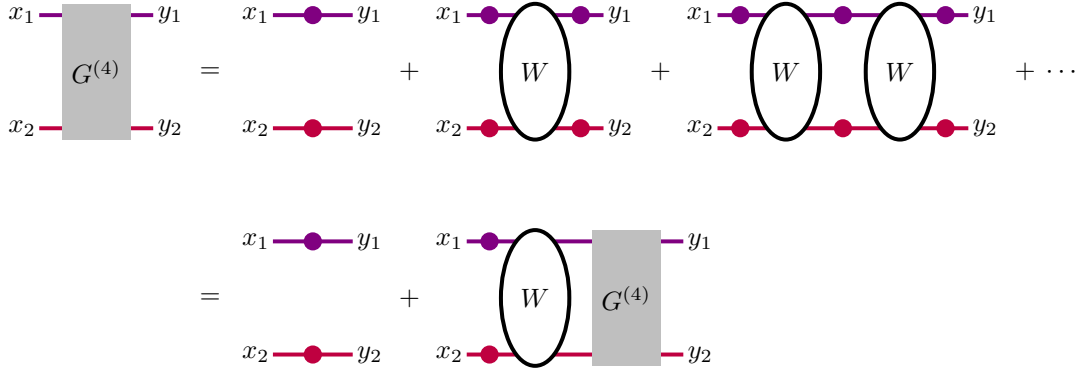

Consider the 4-point Green's function
\beq
G^{(4)}(x_1,x_2,y_1,y_2) = 
\langle \W |T \x_1(x_1) \x_2(x_2) \x_1^\dagger (y_1) \x_2^\dagger(y_2) |\W \rangle \: ,
\label{eq:4pGreens}
\eeq
and let $W(x_1,x_2,y_1,y_2)$ be the perturbative 4-point interaction kernel between $\x_1$ and $\x_2$. Then, $G^{(4)}$ satisfies the Dyson-Schwinger equation,
\begin{multline}
G^{(4)}(x_1,x_2,y_1,y_2) = S_1(x_1 - y_1) S_2(x_2 - y_2) + \\
\int d^4 z_1 \, d^4 z_1' \, d^4 z_2 \, d^4z_2' \:  S_1(x_1 - z_1) S_2(x_2 - z_2) 
W(z_1,z_2;z_1',z_2')  \:  S_1(z_1' - y_1) S_2(z_2' - y_2) 
+ \dots \: ,
\label{eq:DE}
\end{multline}
where $S_1, S_2$ are the full propagators for $\x_1, \x_2$.
Symbolically, the above series can be written as
\begin{align}
G^{(4)} 
&= S_1 S_2 + S_1 S_2 W S_1 S_2 + S_1 S_2 W S_1 S_2 W S_1 S_2 +  \dots \nn \\
&= S_1 S_2 + S_1 S_2 W \(S_1 S_2 + S_1 S_2 W S_1 S_2 +  \dots \) \nn \\
&= S_1 S_2 + S_1 S_2 W G^{(4)} .
\label{eq:DE symb}
\end{align}
Equation~\eqref{eq:DE symb} is sketched in Fig.~\ref{fig:Dyson-Schwinger}.

Due to translational invariance, $W$ and $G^{(4)}$ depend only on coordinate differences. We shall take them to be $x, y, X-Y$, where we used the definitions of \eq{eq:x,X} and assumed analogous definitions for the $y_1, y_2$ variables. Thus
\begin{align}
G^{(4)}(x_1, x_2, y_1, y_2) &= G^{(4)} (x,y; X-Y) \: , \\
W(x_1, x_2, y_1, y_2) &= W (x,y; X-Y) \: ,
\end{align}
where we retained the same symbols to keep the notation simple.
Equation~\eqref{eq:DE} becomes
\begin{align}
G^{(4)} (x,y; X-Y) 
&= S_1 \[ X-Y + \h_2(x-y) \]  S_2 \[ X-Y - \h_1(x-y) \] \nn \\
&+ \int d^4 z \, d^4 Z \, d^4 z' \, d^4 Z' 
\: S_1 \[ X-Z + \h_2(x-z) \]  S_2 \[ X-Z - \h_1(x-z) \] \nn \\
&\times W(z,z';Z-Z') G^{(4)}(z',y;Z'-Y)
\label{eq:DE coord}
\end{align}
We define the Fourier transforms of $G, W, S_1$ and $S_2$,
\begin{align}
\tilde G^{(4)}(p,p';Q) &\equiv
\int d^4 x \, d^4 y \, d^4 (X-Y) \: G^{(4)} (x,y;X-Y)  \: \exp(ipx -ip'y) \: \exp\[i Q (X-Y)\]  \: ,
\label{eq:G_Fourier}
\\
\tilde W(p,p';Q) &\equiv
\int d^4 x \, d^4 y \, d^4 (X-Y) \: W (x,y;X-Y)  \: \exp(ipx -ip'y) \: \exp\[i Q (X-Y)\]  \: ,
\label{eq:W_Fourier}
\end{align}
and
\beq
\tilde S_j(p) = \int d^4 z \: e^{i p z} \: S_j (z) \: ,
\label{eq:S FT}
\eeq
with $\tilde S_j (p)$ being the momentum-space propagator for $\x_j$.
From the above, we deduce the relation between the conjugate momenta of $x_1, x_2$, which we shall call here $p_1, p_2$, and the conjugate momenta of $x, X$, denoted above as $p, Q$:
\begin{align}
Q = p_1 + p_2, \qquad &p = \h_2 p_1 - \h_1 p_2 \: ,
\\
p_1 = \h_1 Q + p, \qquad &p_2 = \h_2 Q - p \: .
\end{align}
Analogous relations hold between the conjugate momenta of $y_1, y_2$ and those of $y,Y$.

For convenience, we also define
\beq
S(p;Q) \equiv \tilde S_1(\h_1 Q+p) \: \tilde S_2(\h_2 Q -p) \: .
\label{eq:S}
\eeq
We may now rewrite the Dyson-Schwinger \eq{eq:DE coord} for the 4-point function, in momentum space
\beq
\tilde G^{(4)}(p,p';Q) 
= (2\p)^4 \d^4 (p-p') \: S(p;Q) 
+ S(p;Q) \int \frac{d^4 k}{(2\p)^4} \: \tilde{W} (p,k;Q) \: \tilde{G}^{(4)}(k,p';Q) \: .
\label{eq:DE mom}
\eeq
We shall use \eq{eq:DE mom} to derive the Bethe-Salpeter equation for the wavefunctions of Sec.~\ref{sec:BSW}.

\subsection{Completeness relation and decomposition of the 4-point function \label{sec:Completeness}}

To compute the Bethe-Salpeter wavefunctions of Sec.~\ref{sec:BSW}, we have to decompose the 4-point Green's function of Sec.~\ref{sec:DE} using the one- and two-particle completeness relation. Then, \eq{eq:DE mom} will yield the equations which the wavefunctions $\Psi_{\vec Q, n}$ and $\F_{\vec Q, \vec q}$ satisfy.

Including the one- and two-particle states with the same quantum numbers as $\x_1$ and $\x_2$, the completeness relation is
\beq
\mathbf{1}
= \sum_n \int \frac{d^3Q}{(2\p)^3 \: 2\w_{\vec Q, n}} |{\cal B}_{\vec Q, n} \>  \< {\cal B}_{\vec Q, n}|
+ \int \frac{d^3 q}{(2\p)^3} \frac{d^3 Q}{(2\p)^3} \frac{1}{2 \w_{\vec Q, \vec q} \, 2\ve_{\vec Q, \vec q}} 
|{\cal U}_{\vec Q, \vec q} \> \< {\cal U}_{\vec Q, \vec q}|  \: ,
\label{eq:complete}
\eeq
where we have assumed the standard relativistic normalization of one-particle momentum eigenstates $\<\vec p|\vec k\> = 2E_{\vec p} \,(2\p)^3 \d^3(\vec p - \vec k)$, with $E_{\vec p}$ being the energy of the state $|\vec p\>$. To lowest (zeroth) order in the interaction strength, 
\beq 
2 \w_{\vec Q, \vec q} 2\ve_{\vec Q, \vec q} \simeq 2E_1(\vec q;\vec Q) \, 2E_2(\vec q;\vec Q) \: ,
\label{eq:2omega2epsilon}
\eeq
where
\beq
E_1(\vec q;\vec Q) \equiv \sqrt{(\h_1 \vec Q + \vec q)^2 + m_1^2} \: , \qquad
E_2(\vec q;\vec Q) \equiv \sqrt{(\h_2 \vec Q - \vec q)^2 + m_2^2} \: .
\label{eq:E1,E2}
\eeq

Next, we insert the unity operator of \eq{eq:complete} in $G^{(4)}$, to obtain the decomposition
\beq
G^{(4)}(x,y;X-Y) =
\sum_n G_n^{(4)} (x,y;X-Y) +  G_{\cal U}^{(4)} (x,y;X-Y) \: ,
\label{eq:G=Gn+GU}
\eeq
where $G_n^{(4)} (x,y;X-Y)$ and $G_{\cal U}^{(4)} (x,y;X-Y)$ are the contributions of the bound and the scattering states, respectively. 
We compute them below. We shall make use of the fact that a non-zero contribution to $G^{(4)}$ from a one- or two-particle state arises only when two annihilation operators act on that state to obtain the quantum numbers of the vacuum. Moreover, in order to extract the poles and the branch-cuts of $G_n^{(4)}$ and $G_{\cal U}^{(4)}$, we will use the integral representation of the $\th$-function,
\beq
\th(z) = \frac{i}{2\p} \int_{-\infty}^{\infty} dk \: \frac{e^{-i k z}}{k+i \e} \: ,
\label{eq:theta fun}
\eeq
and
\beq
\th \[ \min(x_1^0,x_2^0) - \max (y_1^0, y_2^0) \] = \th \[ X^0 - Y^0 + h_-(x^0) -h_+(y^0) \] \: ,
\label{eq:theta res}
\eeq
where
\beq 
h_{\pm}(x^0) \equiv \frac{1}{2} (\h_2 - \h_1) x^0 \pm \frac{1}{2}|x^0| \: .  
\label{eq:h} 
\eeq
(For \eqs{eq:theta res} and \eqref{eq:h}, see Ref.~\cite{Silagadze:1998ri} and appendix~\ref{App:h}.)

\subsubsection*{Contribution of the bound states to the 4-point function}

The contribution of the $n$th bound state to $G^{(4)}$ is
\begin{align}
&G_n^{(4)} (x,y;X-Y) =
\nn \\
&= \int \frac{d^3 K}{(2\p)^3} \frac{1}{2\w_{\vec K, n}}
\< \W | T \x_1(x_1) \x_2 (x_2) | {\cal B}_{\vec K, n}  \>
\< {\cal B}_{\vec K, n} | T \x_1^\dagger(y_1) \x_2^\dagger (y_2) | \W \>  
\: \th \[ \min(x_1^0,x_2^0) - \max (y_1^0, y_2^0) \]  \nn \\
&= \int \frac{d^3 K}{(2\p)^3} \frac{1}{2\w_{\vec K, n}}
\Psi_{\vec K, n} (x) \Psi_{\vec K, n}^\star (y)
e^{-i \w_{\vec K, n} (X^0 - Y^0)}  e^{i \vec K \cdot (\vec X - \vec Y)} 
\: \th \[ X^0 - Y^0 + h_-(x^0) - h_+(y^0) \]  \nn \\
&= \int \frac{d^3 K}{(2\p)^3} \frac{1}{2\w_{\vec K, n}}
\Psi_{\vec K, n} (x) \Psi_{\vec K, n}^\star (y)
e^{-i \w_{\vec K, n} (X^0 - Y^0)}  e^{i \vec K \cdot (\vec X - \vec Y)}   \nn \\
&\times \frac{i}{2\p} \int_{-\infty}^{\infty} dK^0 \ 
\frac{\exp \left\{ -i \[K^0 - \w_{\vec K, n}\] \, \[X^0 - Y^0 + h_-(x^0) - h_+(y^0)\] \right\}}{K^0 - \w_{\vec K, n} + i \e}  \nn \\
&= i \int \frac{d^4 K}{(2\p)^4} \ e^{-i K (X - Y)}   
\ \Psi_{\vec K, n} (x) \: \Psi_{\vec K, n}^\star (y)
\ \frac{\exp \left\{ -i \[K^0 - \w_{\vec K, n}\]  [h_-(x^0) - h_+(y^0)] \right\}}{2\w_{\vec K, n} \[K^0 - \w_{\vec K, n} + i \e\] } \: , \nn
\end{align}
where in the third step we made use of the integral representation of the $\th$ function, given in \eq{eq:theta fun}, which introduces the integration over $K^0$. The Fourier transform of the above is 
\begin{multline}
\tilde G_n^{(4)} (p,p';Q)
=  \int d^4x \, d^4y \, d^4(X-Y) \: e^{i px - i p'y + i Q(X-Y)} \: G_n^{(4)}(x,y;X-Y)  
\\
=i \int d^4x \, d^4y \ e^{ipx - i p'y} \ \Psi_{\vec Q, n} (x) \Psi_{\vec Q, n}^\star (y) 
\: \frac{\exp \left\{ -i \[Q^0 - \w_{\vec Q, n}\] [h_-(x^0) - h_+(y^0)] \right\}}{2\w_{\vec Q, n} \[Q^0 - \w_{\vec Q, n} + i \e\] } \: . 
\label{eq:Gn}
\end{multline}
At $Q^0 \to \w_{\vec Q, n}$, this becomes
\beq
\tilde G_n^{(4)} (p,p';Q)  \to  \frac{ i \tilde \Psi_{\vec Q, n} (p)  \tilde \Psi_{\vec Q, n}^\star (p')} {2\w_{\vec Q, n} \[Q^0 - \w_{\vec Q, n} + i \e\] } \: .
\label{eq:pole contr}
\eeq
Equation~\eqref{eq:Gn} is the contribution of the $n$th bound state to $\tilde {G}^{(4)}(p,p';Q)$. Evidently, the scattering amplitude has a pole at energy equal to the bound-state energy.

\subsubsection*{Contribution of two-particle scattering states to the 4-point function}

Following similar steps, we find the contribution of the two-particle states to $G^{(4)}$,
\begin{multline}
G_{\cal U}^{(4)}(x,y;X-Y) = 
\nn \\
= i \int \frac{d^3k}{(2\p)^3} \: \frac{d^4K}{(2\p)^4} 
\ e^{-i K(X-Y)} \ \F_{\vec K, \vec k} (x) \: \F_{\vec K, \vec k}^\star(y)
\ \frac{\exp \left\{-i \[K^0 - \w_{\vec K, \vec k}\] [h_-(x^0) - h_+(y^0)]\right\} }{2 \w_{\vec K, \vec k} \, 2\ve_{\vec K, \vec k} \: \[K^0 - \w_{\vec K, \vec k} +i \e\]} \: .
\end{multline}
The Fourier transform of $G_{\cal U}^{(4)}(x,y;X-Y)$ is
\begin{multline}
\tilde G_{\cal U}^{(4)} (p,p';Q) =
\\
= i \int \frac{d^3q}{(2\p)^3} \int d^4x \, d^4y \ e^{ipx-ip'y} 
\ \F_{\vec Q, \vec q} (x) \: \F_{\vec Q, \vec q}^\star(y) 
\ \frac{\exp \left\{-i \[Q^0 - \w_{\vec Q, \vec q}\] [h_-(x^0) - h_+(y^0)] \right\} }{2 \w_{\vec Q, \vec q} \, 2\ve_{\vec Q, \vec q} \: \[Q^0 - \w_{\vec Q, \vec q} +i \e\]} \: .
\label{eq:GU}
\end{multline}
Clearly, the contribution of the two-particle states to $\tilde G^{(4)}(p,p';Q)$ gives rise to a branch-cut in the scattering amplitude.

\medskip

Summing the contributions from the bound and the scattering states, we obtain the decomposition of the momentum-space 4-point function
\beq
\tilde G^{(4)}(p,p';Q) =
\sum_n \tilde G_n^{(4)} (p,p';Q) + \tilde G_{\cal U}^{(4)} (p,p';Q) \: .
\label{eq:G=Gn+GU mom}
\eeq
We shall now combine the Dyson-Schwinger \eq{eq:DE mom} and \eq{eq:G=Gn+GU mom}, to obtain the Bethe-Salpeter equation for the wavefunctions $\Ps_{\vec Q, n}$ and $\F_{\vec Q, \vec q}$.

\subsection{The Bethe-Salpeter equation for bound and scattering states \label{sec:BSE}}

We introduce the operator
\beq
A(p,p';Q) \equiv \frac{(2\p)^4 \d^4(p-p')}{S(p;Q)} -  \tilde W(p,p';Q) \: .
\label{eq:A def}
\eeq
Then, the Dyson-Schwinger \eq{eq:DE mom} can be cast in the form
\beq
\int \frac{d^4k}{(2\p)^4}  \: A(p,k;Q) \: \tilde G^{(4)} (k,p';Q) = (2\p)^4\d^4 (p-p') \: .
\label{eq:DE mod}
\eeq
This is formally solved by 
\beq
\tilde G^{(4)} (p,p';Q) = 
\sum_n \frac{1}{c_n(Q)} \: C_n(p;Q) C_n^\dagger (p';Q)
+ \int \frac{da}{f_a(Q)} \: F_a(p;Q) F_a^\dagger (p';Q) \: ,
\label{eq:G sol}
\eeq
where $C_n(p;Q)$ and $F_a(p;Q)$ are the eigenfunctions of the discrete and the continuous spectrum of the operator $A(p,q;Q)$, with eigenvalues $c_n(Q)$ and $f_a(Q)$ respectively,
\begin{align}
\int \frac{d^4 k}{(2\p)^4} \: A(p,k;Q) C_n(k;Q) &= c_n(Q) \, C_n(p;Q) \: ,
\label{eq:eigenvec disc}
\\
\int \frac{d^4 k}{(2\p)^4} \: A(p,k;Q) F_a(k;Q) &= f_a(Q) \, F_a(p;Q) \: ,
\label{eq:eigenvec cont}
\end{align}
normalised according to
\beq
\sum_n C_n(p;Q) C_n^\dagger(p';Q)  +
\int da \: F_a(p;Q) F_a^\dagger (p';Q) = (2\p)^4 \, \d^4(p-p') \: .
\label{eq:eigenvec norm}
\eeq

We may now collect \eqs{eq:Gn}, \eqref{eq:GU}, \eqref{eq:G=Gn+GU mom}, and \eqref{eq:G sol}. Matching the various contributions between \eqref{eq:G=Gn+GU mom} and \eqref{eq:G sol}, we deduce the following.
For the discrete spectrum:
\begin{align}
C_n(p;Q) &\propto \int d^4 x 
\: \Psi_{\vec Q, n}(x) \, e^{ipx} 
\, e^{-i[Q^0-\w_{\vec Q, n}] \, h_-(x^0) }  \: , 
\label{eq:Cn}
\\
C_n^\dagger(p';Q) &\propto \int d^4 y 
\: \Psi_{\vec Q, n}^\star(y) \, e^{-ip'y} 
\, e^{i[Q^0-\w_{\vec Q, n}] \,  h_+(y^0) }  \: , 
\label{eq:Cn dagger}
\\
c_n(Q) &\propto 1-\w_{\vec Q, n}/Q^0  \: .
\label{eq:cn}
\end{align}
For the continuous spectrum, we identify $a \to \vec q$, and deduce
\begin{align}
F_a(p;Q) &\propto \int d^4 x 
\: \F_{\vec Q, \vec q}(x) \, e^{ipx} 
\, e^{-i[Q^0-\w_{\vec Q, n}] \,  h_-(x^0) }  \: ,
\label{eq:Fa} 
\\
F_a^\dagger (p';Q) &\propto \int d^4 y 
\: \F_{\vec Q, \vec q}^\star(x) \, e^{-ip'y} 
\, e^{i[Q^0-\w_{\vec Q, n}] \,  h_+(y^0) }  \: , 
\label{eq:Fa dagger}
\\
f_a(Q) &\propto 1-\w_{\vec Q, \vec q}/Q^0  \: .
\label{eq:fa}
\end{align}
The relations \eqref{eq:Cn}, \eqref{eq:Cn dagger}, \eqref{eq:Fa} and \eqref{eq:Fa dagger} are stipulated because $c_n, f_a$ are independent of the momenta $p,p'$; all the $p,p'$-dependent factors must arise from the eigenfunctions, $C_n$ and $F_a$. The relations \eqref{eq:cn}, \eqref{eq:fa} are warranted so that $C_n$ and $F_a$ are not singular in the limit $Q^0 \to \w_{\vec Q, n}$ and $Q^0 \to \w_{\vec Q, \vec q}$ respectively; the factors $[1-\w_{\vec Q, n}/Q^0]^{-1}, \: [1-\w_{\vec Q, \vec q} /Q^0]^{-1}$ cannot be part of the eigenfunctions, and thus belong to the eigenvalues.

Inserting the above into the eigenvalue equations \eqref{eq:eigenvec disc} and \eqref{eq:eigenvec cont}, and taking the limits $Q^0 \to \w_{\vec Q, n}$ and $Q^0 \to \w_{\vec Q, \vec q}$ respectively, we obtain the  Bethe-Salpeter equations for the bound and the scattering states\footnote{It is possible to obtain \eq{eq:BSE bound} more easily, by taking the residue of both sides of \eq{eq:G=Gn+GU mom} at $Q^0 \to \w_{\vec Q, n}$. However, this is not possible for the two-particle states.}
\beq
\tilde \Psi_{\vec Q, n} (p) =  S(p;Q)
\int \frac{d^4 k}{(2\p)^4} \ \tilde{W}(p,k;Q) \ \tilde \Psi_{\vec Q, n} (k)
\: ,
\label{eq:BSE bound}
\eeq
\beq
\tilde \F_{\vec Q, \vec q} (p) =  S(p;Q)
\int \frac{d^4 k}{(2\p)^4} \ \tilde{W}(p,k;Q) \ \tilde \F_{\vec Q, \vec q} (k)
\: .
\label{eq:BSE free}
\eeq
These are homogeneous equations and do not determine the normalisation of $\Psi_{\vec Q, n}$ and $\F_{\vec Q, \vec q}$. Moreover, because we do not know the exact eigenvalues $c_n$ and $f_a$, we cannot use \eq{eq:eigenvec norm} to obtain the normalisation of $\Psi_{\vec Q, n}$ and $\F_{\vec Q, \vec q}$. We derive their normalisation in the next section.

\subsection{Normalization of the Bethe-Salpeter wavefunctions \label{sec:Norm BS}}

We derive the normalisation of the wavefunctions $\Psi_{\vec Q, n}$ and $\F_{\vec Q, \vec q}$ from the inhomogeneous Dyson-Schwinger \eq{eq:DE mom}, or equivalently from \eq{eq:DE mod}, using the method described in Ref.~\cite{Silagadze:1998ri}.

We define the symbolic multiplication 
\beq
\[ {\cal O}_1 \: {\cal O}_2 \] (p,p';Q) \equiv
\int \frac{d^4k}{(2\p)^4} \: {\cal O}_1 (p,k;Q) \, {\cal O}_2(k,p';Q) \: ,
\eeq
and the unity operator $I(p,p') \equiv (2\p)^4 \, \d^4(p-p')$.
Then, \eq{eq:DE mod} can be expressed in symbolic form
\beq
A \: \tilde G^{(4)}  =  \tilde G^{(4)} \, A  = I \: .
\label{eq:DE mod symb}
\eeq
We differentiate \eq{eq:DE mod symb} over $Q^0$ and re-use it, to obtain
\beq
\tilde G^{(4)} \: \frac{dA}{dQ^0} \: \tilde G^{(4)} = -\frac{d \tilde G^{(4)}}{dQ^0} \: .
\label{eq:norm eq}
\eeq
We shall use \eq{eq:norm eq} to obtain the normalisation of the Bethe-Salpeter wavefunctions.
For later convenience, we define
\begin{align}
\tilde N_n(p,p';\vec Q) &\equiv 
i \[\frac{d A (p,p';Q)}{dQ^0}\]_{Q^0 = \w_{\vec Q, n}} \: , 
\label{eq:Nn tilde}
\\
\tilde N_{\vec q}(p,p';\vec Q) &\equiv 
i \[\frac{d A (p,p';Q)}{dQ^0}\]_{Q^0 = \w_{\vec Q, \vec q}} \: ,
\label{eq:Nq tilde}
\end{align}
and their Fourier transforms,
\begin{align}
N_n(x,x';\vec Q) &\equiv i \[ \frac{d}{dQ^0} \int \frac{d^4 p}{(2\p)^4} \frac{d^4 p'}{(2\p)^4} 
\: e^{- i p x} \: A(p,p';Q) \: e^{i p' x'}
\]_{Q^0 = \w_{\vec Q, n}} \: , 
\label{eq:Nn}
\\
N_{\vec q}(x,x';\vec Q) &\equiv i  
\[ \frac{d}{dQ^0} \int \frac{d^4 p}{(2\p)^4} \frac{d^4 p'}{(2\p)^4} 
\: e^{- i p x} \: A(p,p';Q) \: e^{i p' x'}
\]_{Q^0 = \w_{\vec Q, \vec q}} \: .
\label{eq:Nq}
\end{align}

\subsubsection*{Bound states}

Substituting the contribution to the 4-point function from the $n$th bound state, given in \eq{eq:pole contr}, into \eq{eq:norm eq}, and taking the limit $Q^0 \to \w_{\vec Q, n}$, we obtain the normalisation condition
\beq
\int \frac{d^4 p}{(2\p)^4} \, \frac{d^4 p'}{(2\p)^4} 
\: \tilde\Psi_{\vec Q, n}^\star (p) 
\: \tilde N_n(p,p';\vec Q) 
\: \tilde\Psi_{\vec Q, n'} (p')
= 2 \w_{\vec Q, n} \, \d_{nn'} \: .
\label{eq:norm BS bound}
\eeq
In coordinate space, this becomes
\beq
\int d^4x \, d^4 x' 
\: \Psi_{\vec Q, n}^\star (x)  
\: N_n(x,x';\vec Q) 
\Psi_{\vec Q, n'} (x')
= 2\w_{\vec Q, n} \, \d_{nn'} \: .
\label{eq:norm BS bound coord}
\eeq

\subsubsection*{Two-particle states}

Substituting the contribution to the 4-point function from the two-particle states, \eq{eq:GU}, into \eq{eq:norm eq}, we deduce the normalisation condition\footnote{In fact, from \eqs{eq:GU} and \eqref{eq:norm eq} we obtain the normalisation condition described in \eq{eq:norm BS free} with the functions $\tilde\F_{\vec Q, \vec q}(p)$ replaced by $F_{\vec q}(p;Q) = \int d^4x \, e^{ipx} \, \F_{\vec Q,\vec q}(x) \, e^{-i [Q^0 - \w_{\vec Q, \vec q}] h(x^0)}$  (c.f. \eq{eq:Fa}). Then, taking $Q^0 \to \w_{\vec Q, \vec q}$, we obtain the exact form of \eq{eq:norm BS free}.}
\beq
\int \frac{d^4 p}{(2\p)^4} \, \frac{d^4 p'}{(2\p)^4}
\: \tilde \F_{\vec Q, \vec q}^\star(p)
\: \tilde N_{\vec q}(p,p';\vec Q)
\: \tilde \F_{\vec Q, \vec q'}(p')
= 2 \w_{\vec Q, \vec q} \, 2\ve_{\vec Q, \vec q} \, (2\p)^3 \d^3(\vec q - \vec q') \: .
\label{eq:norm BS free}
\eeq
In coordinate space, this becomes
\beq
\int d^4x \, d^4 x' 
\: \F_{\vec Q, \vec q}^\star (x)  
\: N_{\vec q}(x,x';\vec Q)
\: \F_{\vec Q, \vec q'} (x')
= 2 \w_{\vec Q, \vec q} \, 2\ve_{\vec Q, \vec q} \, (2\p)^3 \d^3(\vec q - \vec q') \: .
\label{eq:norm BS free coord}
\eeq

\bigskip
Note that in the fully relativistic case, the normalisation of the wavefunctions depends in general on the potential.

\subsection{The instantaneous approximation and the Schr\"{o}dinger equation \label{sec:InstApprox}}

In the non-relativistic regime, it is possible to simplify the Bethe-Salpeter equations. The momentum exchange between the two unbound particles is $|\vec q| \sim \mu \vrel$, while between two bound particles $|\vec q| \sim \mu \a$, with $\a$ characterising the interaction strength; in either case, for $\a, \vrel \ll 1$, the energy exchange is $q^0 \sim \vec q^2/(2\mu) \ll |\vec q|$. It is then reasonable to ignore the dependence of the kernel $\tilde{W} (p,p';Q)$ on $p^0,p'^0$. This is the instantaneous approximation.\footnote{For a discussion on relativistic corrections, see e.g. Ref.~\cite{Hoyer:2014gna} and references within.} 
In fact, in the cases of interest, $\tilde{W} (p,p';Q)$ depends only on $|\vec p-\vec p'|$, rather than on $\vec p$ and $\vec p'$ separately, and it does not depend on $Q$ (except perhaps for $Q^2$, which, in the non-relativistic regime, we shall approximate with $Q^2 \simeq m^2$). We shall thus assume that
\beq 
\tilde{W}(p,p';Q) \simeq {\cal W}(|\vec p - \vec p'|) \: .
\label{eq:W cal}
\eeq  
In this approximation, we deduce from the Bethe-Salpeter \eqs{eq:BSE bound}, \eqref{eq:BSE free}, that
$\tilde \Psi_{\vec Q, n} (p) / S(p;Q)$ and $\tilde \F_{\vec Q, \vec q} (p) / S(p;Q)$ are independent of $p^0$. 
We define
\begin{align}
{\cal S}_0 (\vec p;Q) 
&\equiv \int \frac{dp^0}{2\p} \: S(p;Q)
\: ,
\label{eq:S0 cal def}
\\
\tilde\psi_{\vec Q, n}(\vec p) &\equiv
\sqrt{2{\cal N}_{\vec Q}  (\vec p) }
\ \[\frac{{\cal S}_0(\vec p;Q)}{S(p;Q)} \]
\ \tilde \Psi_{\vec Q,n} (p)
\: ,
\label{eq:Schr WF bound def} 
\\
\tilde\f_{\vec Q, \vec q}(\vec p) &\equiv
\sqrt{\frac{2{\cal N}_{\vec Q}  (\vec p)}{2\ve_{\vec Q, \vec q}} }
\ \[\frac{{\cal S}_0(\vec p;Q)}{S(p;Q)} \]
\ \tilde \F_{\vec Q, \vec q} (p)
\: ,
\label{eq:Schr WF free def}
\end{align}
where we choose the normalization factor
\beq
{\cal N}_{\vec Q} (\vec p) \equiv 
\frac{E_1(\vec p;\vec Q) E_2(\vec p;\vec Q)}{E_1(\vec p;\vec Q)+E_2(\vec p;\vec Q)}  \: ,  
\label{eq:N cal}
\eeq
such that we recover the conventional normalisation for $\tilde\psi_{\vec Q, n}$ and $\tilde\f_{\vec Q, \vec q}$, as we shall see in Sec.~\ref{sec:Norm Schr}. We calculate ${\cal S}_0 (\vec p;Q)$ in appendix \ref{App:S cal}. Multiplying both sides of \eqs{eq:Schr WF bound def}, \eqref{eq:Schr WF free def} with $S(p;Q)$, integrating over $p^0$, and using \eq{eq:S0 cal def}, it follows that
\begin{align}
\tilde \psi_{\vec Q, n} (\vec p) 
&= \sqrt{2{\cal N}_{\vec Q} (\vec p)} \ \int \frac{dp^0}{2\p} \ \tilde \Psi_{\vec Q, n} (p)
 = \sqrt{2{\cal N}_{\vec Q} (\vec p)} \ \int d^3x \ \Psi_{\vec Q, n} (\{x^0=0, \vec x\})  
\: e^{-i \vec p \cdot \vec x}
\: ,
\label{eq:Schr WF bound}
\\
\tilde\f_{\vec Q, \vec q}(\vec p) 
&= \sqrt{\frac{2{\cal N}_{\vec Q}  (\vec p)}{2\ve_{\vec Q, \vec q}} }
\ \int \frac{dp^0}{2\p} \ \tilde \F_{\vec Q, \vec q} (p)
= \sqrt{\frac{2{\cal N}_{\vec Q}  (\vec p)}{2\ve_{\vec Q, \vec q}} }
\ \int d^3x \ \F_{\vec Q, \vec q} (\{x^0=0, \vec x\}) \: e^{-i \vec p \cdot \vec x}
\: .
\label{eq:Schr WF free}
\end{align}
$\tilde \ps_{\vec Q, n}(\vec p)$ and $\tilde \f_{\vec Q,\vec q} (\vec p)$ are sometimes called the ``equal-time wavefunctions". From the above, and recalling \eqs{eq:Psi star=Tbar cc}, \eqref{eq:Phi star=Tbar cc} we see that
\begin{align}
\tilde \psi_{\vec Q, n}^\star (\vec p) = \tilde \psi_{\vec Q, n}^* (\vec p) 
\, , \qquad
\tilde \f_{\vec Q, \vec q}^\star (\vec p) = \tilde \f_{\vec Q, \vec q}^* (\vec p) \: .
\label{eq:star=cc}
\end{align}

Given the definitions \eqref{eq:Schr WF bound def}, \eqref{eq:Schr WF free def} and \eqs{eq:Schr WF bound}, \eqref{eq:Schr WF free}, the Bethe-Salpeter \eqs{eq:BSE bound} and \eqref{eq:BSE free} become
\begin{align}
\frac{ \tilde\psi_{\vec Q, n}(\vec p) }{\sqrt{2 {\cal N}_{\vec Q}(\vec p)} \: {\cal S}_0(\vec p;Q)}  &= 
\int \frac{d^3 k}{(2\p)^3} 
\frac{{\cal W} (|\vec p - \vec k|) }{\sqrt{2 {\cal N}_{\vec Q}(\vec k)}}
\: \tilde\psi_{\vec Q, n}(\vec k) 
\: , \quad {\rm with} \ Q^0 = \w_{\vec Q, n} \: ,
\label{eq:SE mom bound} 
\\
\frac{ \tilde\f_{\vec Q, \vec q}(\vec p) }{\sqrt{2 {\cal N}_{\vec Q}(\vec p)} \: {\cal S}_0(\vec p;Q)} &= 
\int \frac{d^3 k}{(2\p)^3} 
\frac{{\cal W} (|\vec p - \vec k|) }{\sqrt{2 {\cal N}_{\vec Q}(\vec k)}}
\: \tilde\f_{\vec Q, \vec q}(\vec k) 
\: , \quad {\rm with} \ Q^0 = \w_{\vec Q, \vec q} \: . 
\label{eq:SE mom free} 
\end{align}

\subsubsection*{Non-relativistic approximation}

Using the non-relativistic approximation described in appendix~\ref{App:S cal},  \eqs{eq:E1 NR} -- \eqref{eq:S0 cal NR}, and setting, in accordance to \eq{eq:P0}, 
\begin{align}
\w_{\vec Q, n} &= m + \vec Q^2/2m + {\cal E}_n \: ,
\label{eq:omega n}
\\
\w_{\vec Q, \vec q} &= m + \vec Q^2/2m + {\cal E}_{\vec q} \: ,
\label{eq:omega q}
\end{align}
equations~\eqref{eq:SE mom bound} and \eqref{eq:SE mom free} become
\begin{align}
\(- \frac{\vec p^2}{2\mu} + {\cal E}_n \) \tilde\psi_n(\vec p) 
&= -\frac{1}{i\,4m\mu} \int \frac{d^3 k}{(2\p)^3} \ {\cal W}(|\vec p - \vec k|) \ \tilde\psi_n(\vec k) \: ,
\label{eq:SE mom bound NR} 
\\
\(- \frac{\vec p^2}{2\mu} + {\cal E}_{\vec q} \) \tilde\f_{\vec q}(\vec p) 
&= -\frac{1}{i\,4m\mu} \int \frac{d^3 k}{(2\p)^3} \ {\cal W}(|\vec p - \vec k|) \ \tilde\f_{\vec q}(\vec k) \: .
\label{eq:SE mom free NR} 
\end{align}
These are the Schr\"{o}dinger equations for the bound and the scattering states in momentum space. They are eigenvalue equations, and as such, their solutions determine ${\cal E}_n$ and ${\cal E}_{\vec q}$. Because in \eqs{eq:SE mom bound NR}, \eqref{eq:SE mom free NR}, all dependence on the CM momentum $\vec Q$ has been eliminated, we have dropped this subscript from the $\psi, \f$ wavefunctions, but kept the same symbols in order to avoid cluttering the notation. Note that from \eq{eq:omega n}, it follows that the mass of the bound state is 
\beq M_n = m + {\cal E}_n \: . \label{eq:Mn} \eeq

It is convenient to Fourier-transform \eqs{eq:SE mom bound NR} and \eqref{eq:SE mom free NR} to coordinate space. We set
\begin{align}
\ps_n ({\bf r}) = \int \frac{d^3 p}{(2\p)^3} \: \tilde{\ps}_n ({\bf p}) \: e^{i {\bf p \cdot r}} \, , \qquad
\tilde \ps_n ({\bf p}) = \int d^3 r \: \ps_n ({\bf r}) \: e^{-i {\bf p \cdot r}} \: , 
\label{eq:psi FT}
\\
\f_{\vec q} ({\bf r}) = \int \frac{d^3 p}{(2\p)^3} \: \tilde{\f}_{\vec q} ({\bf p}) \: e^{i {\bf p \cdot r}} \, , \qquad
\tilde \f_{\vec q} ({\bf p}) = \int d^3 r \: \f_{\vec q} ({\bf r}) \: e^{-i {\bf p \cdot r}} \: .
\label{eq:phi FT}
\end{align}
Acting on both sides of \eqs{eq:SE mom bound NR}, \eqref{eq:SE mom free NR} with $\int \frac{d^3p}{(2\pi)^3} e^{i {\bf p\cdot r}}$, we obtain the Schr\"{o}dinger equations in coordinate space
\begin{align}
\[-\frac{\nabla^2}{2\m} + V({\bf r})\]\ps_n ({\bf r}) &= {\cal E}_n \ps_n ({\bf r}) \: ,
\label{eq:SE coord bound}
\\
\[-\frac{\nabla^2}{2\m} + V({\bf r})\]\f_{\vec q} ({\bf r}) &= {\cal E}_{\vec q} \f_{\vec q} ({\bf r}) \: ,
\label{eq:SE coord free}
\end{align}
where $V(\vec r)$ is the non-relativistic potential,
\beq
V({\vec r}) \equiv -\frac{1}{i\,4m\m} \int \frac{d^3k}{(2\p)^3} \ {\cal W} ({\bf k}) \ e^{i {\bf k \cdot r}} \, .
\label{eq:U def}
\eeq
We quote the bound-state and scattering-state solutions of the Schr\"{o}dinger equation for a Coulomb potential, in appendix~\ref{App:WaveFun}, and use them in our computations in Sec.~\ref{Sec:Interactions}.

\subsection{Normalization of the Schr\"{o}dinger wavefunctions  \label{sec:Norm Schr}}

To find the normalization of $\psi_n$ and $\f_{\vec p}$, it is easiest to follow a similar procedure to that of Sec.~\ref{sec:Norm BS}.  We first define
\begin{align}
{\cal G}^{(4)} (\vec p, \vec p'; Q) &\equiv 
\int \frac{dp^0}{2\p} \frac{dp'^0}{2\p} \tilde{G}^{(4)} (p,p';Q) 
\, ,
\label{eq:G cal}
\\
{\cal G}_n^{(4)} (\vec p, \vec p'; Q) 
&\equiv \int \frac{dp^0}{2\p} \frac{dp'^0}{2\p} \tilde{G}_n^{(4)} (p, p'; Q) 
= \frac{1}{\sqrt{2{\cal N}_{\vec Q}(\vec p) 2{\cal N}_{\vec Q}(\vec p')}} 
\: \frac{i \tilde\psi_n (\vec p) \tilde\psi_n^\star (\vec p')}{2\w_{\vec Q,n} \[Q^0 - \w_{\vec Q, n} +i \e\]} \, .
\label{eq:cal Gn}
\\
{\cal G}_{\cal U}^{(4)} (\vec p, \vec p'; Q) 
&\equiv  \int \! \frac{dp^0}{2\p} \frac{dp'^0}{2\p} \tilde{G}_{\cal U}^{(4)} (p, p'; Q) 
=  \frac{1}{\sqrt{2{\cal N}_{\vec Q}(\vec p) 2{\cal N}_{\vec Q}(\vec p')}} 
\int \! \frac{d^3q}{(2\p)^3} \frac{i \tilde\f_{\vec q} (\vec p) \tilde\f_{\vec q}^\star (\vec p')}
{2\w_{\vec Q, \vec q} \[Q^0 - \w_{\vec Q, \vec q} +i \e\]} \, . 
\label{eq:cal GU}
\end{align}
Integrating \eq{eq:DE mom} with respect to $p^0, k^0$ yields
\begin{align}
{\cal G}^{(4)} (\vec p, \vec p'; Q) 
= (2\p)^3 \d^3 (\vec p - \vec p') \, {\cal S}_0(\vec p;Q)  
+ {\cal S}_0(\vec p;Q) \int \frac{d^3k}{(2\p)^3} \: {\cal W} (|\vec p-\vec k|) 
\: {\cal G}^{(4)} (\vec k,\vec p';Q) \, .
\label{eq:DE inst}
\end{align}
Following the steps of Sec.~\ref{sec:Norm BS}, we find the equivalent of \eq{eq:norm eq},
\beq
\int \frac{d^3 k}{(2\p)^3} 
\: {\cal G}^{(4)} (\vec p, \vec k; Q) 
\: \frac{d}{dQ^0} \[\frac{1}{{\cal S}_0(\vec k; Q)}\]
\: {\cal G}^{(4)} (\vec k,\vec p';Q)
= - \frac{d}{dQ^0} {\cal G}^{(4)}(\vec p, \vec p'; Q) \: .
\label{eq:norm Schr eq}
\eeq
From \eq{eq:S0 cal},
\beq
\frac{d}{dQ^0} \[\frac{1}{{\cal S}_0(\vec k; Q)}\]   
= - i \, 2 {\cal N}_{\vec Q} (\vec k) \, 2 Q^0 \: .
\eeq
We may now obtain the normalisation conditions for the wavefunctions $\psi_n$ and $\f_{\vec q}$.

\subsubsection*{Bound states}

Close to the pole, at $Q^0 \to \w_{\vec Q, n}$, we may substitute the contribution from the $n$th bound state, \eq{eq:cal Gn}, into \eqref{eq:norm Schr eq}. We obtain
\beq
\int \frac{d^3 p}{(2\p)^3} \: \tilde\psi_n^\star (\vec p)   \: \tilde\psi_n (\vec p) 
= \int d^3 r \: \psi_n^\star (\vec r)   \: \psi_n (\vec r) = 1 \: .
\label{eq:norm Schr bound}
\eeq

\subsubsection*{Two-particle states}

Substituting the contribution from the two-particle states, \eq{eq:cal GU} into \eqref{eq:norm Schr eq}, and for $Q^0 \to \w_{\vec Q, \vec q}$, we deduce the normalisation condition
\beq
\int \frac{d^3 p}{(2\p)^3} \: \tilde \f_{\vec q}^\star(\vec p) \: \tilde \f_{\vec q'}(\vec p)
=\int d^3 r \: \f_{\vec q}^\star(\vec r) \: \f_{\vec q'}(\vec r)
= (2\p)^3 \d^3(\vec q - \vec q') \: .
\label{eq:norm Schr free}
\eeq

\bigskip

Note that in obtaining the normalisation conditions~\eqref{eq:norm Schr bound} and \eqref{eq:norm Schr free}, we did not make use of the non-relativistic expansions of the factors ${\cal N}_{\vec Q}(\vec p)$ and $\ve_{\vec Q, \vec q}$,  given in \eqs{eq:Norm NR} and \eqref{eq:epsilon NR}.


\section{Radiative level transitions \label{Sec:TransAmpl}}

In this section, we determine the amplitudes for the radiative BSF and de-excitation processes
\begin{align}
\x_1 + \x_2 \ &\to \ (\x_1\x_2)_{\rm bound} + \vf \: ,
\label{eq:BSF} \\
(\x_1\x_2)_{{\rm bound}, n'}  \ &\to \ (\x_1\x_2)_{{\rm bound}, n} + \vf  \: ,
\label{eq:de-excitation}
\end{align}
in terms of the bound-state and scattering-state wavefunctions computed in Sec.~\ref{Sec:Wavefunctions} and a perturbative interaction which describes the emission of the force mediator. The $\mathsf{S}$-matrix elements of interest are
\begin{align}
{}_{\rm out}\langle {\cal B}_{\vec P, n}  \,;\, \vf_{_{\vec P_\vf}} \,|\, {\cal U}_{\vec K, \vec k} \rangle_{\rm in}
\: &= \: \langle {\cal B}_{\vec P, n} \,;\, \vf_{_{\vec P_\vf}} \,|\, \mathsf{S} \,|\, {\cal U}_{\vec K, \vec k} \rangle  
\: , \label{eq:S matrix BSF}
\\
{}_{\rm out}\langle {\cal B}_{\vec P, n} \,;\, \vf_{_{\vec P_\vf}} \,|\, {\cal B}_{\vec K, n'} \rangle_{\rm in}
\: &= \: \langle {\cal B}_{\vec P, n} \,;\,  \vf_{_{\vec P_\vf}} \,|\, \mathsf{S} \,|\, {\cal B}_{\vec K, n'} \rangle  \: ,
\label{eq:S matrix de-excitation}
\end{align}
where the indices stand for the momenta and the quantum numbers of the corresponding states, as stated in the beginning of Sec.~\ref{Sec:Wavefunctions}.

\subsection{The 5-point Green's function \label{sec:Greens 5p}}

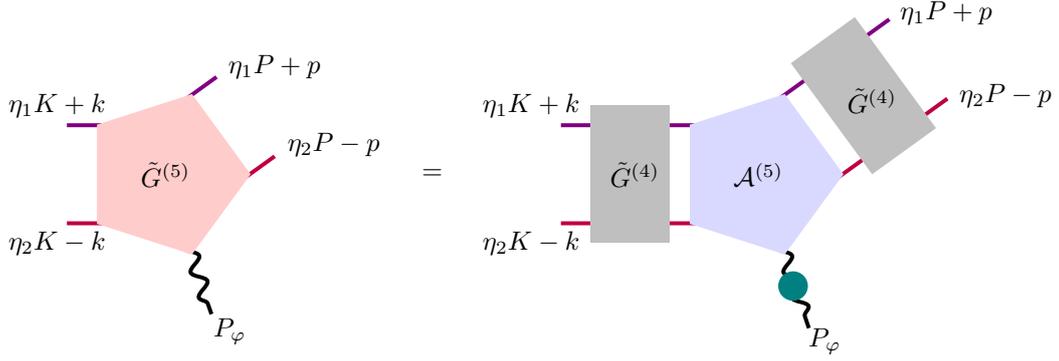
\begin{figure}[t]
\centering
\begin{tikzpicture}[line width=1.5 pt, scale=1.3]
\begin{scope}
\node at (-1.1, 0.7){$\h_1 K + k$};
\node at (-1.1,-0.7){$\h_2 K - k$};
\draw[violet] (-1, 0.5)--(0, 0.5);
\draw[purple] (-1,-0.5)--(0,-0.5);
\draw[rotate=36,violet] (0, 0.5)--(1, 0.5);
\draw[rotate=36,purple] (0,-0.5)--(1,-0.5);
\draw[vector,rotate=-72] (0,0)--(1.5,0);
\node at (1.1,1.1){$\h_1 P + p$};
\node at (1.7,0.3){$\h_2 P - p$};
\node at (0.65,-1.6){$P_\vf$};
\draw[fill=red!20,draw=none] (0,0) 
+(0:.86) -- +(72:0.86) -- +(144:0.86) -- +(216:0.86) -- +(288:0.86) -- cycle;
\node at (0,0){$\tilde G^{(5)}$};
\end{scope}
\node at (2.7,0){$=$};
\begin{scope}[shift={(6,0)}]
\node at (-2.3, 0.7){$\h_1 K+k$};
\node at (-2.3,-0.7){$\h_2 K-k$};
\draw[violet] (-2, 0.5)--(0, 0.5);
\draw[purple] (-2,-0.5)--(0,-0.5);
\draw[rotate=36,violet] (0, 0.5)--(2, 0.5);
\draw[rotate=36,purple] (0,-0.5)--(2,-0.5);
\draw[vector,rotate=-72] (0,0)--(1.65,0);
\draw[fill=lightgray,draw=none,shift={(-1.7,-0.7)}] 
(0,0) rectangle (0.8,1.4);
\draw[fill=lightgray,draw=none,shift={(1.15,0)},rotate=36] 
(0,0) rectangle (0.8,1.4);
\draw[fill=teal,draw=none] (0.35,-1.14) circle (0.15);
\draw[fill=blue!15,draw=none] (0,0) 
+(0:.86) -- +(72:0.86) -- +(144:0.86) -- +(216:0.86) -- +(288:0.86) -- cycle;
\node at (-1.25, 0){$\tilde G^{(4)}$};
\node at (0,0){${\cal A}^{(5)}$};
\node at (1.15,0.75){$\tilde G^{(4)}$};
\node at (0.67,-1.7){$P_\vf$};
\node at (1.9,1.65){$\h_1 P + p$};
\node at (2.5,0.8){$\h_2 P - p$};
\end{scope}
\end{tikzpicture}
\caption{Diagrammatic representation of the equation~\eqref{eq:G tilde 5p} for the 5-point function $\tilde G^{(5)}$.
\protect\PropMed stands for the full propagator of the force mediator $\vf$, which may be either a scalar or a vector boson.}
\label{fig:G tilde 5p}
\end{figure}

Since ${\cal B}_{\vec Q, n}$ and ${\cal U}_{\vec Q, \vec q}$ are generated by the action of $\x_1^\dagger$ and $\x_2^\dagger$ on the vacuum (c.f. \eqs{eq:BSW* bound}, \eqref{eq:BSW* free}), in order to compute the $\mathsf{S}$-matrix elements of \eqs{eq:S matrix BSF} and \eqref{eq:S matrix de-excitation} we need to consider the 5-point function
\beq
G^{(5)} (X_\vf,x_1,x_2; \: y_1,y_2) \equiv 
\<\W| T \vf(X_\vf) \x_1(x_1) \x_2(x_2)  \x_1^\dagger(y_1) \x_2^\dagger(y_2) |\W\> \: .
\label{eq:Greens 5p}
\eeq
We define the Fourier transform
\begin{multline}
\tilde G^{(5)}(P_\vf,p_1,p_2; \: k_1,k_2) =
\\
= \int
d^4 X_\vf \, d^4 x_1 \, d^4 x_2 \, d^4 y_1 \, d^4 y_2
\ e^{i (P_\vf X_\vf + p_1 x_1 + p_2 x_2 - k_1 y_1 - k_2 y_2)}
\ G^{(5)} (X_\vf, x_1, x_2; \: y_1,y_2) \: .
\end{multline}
As in \eq{eq:x,X}, we set
\begin{align}
 x \equiv  x_1 - x_2 \, , \qquad &X \equiv \h_1 x_1 + \h_2 x_2 \\
 y \equiv  y_1 - y_2 \, , \qquad &Y \equiv \h_1 y_1 + \h_2 y_2 \: ,
\end{align}
and rewrite the above as
\begin{multline}
\tilde G^{(5)} (P_\vf, \h_1 P + p, \h_2 P - p;  \: \h_1 K + k,  \h_2 K-k) =
\\
\int d^4 X_\vf \, d^4 X \, d^4 x \, d^4 Y \, d^4 y
\ e^{i (P_\vf X_\vf + P X + p x - K y - k y)}
\ G^{(5)} (X_\vf, X+ \h_2 x, X - \h_1 x; \, Y + \h_2 y, Y -\h_1 y) \: , 
\end{multline}
i.e.~the conjugate momenta of $X, x$ are $P, p$, and the conjugate momenta of $Y, y$ are $K, k$ defined as
\begin{align}
P \equiv p_1 + p_2 \: , \qquad p \equiv \h_2 p_1 - \h_1 p_2 \: ,
\\
K \equiv k_1 + k_2 \: , \qquad k \equiv \h_2 k_1 - \h_1 k_2 \: .
\label{eq:P,p,K,k}
\end{align}

\begin{figure}[t]
\centering
\begin{tikzpicture}[line width=1.5 pt, scale=1.3]
\begin{scope}
\node at (-1.15, 0.5){$k_1$};
\node at (-1.15,-0.5){$k_2$};
\draw[violet] (-1, 0.5)--(0, 0.5);
\draw[purple] (-1,-0.5)--(0,-0.5);
\draw[rotate=36,violet] (0, 0.5)--(1, 0.5);
\draw[rotate=36,purple] (0,-0.5)--(1,-0.5);
\node at (0.7, 1){$p_1$};
\node at (1.28,.2){$p_2$};
\draw[vector,rotate=-72] (0,0)--(1.5,0);
\node at (0.75,-1.6){$P_\vf$};
\draw[fill=blue!45,draw=none] (0,0)
+(0:.86) -- +(72:0.86) -- +(144:0.86) -- +(216:0.86) -- +(288:0.86) -- cycle;
\node at (0,0){${\cal C}^{(5)}$};
\end{scope}
\node at (1.75,0) {$=$};
\begin{scope}[shift={(3.5,0)}]
\node at (-1.15, 0.5){$k_1$};
\node at (-1.15,-0.5){$k_2$};
\draw[violet] (-1, 0.5)--(0, 0.5);
\draw[purple] (-1,-0.5)--(0,-0.5);
\draw[rotate=36,violet] (0, 0.5)--(1, 0.5);
\draw[rotate=36,purple] (0,-0.5)--(1,-0.5);
\node at (0.7, 1){$p_1$};
\node at (1.28,.2){$p_2$};
\draw[vector,rotate=-72] (0,0)--(1.65,0);
\node at (0.75,-1.6){$P_\vf$};
\draw[fill=blue!30,draw=none] (0,0)
+(0:.86) -- +(72:0.86) -- +(144:0.86) -- +(216:0.86) -- +(288:0.86) -- cycle;
\draw[fill=teal,  draw=none] ( 0.38,-1.18) circle (0.15);
\node at (0,0){${\cal C}_{\vf-{\rm amp}}^{(5)}$};
\end{scope}
\node at (5.4,0) {$=$};
\begin{scope}[shift={(7.6,0)}]
\node at (-1.65, 0.5){$k_1$};
\node at (-1.65,-0.5){$k_2$};
\draw[violet] (-1.5, 0.5)--(0, 0.5);
\draw[purple] (-1.5,-0.5)--(0,-0.5);
\draw[rotate=36,violet] (0, 0.5)--(1.5, 0.5);
\draw[rotate=36,purple] (0,-0.5)--(1.5,-0.5);
\node at (1.1, 1.3){$p_1$};
\node at (1.7,.5){$p_2$};
\draw[vector,rotate=-72] (0,0)--(1.65,0);
\node at (0.75,-1.6){$P_\vf$};
\draw[fill=blue!15,draw=none] (0,0) 
+(0:.86) -- +(72:0.86) -- +(144:0.86) -- +(216:0.86) -- +(288:0.86) -- cycle;
\draw[fill=violet,draw=none] (-1.1, 0.5)  circle (0.15);
\draw[fill=purple,draw=none] (-1.1,-0.5)  circle (0.15);
\draw[fill=violet,draw=none] ( 0.6, 1.05) circle (0.15);
\draw[fill=purple,draw=none] ( 1.15,0.24) circle (0.15);
\draw[fill=teal,  draw=none] ( 0.38,-1.18) circle (0.15);
\node at (0,0){${\cal A}^{(5)}$};
\end{scope}
\end{tikzpicture}
\caption{Diagrammatic representation of the equations~\eqref{eq:C and A} and \eqref{eq:C varphi}. 
${\cal C}^{(5)}$ stands for the sum of all connected diagrams with no legs amputated; this includes not fully connected diagrams. ${\cal C}_{\vf-\rm amp}^{(5)}$ is equal to ${\cal C}^{(5)}$ with only the $\vf$-leg amputated. 
\protect\PropV and \protect\PropP stand for the $\x_1$ and $\x_2$ full propagators, respectively.
\protect\PropMed stands for the full propagator of the force mediator $\vf$, which may be either a scalar or a vector boson.
}
\label{fig:amputation}
\end{figure}
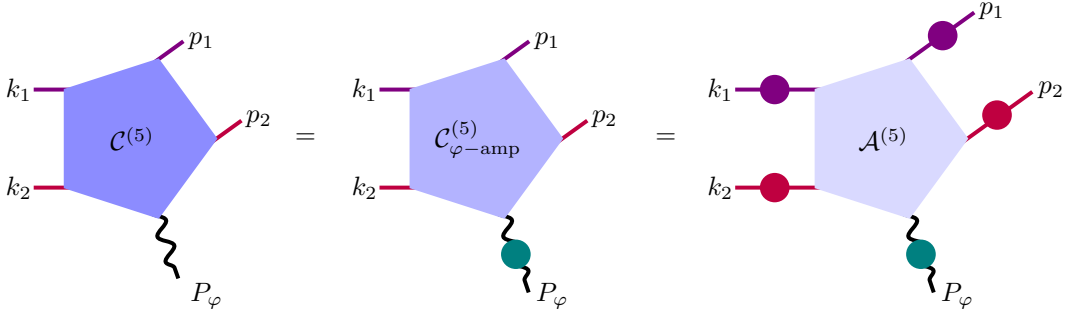

The 5-point Green's function $G^{(5)}(X_\vf,x_1,x_2;y_1,y_2)$ is equal to the sum of all connected diagrams with five external points. The momentum-space $\tilde G^{(5)}$ is sketched in Fig.~\ref{fig:G tilde 5p}, and can be written as
\begin{multline}
\tilde G^{(5)} (P_\vf, \h_1 P + p, \h_2 P - p;  \: \h_1 K + k, \h_2 K-k) = 
\tilde S_\vf(P_\vf) \int \frac{d^4 p'}{(2\p)^4} \frac{d^4 k'}{(2\p)^4} \: \tilde G^{(4)} (p,p';P) \times 
\\
(2\p)^4 \d^4 (K-P-P_\vf) \: i {\cal A}^{(5)} (P_\vf, \h_1 P + p', \h_2 P - p'; \: \h_1 K + k', \h_2 K - k')
\: \tilde G^{(4)} (k',k;K) \, ,
\label{eq:G tilde 5p}
\end{multline}
where
\beq 
\tilde S_\vf(P_\vf) = \frac{i Z_\vf (\vec P_\vf) }{P_\vf^2 - m_\vf^2 + i\e}
\label{eq:S varphi}
\eeq 
is the $\vf$ propagator, with 
\beq
Z_\vf (\vec q) \equiv |\<\W|\vf(0)|\vf_{\vec q}\>|^2
\label{eq:Z varphi} 
\eeq
being the field-strength renormalisation parameter for $\vf$.  ${\cal A}^{(5)}$ is defined via the relation
\beq
i\, {\cal C}^{(5)}(P_\vf,p_1,p_2; \: k_1,k_2) =
\tilde S_\vf(P_\vf) \tilde S_1(p_1) \tilde S_2(p_2) \tilde S_1(k_1) \tilde S_2(k_2)
\, i {\cal A}^{(5)}(P_\vf,p_1,p_2; \: k_1,k_2) \: , 
\label{eq:C and A}
\eeq
where 
\beq
i\,{\cal C}^{(5)}(P_\vf,p_1,p_2; \: k_1,k_2) = 
\text{sum of all connected diagrams.}
\label{eq:C 5p}
\eeq
Note that ${\cal C}^{(5)}$ may include diagrams that are \emph{not} fully connected, i.e. diagrams in which external legs are disconnected from each other,\footnote{In fact, the lowest-order contribution to ${\cal C}^{(5)}$, for the transition processes considered in Sec.~\ref{Sec:Interactions}, arises from diagrams that are not fully connected, as shown in Fig.~\ref{fig:C phi-amp}. However, the entire transition processes are described by fully connected diagrams, shown in Fig.~\ref{fig:Transition}.}  
but it does not, of course, include vacuum bubble diagrams. (If only fully connected diagrams contributed to ${\cal C}^{(5)}$, then ${\cal A}^{(5)}$ would simply be the sum of all connected and amputated diagrams, as conventionally defined.) For later convenience, we also define ${\cal C}_{\vf-{\rm amp}}^{(5)}$ as the sum of all connected diagrams with only the $\vf$-leg amputated,
\beq
i\,{\cal C}^{(5)}(P_\vf,p_1,p_2; \: k_1,k_2) =
\tilde S_\vf(P_\vf) \: i\,{\cal C}_{\vf-{\rm amp}}^{(5)}(P_\vf,p_1,p_2; \: k_1,k_2) \: .
\label{eq:C varphi}
\eeq
Then, ${\cal A}^{(5)}$ appearing in \eq{eq:G tilde 5p}, becomes
\beq
i{\cal A}^{(5)} (P_\vf, \h_1 P + p, \h_2 P - p; \: \h_1 K + k, \h_2 K - k)
= \frac{i\,{\cal C}^{(5)}_{\vf-\rm amp} (P_\vf, \h_1 P + p, \h_2 P - p; \: \h_1 K + k, \h_2 K - k)}
{S(p;P)S(k;K)} \: ,
\label{eq:A & C varphi}
\eeq
where we remind that $S(p;P) \equiv \tilde S_1(\h_1 P + p) \, \tilde S_2(\h_2 P - p)$  (c.f. \eq{eq:S}). 
We sketch \eqs{eq:C and A} and \eqref{eq:C varphi} in Fig.~\ref{fig:amputation}.

\subsection{Transition amplitudes \label{sec:LSZ}}

We now extract the $\mathsf{S}$-matrix elements of \eqs{eq:S matrix BSF} and \eqref{eq:S matrix de-excitation} from the 5-point Green's function of \eq{eq:Greens 5p}. Our analysis follows closely Sec.~7.2 of Ref.~\cite{PeskinSchroeder}.

Let us first focus on the BSF amplitude of \eq{eq:S matrix BSF}, for which 
\beq
P_\vf^0 \to \w_\vf(\vec P_\vf), 
\ P^0 \to \w_{\vec P, n}, 
\ K^0 \to \w_{\vec K, \vec k} \: .
\label{eq:limit}
\eeq
In this limit, the Lehmann-Symanzik-Zimmermann reduction formula yields
\begin{multline}
\int d^4 X_\vf \: e^{i P_\vf X_\vf}   
\int d^4 X \: e^{i P X}    
\int d^4 Y \: e^{-i K Y} 
\: G^{(5)}(X_\vf, X+\h_2 x, X-\h_1 x; \: Y +\h_2 y, Y - \h_1 y) 
\sim \\
\sim
\[\frac{i \, \<\W|\vf(0)|\vf_{_{\vec P_\vf}}\>}{2 \w_\vf(\vec P_\vf) (P_\vf^0 - \w_\vf(\vec P_\vf) +i \e )} \]
\[\frac{i \, \<\W|T \x_1(\h_2 x) \x_2(-\h_1 x)|{\cal B}_{\vec P,n}\> }{2\w_{\vec P, n} ( P^0 - \w_{\vec P, n} +i \e )} \]
\times 
\\
\int \frac{d^3 k'}{(2\p)^3 2\ve_{\vec K, \vec k'}} 
\frac{i \, \<{\cal U}_{\vec K, \vec k'}|T \x_1^\dagger (\h_2 y) \x_2^\dagger (-\h_1 y) |\W\> }{2\w_{\vec K, \vec k'}(K^0 - \w_{\vec K, \vec k'} +i \e)} \ 
\langle {\cal B}_{\vec P, n}, \vf_{_{\vec P_\vf}} | {\sf S} | {\cal U}_{\vec K, \vec k'} \rangle \: .
\label{eq:LSZ}
\end{multline}
Here, the $\sim$ sign means that the two sides have the same singularities in the limit \eqref{eq:limit}; to compute the $\mathsf{S}$-matrix element, we need to extract the residues of these singularities from both sides of \eq{eq:LSZ}.

In the above expression, the correlation functions involving the $\x_1$ and $\x_2$ fields correspond to the bound and scattering state wavefunctions (c.f. \eqs{eq:Psi}, \eqref{eq:Phi star}). The correlation function involving the $\vf$ field is the $\vf$ field-strength renormalisation parameter (c.f. \eq{eq:Z varphi}). We Fourier-transform \eq{eq:LSZ} with respect to $x,y$, to obtain
\begin{multline}
\int d^4 X_\vf \: d^4 X  \: d^4 Y \: d^4 x \: d^4 y 
\: e^{i (P_\vf X_\vf + P X - K Y)} 
\: e^{ i (p x - q y)}    
\: G^{(5)}(X_\vf, X+\h_2 x, X-\h_1 x; \: Y +\h_2 y, Y - \h_1 y) 
\\
\sim 
\[\frac{i \, \sqrt{Z_\vf (\vec P_\vf)} }{2 \w_\vf(\vec P_\vf) (P_\vf^0 - \w_\vf(\vec P_\vf) +i \e )} \]
\[\frac{i \, \tilde\Psi_{\vec P,n}(p) }{2\w_{\vec P, n} ( P^0 - \w_{\vec P, n} +i \e )} \]
\times 
\\
\int \frac{d^3 k'}{(2\p)^3 2\ve_{\vec K, \vec k'}} 
\frac{i \, \tilde\F_{\vec K,\vec k'}^\star(q) }{2\w_{\vec K, \vec k'}(K^0 - \w_{\vec K, \vec k'} +i \e)} \
\langle {\cal B}_{\vec P, n}, \vf_{_{\vec P_\vf}} | {\sf S} | {\cal U}_{\vec K, \vec k'} \rangle \: .
\label{eq:LSZ FT}
\end{multline}
The left side of the above equation is 
$\tilde G^{(5)}(P_\vf, \h_1 P+p, \h_2 P - p  ;\: \h_1 K + q, \h_2 K - q)$, which may be decomposed according to \eq{eq:G tilde 5p}. Recalling \eqs{eq:Gn}, \eqref{eq:GU} and \eqref{eq:G=Gn+GU mom} for the 4-point function $\tilde G^{(4)}$, and keeping only the leading singularities in the limit \eqref{eq:limit}, the left side of \eq{eq:LSZ FT} becomes
\begin{multline}
\tilde G^{(5)}(P_\vf, \h_1 P+p, \h_2 P - p  ;\: \h_1 K + q, \h_2 K - q) \sim 
\\
\sim \frac{i Z_\vf (\vec P_\vf) }{P_\vf^2 - m_\vf^2 + i\e}
\: \int \frac{d^4 p'}{(2\p)^4} \frac{d^4 q'}{(2\p)^4} 
\: \frac{i \tilde \Psi_{\vec P, n}(p) \tilde \Psi_{\vec P, n}^\star (p')}{2\w_{\vec P, n} \[P^0 - \w_{\vec P, n} + i\e\]} 
\int \frac{d^3 k'}{(2\p)^3} \frac{i \tilde \F_{\vec K, \vec k'}(q') \tilde \F_{\vec K, \vec k'}^\star (q)}{2\w_{\vec K, \vec k'} 2\ve_{\vec K, \vec k'} [K^0 - \w_{\vec K, \vec k'} +i\e]}
\\ 
\times (2\p)^4 \d^4(K-P-P_\vf) \: i {\cal A}^{(5)} (P_\vf, \h_1 P + p', \h_2 P - p'; \: \h_1 K + q', \h_2 K - q')  
\, . \nn
\end{multline}
At $P_\vf^0 \to \w_\vf(\vec P_\vf)$ and $P^0 \to \w_{\vec P, n}$, this expression has the same poles as the right side of \eq{eq:LSZ FT}. Identifying their residues, we obtain
\begin{multline}
\sqrt{Z_\vf (\vec P_\vf)} \int \frac{d^4p'}{(2\p)^4} \frac{d^4q'}{(2\p)^4}
\tilde \Psi_{\vec P, n}^\star (p') 
\ \int \frac{d^3 k'}{(2\p)^3} \frac{i \tilde \F_{\vec K, \vec k'}(q') \tilde \F_{\vec K, \vec k'}^\star (q)}{2\w_{\vec K, \vec k'} 2\ve_{\vec K, \vec k'} [K^0 - \w_{\vec K, \vec k'} +i\e]} 
\times  \\ 
(2\p)^4 \d^4(K-P-P_\vf) \: i {\cal A}^{(5)} (P_\vf, \h_1 P + p', \h_2 P - p'; \: \h_1 K + q', \h_2 K - q') \sim
\\ 
\sim
\int \frac{d^3 k'}{(2\p)^3 2 \ve_{\vec K, \vec k'}}  
\frac{i \, \tilde\F_{\vec K,\vec k'}^\star(q) }{2\w_{\vec K, \vec k'}[K^0 - \w_{\vec K, \vec k'} +i \e]} \
\< {\cal B}_{\vec P, n}, \vf_{_{\vec P_\vf}} | {\sf S} | {\cal U}_{\vec K, \vec k'} \>   \: .
\nn
\end{multline}
We still have to extract the leading singularity at $K^0 \to \w_{\vec K, \vec k}$. We multiply both sides of the above expression with $\tilde{N}_{\vec k}(q,q'';\vec K) \F_{\vec K, \vec k}(q'')$, integrate over $q$ and $q''$, and use the orthonormality condition \eqref{eq:norm BS free}, to obtain the $\mathsf{S}$-matrix element for BSF
\begin{multline}
\< {\cal B}_{\vec P, n}  \,;\,  \vf_{_{\vec P_\vf}}   \,|\,  {\sf S}  \,|\,  {\cal U}_{\vec K, \vec k} \>
= \sqrt{Z_\vf (\vec P_\vf)} \int \frac{d^4p}{(2\p)^4} \frac{d^4q}{(2\p)^4} 
\: \tilde \Psi_{\vec P, n}^\star (p)
\: \tilde \F_{\vec K, \vec k}(q) 
\\
\times (2\p)^4 \d^4(K-P-P_\vf) \: i {\cal A}^{(5)} (P_\vf, \h_1 P + p, \h_2 P - p; \: \h_1 K + q, \h_2 K - q)
\: .
\label{eq:S-matrix BSF}
\end{multline}
Following similar steps, we obtain the $\mathsf{S}$-matrix element for transition between discrete energy levels,
$\< {\cal B}_{\vec P, n}  \,;\,  \vf_{_{\vec P_\vf}}   \,|\,  {\sf S}  \,|\,  {\cal B}_{\vec K, n'} \>$.

In standard notation, we write the $\mathsf{S}$-matrix elements as
\begin{align}
\< {\cal B}_{\vec P, n} \,;\, \vf_{_{\vec P_\vf}}  \,|\,  {\sf S} \,|\, {\cal U}_{\vec K, \vec k} \> 
&= (2\p)^4 \, \d^4(K - P - P_\vf) \: i \M_{\vec k \to n} \: ,
\label{eq:S BSF = delta x M}
\\
\< {\cal B}_{\vec P, n} \,;\, \vf_{_{\vec P_\vf}}  \,|\,  {\sf S} \,|\, {\cal B}_{\vec K, n'} \> 
&= (2\p)^4 \, \d^4(K - P - P_\vf) \: i \M_{n' \to n} \: ,
\label{eq:S LevelTrans = delta x M}
\end{align}
with 
\begin{multline}
\M_{\vec k \to n} = \sqrt{Z_\vf (\vec P_\vf)} \int \frac{d^4p}{(2\p)^4} \frac{d^4q}{(2\p)^4} 
\: \tilde \Psi_{\vec P, n}^\star (p) \: \tilde \F_{\vec K, \vec k}(q) 
\\
\times {\cal A}^{(5)} (P_\vf, \h_1 P + p, \h_2 P - p; \: \h_1 K + q, \h_2 K - q) \: ,
\label{eq:Ampl BSF}
\end{multline}
\vspace{-7mm}
\begin{multline}
\M_{n' \to n} = \sqrt{Z_\vf (\vec P_\vf)} \int \frac{d^4p}{(2\p)^4} \frac{d^4q}{(2\p)^4} 
\: \tilde \Psi_{\vec P, n}^\star (p) \: \tilde \Psi_{\vec K, n'}(q) 
\\
\times {\cal A}^{(5)} (P_\vf, \h_1 P + p, \h_2 P - p; \: \h_1 K + q, \h_2 K - q) \: .
\label{eq:Ampl LevelTrans}
\end{multline}

If non-fully connected diagrams contribute to the perturbative part of the transition  amplitudes, then ${\cal A}^{(5)}$ should be replaced by ${\cal C}_{\vf-{\rm amp}}^{(5)}$ using \eq{eq:A & C varphi}. In this case, we obtain
\begin{multline}
\M_{\vec k \to n}  =
\sqrt{Z_\vf (\vec P_\vf)} \int \frac{d^4p}{(2\p)^4} \frac{d^4q}{(2\p)^4} 
\: \frac{\tilde \Psi_{\vec P, n}^\star (p)}{S(p;P)}
\: \frac{\tilde \F_{\vec K, \vec k}(q)}{S(q;K)} 
\\
\times {\cal C}_{\vf-{\rm amp}}^{(5)} (P_\vf, \h_1 P + p, \h_2 P - p; \: \h_1 K + q, \h_2 K - q) \: ,
\label{eq:Ampl BSF mod}
\end{multline}
\vspace{-7mm}
\begin{multline}
\M_{n' \to n} =
\sqrt{Z_\vf (\vec P_\vf)} \int \frac{d^4p}{(2\p)^4} \frac{d^4q}{(2\p)^4} 
\: \frac{\tilde \Psi_{\vec P, n}^\star (p)}{S(p;P)}
\: \frac{\tilde \Psi_{\vec K, n'}(q)}{S(q;K)} 
\\
\times {\cal C}_{\vf-{\rm amp}}^{(5)} (P_\vf, \h_1 P + p, \h_2 P - p; \: \h_1 K + q, \h_2 K - q) \: .
\label{eq:Ampl LevelTrans mod}
\end{multline}
In the case of a vector mediator $\vf_\mu$, $Z_\vf$ becomes the charge-renormalisation parameter and the amplitudes contain the polarisation vector $\e^\mu$, i.e.~${\cal A}^{(5)} = \e^\mu {\cal A}^{(5)}_\mu$ and ${\cal C}_{\vf-{\rm amp}}^{(5)} = \e^\mu \, {\cal C}_{\vf-{\rm amp},  \, \mu}^{(5)}$.

\subsection{Instantaneous approximation \label{sec:trans ampl inst}}

In the instantaneous and non-relativistic approximations, we may express the transition amplitudes in terms of the Schr\"{o}dinger wavefunctions defined in \eqs{eq:Schr WF bound def}, \eqref{eq:Schr WF free def}, as follows
\begin{align}
\M_{\vec k \to n}  &\simeq \sqrt{2\ve_{\vec K,\vec k}}
\int \frac{d^3p}{(2\p)^3} \frac{d^3q}{(2\p)^3} 
\: \frac{\tilde \psi_n^\star (\vec p)  \: \tilde \f_{\vec k}(\vec q)}
{\sqrt{2{\cal N}_{\vec P}(\vec p) \, 2{\cal N}_{\vec K}(\vec q)}}
\ \M_{\rm trans} (\vec q; \vec p) \: ,
\label{eq:Ampl BSF Inst}
\\
\M_{n' \to n}  &\simeq
\int \frac{d^3p}{(2\p)^3} \frac{d^3q}{(2\p)^3} 
\: \frac{\tilde \psi_n^\star (\vec p)  \: \tilde \psi_{n'}(\vec q)}
{\sqrt{2{\cal N}_{\vec P}(\vec p) \, 2{\cal N}_{\vec K}(\vec q)}}
\ \M_{\rm trans} (\vec q; \vec p) \: ,
\label{eq:Ampl LevelTrans Inst}
\end{align}
where we took $Z_\vf (\vec P_\vf) \simeq 1$ to lowest order, and set
\beq
\M_{\rm trans} (\vec q; \vec p) \equiv \frac{1}{{\cal S}_0(\vec q;K) \, {\cal S}_0(\vec p;P)}
\int \frac{dp^0}{2\p} \frac{dq^0}{2\p} \ 
{\cal C}_{\vf-{\rm amp}}^{(5)} (P_\vf, \h_1 P + p, \h_2 P - p; \: \h_1 K + q, \h_2 K - q)
\: .
\label{eq:M trans}
\eeq

\medskip

It is sufficient for our purposes, and consistent with our approximation (see footnote~\ref{foot:corrections}), to expand the normalisation factors up to first order in $\vec p^2, \vec q^2$, as follows
\begin{align}
{\cal N}_{\vec Q} (\vec p) &\simeq \mu \[1+\frac{\vec p^2}{2\mu^2} \(1-\frac{3\mu}{m}\)\] \: ,
\label{eq:Norm NR}
\\
\frac{1}{\sqrt{2{\cal N}_{\vec P}(\vec p) \, 2{\cal N}_{\vec K}(\vec q)}} &\simeq
\frac{1}{2\mu} \[1-\frac{\vec p^2 + \vec q^2}{4\mu^2} \(1-\frac{3\mu}{m}\)\] \, .
\label{eq:Norm trans NR}
\end{align}
The $\vec p^2, \, \vec q^2$ terms in \eq{eq:Norm trans NR} introduce corrections of order $\a^2$ and $\vrel^2$ (see appendix~\ref{App:WaveFun}), where $\a$ parametrises the strength of the interaction (for a Coulomb potential, it is the fine-structure constant) and gives the expectation value of the relative velocity inside the bound state. Similar corrections arise also in $\M_{\rm trans}$ (see appendix~\ref{App:Ksi}). We shall retain such corrections only where the dominant term in the respective expansion vanishes, as in the case of degenerate particles interacting via scalar boson exchange (see Sec.~\ref{sec:SSS}). 
Moreover, from \eqs{eq:2omega2epsilon} and \eqref{eq:omega q}, we find that to zeroth order in the relative velocity,
\beq \ve_{\vec K, \vec k} \simeq \mu \, . \label{eq:epsilon NR} \eeq
Because $\ve_{\vec K, \vec k}$ factors out of the integrals, as seen in \eq{eq:Ampl BSF Inst}, we neglect $\vec k^2$ corrections, which always produce subdominant terms in $\vrel^2$. In addition, in our computations, we consistently ignore corrections that involve at least one power of the total momentum of any of the initial or final states (denoted typically with capital letters). In the CM frame, these momenta are of order $\sim \O (\a^2 + \vrel^2)$, which renders their scalar products with any other momenta, of higher order in $\a$ and $\vrel$ than the $\vec p^2, \vec q^2$ corrections.

We shall employ \eqs{eq:Ampl BSF Inst} -- \eqref{eq:Norm trans NR} to evaluate the transition amplitudes of Sec.~\ref{Sec:Interactions}.

\subsection{On-shell approximation}

Let us now consider the case when the perturbative part of the transition amplitude ${\cal C}^{(5)}$ consists only of fully connected diagrams.\footnote{This is the case if the initial-state particles and the particles participating in the bound state are different.} Then, ${\cal A}^{(5)} (P_\vf, \h_1 P + p, \h_2 P - p; \: \h_1 K + q, \h_2 K - q)$ is the perturbative amplitude for the $2\to3$ transition (with no on-shell conditions imposed). Equation~\eqref{eq:M trans} becomes
\begin{align}
\M_{\rm trans} (\vec q;\vec p) \simeq \!
\int \! \frac{dp^0}{2\p} \frac{dq^0}{2\p} 
\frac{S(p;P) \, S(q;K)}{{\cal S}_0(\vec p;P) \, {\cal S}_0(\vec q;K)}
\: {\cal A}^{(5)} (P_\vf, \h_1 P + p, \h_2 P - p; \: \h_1 K + q, \h_2 K - q) .
\label{eq:M BSF approx}
\end{align}

Provided that ${\cal A}^{(5)}$ has no singularities in $p^0$ and $q^0$,\footnote{${\cal A}^{(5)}(P_\vf, \h_1 P + p, \h_2 P - p; \: \h_1 K + q, \h_2 K - q)$ may have singularities in $p^0$ and $q^0$ (for the energy of interest, $K^0 = \w_{\vec K, \vec k}$), if the initial-state particles and the particles participating in the bound state are different.}  
the integrations over $p^0,q^0$ force the evaluation of ${\cal A}^{(5)}$ on the poles of $S(p;P)$ and $S(q;K)$ that are located in either the lower or upper $p^0$ and $q^0$ complex planes (depending on the choice of integration contours). As described in appendix~\ref{App:S cal}, each integration picks out two poles: One physical pole, which corresponds to setting one of the particles on-shell, and one unphysical pole, where the energy of the other particle is negative (c.f. \eqs{eq:poles t>0}, \eqref{eq:poles t<0}). In the non-relativistic regime, the contribution from the physical pole dominates. For concreteness, let us take these poles to be in the lower $p^0, q^0$ complex planes (as in \eq{eq:poles t<0}),
\begin{align}
p^0 &= -\h_1 P^0 + E_1(\vec p; \vec P) - i\e \, , \nn \\
q^0 &= -\h_1 K^0 + E_1(\vec q; \vec K) - i\e \, . \nn
\end{align}
Fixing $p^0$ and $q^0$ to the pole values means that the energies of the $\x_1, \x_2$ particles in the bound state and in the two-particle states are specified as functions of the 3-momenta $\vec p, \vec q, \vec P, \vec K$ and the quantum numbers $n$ and $\vec k$ (note that $P^0 = \w_{\vec P, n}$ and $K^0 = \w_{\vec K, \vec k}$), as follows
\begin{align}
p_1^0 = \h_1 P^0 + p^0 = E_1(\vec p; \vec P) \, , \qquad
&p_2^0 = \w_{\vec P, n} - E_1(\vec p; \vec P) \simeq 
E_2(\vec p; \vec P) + {\cal E}_n - \vec p^2 / (2\mu)  \, , 
\label{eq:momenta BS}
\\
q_1^0 = \h_1 K^0 + q^0 = E_1(\vec q; \vec K) \, , \qquad
&q_2^0 = \w_{\vec K, \vec k} - E_1(\vec q; \vec K) \simeq 
E_2(\vec q; \vec K) + {\cal E}_{\vec k} - \vec q^2/(2\mu)  \, , 
\label{eq:momenta US}
\end{align}
where we used \eqs{eq:omega n}, \eqref{eq:omega q} and \eqref{eq:E1 NR}, \eqref{eq:E2 NR}.  Evidently, in both the bound state and the two-particle state, the $\x_2$ degree of freedom is off-shell, by ${\cal E}_n - \vec p^2/2\mu$ and ${\cal E}_{\vec k}-\vec q^2/2\mu$ respectively. However, $\langle \vec p^2 \rangle  / (2\mu) \sim -{\cal E}_n$ and $\langle \vec q^2 \rangle  / (2\mu) \sim {\cal E}_{\vec k}$; provided that ${\cal E}_n, {\cal E}_{\vec k} \ll \mu$,\footnote{For a Coulomb or Yukawa potential with fine structure constant $\a$, these conditions are equivalent to $\a, \vrel \ll 1$ (c.f. \eq{eq:En,Ek}).}  
we may ignore this small deviation from the on-shell condition and evaluate ${\cal A}^{(5)}$ on-shell. Then, from \eq{eq:M trans}, we obtain
\begin{align}
\M_{\rm trans} (\vec q;\vec p) \simeq 
\[ {\cal A}^{(5)} (P_\vf, \h_1 P + p, \h_2 P - p; \: \h_1 K + q, \h_2 K - q) \]_{\rm on-shell}
\: .
\label{eq:M trans on-shell}
\end{align}
This is the approximation presented in Sec.~5.3 of Ref.~\cite{PeskinSchroeder}. Note that, for consistency, when using \eq{eq:M trans on-shell} inside \eqs{eq:Ampl BSF Inst} and \eqref{eq:Ampl LevelTrans Inst}, the normalisation factor of \eq{eq:Norm trans NR} should be approximated to zeroth order in $\vec p^2, \vec q^2$. Indeed, corrections of the order $\vec p^2, \vec q^2$ arise not only due to the normalisation factor, but also due to the off-shellness of the amplitude ${\cal A}^{(5)}$. When important, such corrections should be included self-consistently, by making use both of the full expansion of \eq{eq:Norm trans NR}, and of the off-shell momenta of \eqs{eq:momenta BS}, \eqref{eq:momenta US} instead of the on-shell conditions. 

We will not make use of \eq{eq:M trans on-shell} in our computations in Sec.~\ref{Sec:Interactions}.

\subsection{Bound-state formation cross-sections \label{sec:sigma BSF general}}

In the CM frame ($\vec K = 0$), the differential cross-section for radiative BSF, ${\cal U}_{\vec K=0, \vec k} \to {\cal B}_{\vec P, n} + \vf_{_{\!-\vec P}}$, is
\beq
\frac{d\s\BSF^{(n)}}{d\W} = 
\frac{1}{2\sqrt{(s-m_1^2 -m_2^2)^2 - 4 m_1^2 m_2^2}} 
\: \frac{|\vec P|}{16\p^2 \, \sqrt{s}} \: |\M_{\vec k \to n}|^2 
\: ,
\label{eq:diff sigma 0}
\eeq
where $s = \w_{\vec K=0, \vec k}^2 \simeq (m+{\cal E}_{\vec k})^2$ (c.f. \eq{eq:omega q}), and $\M_{\vec k \to n}$ is found from \eq{eq:Ampl BSF Inst}. The bound-state and mediator momenta are
\beq
|\vec P| = 
\[\frac{(s-M_n^2-m_\vf^2)^2 - 4 M_n^2 m_\vf^2}{4s}\]^{1/2} 
\simeq ({\cal E}_{\vec k} - {\cal E}_n) \[1-\frac{m_\vf^2}{({\cal E}_{\vec k} - {\cal E}_n)^2}\]^{1/2}
\: ,
\label{eq:P CM}
\eeq
where we used $M_n = m + {\cal E}_n$ (c.f. \eq{eq:Mn}). In addition, 
\beq
2\sqrt{(s-m_1^2 -m_2^2)^2 - 4 m_1^2 m_2^2} 
\: \simeq \: 4 m k  \: = \:  4 m \mu \vrel \: .
\eeq
Then
\beq
\frac{d\s\BSF^{(n)}}{d\W} 
\ = \ \frac{({\cal E}_{\vec k} - {\cal E}_n)}{64 \p^2 m^2\mu\vrel } 
\[1-\frac{m_\vf^2}{({\cal E}_{\vec k} - {\cal E}_n)^2}\]^{1/2}
\ |\M_{\vec k \to n}|^2
\: .
\label{eq:diff sigma}
\eeq

\subsubsection*{Partial-wave decomposition and unitarity}

It will be useful to decompose the amplitude $\M_{\vec k \to n}$ in partial waves
\beq
\M_{\vec k \to n} (\W) = \sum_J  \(\frac{2J + 1}{4\p}\) \M_J P_J (\cos\theta)
\label{eq:M decomp}
\eeq
where $P_J$ are the Legendre polynomials, and  
\beq
\M_J = \int d\W  \, P_J(\cos \theta) \, \M_{\vec k \to n} (\W) \: .
\label{eq:MJ}
\eeq
Then, \eq{eq:diff sigma} gives
\beq
\s\BSF^{(n)} = \sum_J \s_{_{{\rm BSF},J}}^{(n)} \: .
\label{eq:sigma = sumJ sigmaJ}
\eeq
with the partial-wave cross-section given by
\beq
\s_{_{{\rm BSF},J}}^{(n)} =
\frac{({\cal E}_{\vec k} - {\cal E}_n)}{64 \p^2 m^2\mu \vrel} \[1-\frac{m_\vf^2}{({\cal E}_k - {\cal E}_n)^2}\]^{1/2} 
\frac{2J+1}{4\p} \ |\M_J|^2 \: .
\label{eq:sigmaJ}
\eeq

Unitarity implies an upper limit on the partial-wave inelastic cross-sections. In the non-relativistic regime, for the $J$th partial wave, this is~\cite{Griest:1989wd}
\beq
\s_{{\rm inel}, J}  \leqslant (\s_{{\rm inel}, J})_{\rm max} = \frac{(2J+1)\p}{\mu^2 \vrel^2} \, .
\label{eq:unitarity}
\eeq
For a given inelastic process and associated cross-section, this bound yields an estimate for the value of the coupling at which the probability for inelastic scattering saturates. In Sec.~\ref{Sec:Interactions}, we use the unitarity bound to deduce the range of validity of our calculation.

\subsection{Bound-state de-excitation rates}

The differential rate for the radiative de-excitation of a bound state, 
${\cal B}_{\vec K=0,n'} \to {\cal B}_{\vec P,n} + \vf_{_{\!-\vec P}}$, is
\beq
\frac{d\G_{n' \to n}}{d\W} = \frac{|\vec P|}{32 \p^2 M_{n'}^2} |\M_{n'\to n}|^2 \: ,
\nn 
\eeq
where $ |\M_{n'\to n}|^2$ is found from \eq{eq:Ampl LevelTrans Inst} and $|\vec P|$ is given by \eq{eq:P CM} with the replacement ${\cal E}_{\vec k} \to {\cal E}_{n'}$. 
Setting $M_n' = m +{\cal E}_n \simeq m$, we obtain 
\beq
\G_{n' \to n} \simeq  
\frac{({\cal E}_{n'} - {\cal E}_n)}{32 \p^2 m^2} 
\[1-\frac{m_\vf^2}{({\cal E}_{n'} - {\cal E}_n)^2}\]^{1/2} 
\int |\M_{n'\to n}|^2 \, d\W\: .
\label{eq:de-excitation rate}
\eeq
%


\section{Decay of unstable bound states and (co-)annihilation of unbound pairs \hspace{-3mm}  \label{Sec:AnnDec}}

\subsection{Non-perturbative amplitude}

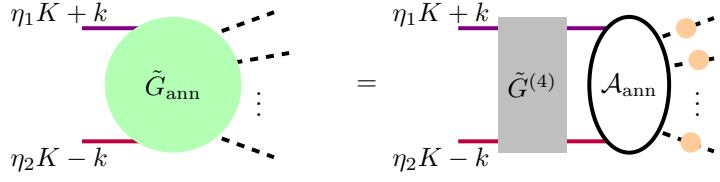
\begin{figure}[t]
\centering
\begin{tikzpicture}[line width=1.5 pt, scale=1.5]
\begin{scope}
\node at (-1, 0.62){$\h_1 K + k$};
\node at (-1,-0.66){$\h_2 K - k$};
\draw[violet] (-0.8, 0.5)--(0, 0.5);
\draw[purple] (-0.8,-0.5)--(0,-0.5);
\draw[scalarnoarrow,rotate= 20] (0, 0.3)--(1.05, 0.3);
\draw[scalarnoarrow,rotate= 10] (0, 0.1)--(1.05, 0.1);
\draw[scalarnoarrow,rotate=-20] (0,-0.3)--(1.05,-0.3);
\node at (0.75,-0.1){$\vdots$};
\draw[fill=green!30,draw=none] (0,0) circle (0.6cm);
\node at (0,0){$\tilde G\ann$};
\node at (1.7,0){$=$};
\end{scope}
\begin{scope}[shift={(4,0)}]
\node at (-1.65, 0.62){$\h_1 K + k$};
\node at (-1.65,-0.66){$\h_2 K - k$};
\draw[violet] (-1.5, 0.5)--(0, 0.5);
\draw[purple] (-1.5,-0.5)--(0,-0.5);
\draw[scalarnoarrow,rotate= 20] (0, 0.3)--(0.9, 0.3);
\draw[scalarnoarrow,rotate= 10] (0, 0.1)--(0.9, 0.1);
\draw[scalarnoarrow,rotate=-20] (0,-0.3)--(0.9,-0.3);
\node at (0.6,-0.1){$\vdots$};
\draw[fill=lightgray,draw=none] (-1.15,-0.6) rectangle (-0.55,0.6);
\draw[fill=white] (0,0) ellipse (0.35cm and 0.6cm);
\draw[fill=orange!40,draw=none] (0.5,0.5) circle (0.09cm);
\draw[fill=orange!40,draw=none] (0.61,0.22) circle (0.09cm);
\draw[fill=orange!40,draw=none] (0.54,-0.51) circle (0.09cm);
\node at (-.85,0){$\tilde G^{(4)}$};
\node at ( 0,0){${\cal A}\ann$};
\end{scope}
\end{tikzpicture}
\caption{Diagrammatic representation of equation~\eqref{eq:Gann decomp}. 
\protect\PropAnn represent the full propagators of annihilation/decay products $f_j$.
}
\label{fig:AnnDec}
\end{figure}

If an unbound $\x_1, \, \x_2$ pair can (co-)annihilate into a number of light particles $f_1, \cdots f_N$, then the $\x_1 -\x_2$ bound states are unstable against decay into the same final states (provided that this is allowed by angular momentum conservation). For example, unbound and bound particle-antiparticle pairs can annihilate and decay, respectively, into force mediators. (Co-)annihilation and bound-state decay have the same diagrammatic representation, shown in Fig.~\ref{fig:AnnDec}; the difference in evaluating these two processes is which initial state is singled out from the $G^{(4)}$ function. In this section, we express the (co-)annihilation and the bound-state decay amplitudes, 
\begin{align}
{}_{\rm out}\langle f_1 f_2 \cdots f_N \,|\, {\cal U}_{\vec K, \vec k} \rangle_{\rm in} 
\ = \
\langle f_1 f_2 \cdots f_N \,|\, {\sf S} \,|\, {\cal U}_{\vec K, \vec k} \rangle \: ,
\label{eq:S matrix ann}
\\
{}_{\rm out}\langle f_1 f_2 \cdots f_N \,|\, {\cal B}_{\vec K, n} \rangle_{\rm in} 
\ = \ 
\langle f_1 f_2 \cdots f_N \,|\, {\sf S} \,|\, {\cal B}_{\vec K, n} \rangle \: ,
\label{eq:S matrix dec}
\end{align}
in terms of the initial state wavefunction and the perturbative interaction that gives rise to these processes.\footnote{As already mentioned, for annihilation processes, similar analyses have been carried out in  previous works,  e.g.~\cite{Hisano:2002fk, Hisano:2003ec, Cirelli:2007xd, Iengo:2009ni, Cassel:2009wt, Hryczuk:2011tq, Hryczuk:2014hpa,Beneke:2014hja}.}

To calculate the $\mathsf{S}$-matrix elements \eqref{eq:S matrix ann} and \eqref{eq:S matrix dec}, we need to consider the Green's function
\beq
G\ann (x_1, x_2, ... x_N;\, y_1, y_2) \equiv \<\W| T f(x_1) f(x_2) ... f(x_N) \, \x_1^\dagger (y_1) \x_2^\dagger (y_2)|\W\> \: .
\label{eq:Gann}
\eeq
and its Fourier transform
\begin{align}
\tilde G\ann (p_1, ... p_N;\, k_1, k_2)  
&= \prod_{j=1}^N \int d^4x_j \ e^{i p_j x_j} 
\int d^4y_1 \, d^4y_2 \ e^{-i (k_1 y_1 + k_2 y_2)}
\ G\ann (x_1, \cdots, x_N; y_1,y_2) \: .
\label{eq:Gann FT}
\end{align}
Let ${\cal A}\ann (p_1, \cdots p_N; \: k_1, k_2)$ be the sum of all connected and amputated diagrams contributing the (co-)annihilation of a $\x_1, \,\x_2$ pair with momenta $k_1, k_2$, into $f_1, \cdots, f_N$ particles with momenta $p_1, \cdots, p_N$.\footnote{Note that, in contrast to level transitions processes, all diagrams contributing to the perturbative part of the (co-)annihilation processes are fully connected. The amputation of fully connected diagrams is well defined, and it is thus sensible to express the full amplitudes for the (co-)annihilation and decay processes of interest, in terms of the sum of amputated diagrams.} 
Then, as sketched in Fig.~\ref{fig:AnnDec},
\begin{multline}
\tilde G\ann (p_1, \cdots, p_N; \h_1 K +k, \h_2 K-k) =
\\
= \prod_{j=1}^N \tilde{S}_{f_j}(p_j)
\int \frac{d^4 k'}{(2\p)^4} \ {\cal A}\ann (p_1, \cdots, p_N; \h_1 K +k', \h_2 K-k') \ \tilde G^{(4)} (k',k;K) \: ,
\label{eq:Gann decomp}
\end{multline}
where $\tilde{S}_{f_j}(p_j)$ is the propagator of the $f_j$ particle with momentum $p_j$. As always, energy-momentum conservation implies that
\beq
{\cal A}\ann (p_1, \cdots, p_N; \: k_1, k_2) = 
i (2\p)^4\, \d^4(k_1 + k_2 - p_1 \cdots - p_N) \: \M\ann^{\rm pert} (k_1, k_2; \, p_1, \cdots, p_N) \, ,
\eeq
where $\M\ann^{\rm pert}$ is the perturbative annihilation amplitude, with no on-shell conditions imposed on the incoming and outgoing degrees of freedom.  We follow a similar procedure as in Sec.~\ref{sec:LSZ}, and determine the $\mathsf{S}$-matrix elements of interest to be
\begin{align}
\langle f_1 \cdots f_N \,|\, {\sf S} \,|\, {\cal U}_{\vec K, \vec k} \rangle 
& =  (2\p)^4 \, \d^4 (K - p_1 \cdots - p_N) \ i \M_{\rm ann} \: ,
\label{eq:S ann = delta x M }
\\
\langle f_1 \cdots f_N \,|\, {\sf S} \,|\, {\cal B}_{\vec K, n} \rangle 
& =  (2\p)^4 \, \d^4 (K - p_1 \cdots - p_N) \ i \M\dec^{(n)} \: ,
\label{eq:S dec = delta x M }
\end{align}
with
\begin{multline}
\M_{\rm ann} = \prod_{j=1}^N  \sqrt{Z_{f_j} (\vec p_j)}
\int \frac{d^4 q}{(2\p)^4} \ \tilde \F_{\vec K, \vec k}(q)  
\ \M\ann^{\rm pert} (\h_1 K + q, \h_2 K - q; \, p_1, \cdots, p_N)  
\\
\simeq \sqrt{2\ve_{\vec K, \vec k}}
\int \! \frac{d^3 q}{(2\p)^3}
\, \frac{\tilde \f_{\vec k}(\vec q)}{\sqrt{2{\cal N}_{\vec K}(\vec q)}}
\int \! \frac{dq^0}{2\p}  \, \frac{S (q;K)}{{\cal S}_0 (\vec q; K)}
\ \M\ann^{\rm pert} (\h_1 K + q, \h_2 K - q; \, p_1, \cdots, p_N) \: ,
\label{eq:Ampl ann}
\end{multline}
\begin{multline}
\M\dec^{(n)} = \prod_{j=1}^N  \sqrt{Z_{f_j} (\vec p_j)}
\int \frac{d^4 q}{(2\p)^4} \ \tilde \Ps_{\vec K, n}(q)  
\ \M\ann^{\rm pert} (\h_1 K + q, \h_2 K - q; \, p_1, \cdots, p_N) 
\\
\simeq 
\int \! \frac{d^3 q}{(2\p)^3} 
\, \frac{\tilde \ps_n(\vec q)}{\sqrt{2{\cal N}_{\vec K}(\vec q)}}
\int \! \frac{dq^0}{2\p}  \, \frac{S (q;K)}{{\cal S}_0 (\vec q; K)}
\, \M\ann^{\rm pert} (\h_1 K + q, \h_2 K - q; \, p_1, \cdots, p_N)  \: ,
\label{eq:Ampl dec}
\end{multline}
where  $K^0 = \w_{\vec K, \vec k}$ and $K^0 = \w_{\vec K, n}$, respectively. In the above, $Z_{f_j} (\vec p) \equiv |\langle \W|f_j(0)|f_{j,\vec p}\rangle|^2$, with $f_{j,\vec p}$ being a $f_j$ particle with momentum $\vec p$. In the second line \eqs{eq:Ampl ann} and \eqref{eq:Ampl dec}, we have used the instantaneous approximation for the wavefunctions, and set $Z_{f_j} (\vec p)\simeq 1$.

\subsection{On-shell approximation}

Following the same arguments as in Sec.~\ref{sec:trans ampl inst}, we may evaluate the perturbative amplitude $\M\ann^{\rm pert}$ on-shell.  This enables us to express the annihilation and decay amplitudes as follows
\begin{align}
\M\ann &\simeq
\int \frac{d^3 q}{(2\p)^3} \ \tilde \f_{\vec k}(\vec q)
\ \hat\M\ann^{\rm pert} (\h_1 \vec K + \vec q, \h_2 \vec K - \vec q; \, \vec p_1, \cdots, \vec p_N)  \: ,
\label{eq:Ampl ann NR}
\\
\M\dec^{(n)} &\simeq 
\frac{1}{\sqrt{2\mu}} \int \frac{d^3 q}{(2\p)^3} \ \tilde \psi_n(\vec q)
\ \hat\M\ann^{\rm pert} (\h_1 \vec K + \vec q, \h_2 \vec K - \vec q; \, \vec p_1, \cdots, \vec p_N)  \: ,
\label{eq:Ampl dec NR}
\end{align}
where $\hat\M\ann^{\rm pert}$ is the on-shell perturbative (co-)annihilation amplitude. Note that, as discussed below \eq{eq:M trans on-shell}, the integrands in \eqs{eq:Ampl ann NR}, \eqref{eq:Ampl dec NR} admit $\vec q^2$ and higher order corrections from the normalisation factor of \eq{eq:Norm NR} and from the off-shellness of the perturbative amplitude $\M\ann^{\rm pert}$.

\subsection{Two-body (co-)annihilation cross-sections and bound-state decay rates}

Let us now focus on the case of decays and (co-)annihilations into two final-state particles. In the CM frame ($\vec K=0$), the momenta of the final particles are $|\vec p_1| = |\vec p_2| =|\vec p|$, with $|\vec p| \simeq \w_{\vec K = 0, \vec k} / 2 = (m+{\cal E}_{\vec k})/2 \simeq m/2$ in the case of (co-)annihilation, and $|\vec p|  \simeq M_n/2 = (m+{\cal E}_n)/2 \simeq m/2$ in the case of decay. We ignore the masses of the final-state particles for simplicity. The (co-)annihilation and decay amplitudes can be expanded in partial waves as follows
\begin{align}
\M\ann (\W_{\vec p}) = \sum_{\ell = 0}^\infty \(\frac{2 \ell +1}{4\p}\) 
P_\ell (\cos\theta_{\vec p}) \, \M_{{\rm ann}, \ell} \, ,
\label{eq:Mann ell}
\\
\M\dec^{(n)} (\W_{\vec p}) = \sum_{\ell = 0}^\infty \(\frac{2 \ell +1}{4\p}\) 
P_\ell (\cos\theta_{\vec p}) \, \M_{{\rm dec}, \ell}^{(n)} \, ,
\label{eq:Mdec ell}
\end{align}
where
\begin{align}
\M_{\rm ann, \ell} \equiv \int d\W_{\vec p} \, P_\ell (\cos\theta_{\vec p}) \M\ann (\W_{\vec p}) ,
\\
\M_{\rm dec, \ell}^{(n)} \equiv \int d\W_{\vec p} \, P_\ell (\cos\theta_{\vec p}) \M\dec^{(n)} (\W_{\vec p}) .
\end{align}
Here and in the following, the indices in the angle variables specify the vector to which this angle refers; a double index denotes the angle between the two vectors. The (co-)annihilation cross-section times relative velocity and the decay rate are
\begin{align} 
\s\ann \vrel &
= \frac{f_s}{128 \p^2 m \mu} \, \int |\M\ann (\W_{\vec p})|^2 \, d\W_{\vec p} 
= \frac{f_s}{128 \p^2 m \mu} \, \sum_\ell \frac{2\ell +1}{4\p} \, |\M_{\rm ann, \ell}|^2
\, ,
\label{eq:sigma ann 2-body}
\\
\Gamma\dec^{(n)} &
= \frac{ f_s }{64\p^2 m} \, \int |\M\dec^{(n)} (\W_{\vec p})|^2 \, d\W_{\vec p}
= \frac{ f_s }{64\p^2 m} \, \sum_\ell \frac{2\ell +1}{4\p} |\M_{\rm dec, \ell}^{(n)}|^2 
\, ,
\label{eq:decay rate 2-body}
\end{align}
where $f_s =1/2$ if the final-state particles are identical, or $f_s=1$ otherwise.  We shall now express $\M_{\rm ann, \ell}$ and $\M_{\rm dec, \ell}^{(n)}$ in terms of the perturbative on-shell annihilation amplitude.

We expand $\hat \M\ann^{\rm pert}$ in partial waves, as follows
\beq 
\hat\M\ann^{\rm pert} (\vec q, -\vec q; \vec p, - \vec p) = 
\sum_{\ell=0}^\infty \frac{\tilde a_\ell}{(m\mu)^\ell} \, |\vec p|^\ell \, |\vec q|^\ell \, P_\ell(\cos\theta_{\vec q, \vec p}) \: .
\label{eq:M ann expansion}
\eeq
In general, the expansion coefficients $\tilde a_J$ may depend on $\vec q$; in the non-relativistic regime, they can be expanded as
\beq
\tilde a_\ell (\vec q) \simeq a_\ell + {\cal F}_\ell(\vec q^2, \ \boldsymbol{\e}_A \cdot \vec q) \, ,
\label{eq:a ell expansion}
\eeq
where $\boldsymbol{\e}_A$ stands for the polarisation vectors of possible final-state vector bosons, and ${\cal F}_\ell$ is a polynomial function of the scalar products $\vec q^2$ and $\boldsymbol{\e}_A \cdot \vec q$ that vanishes at $\vec q=0$. Note that $|\vec p|$ is determined by energy conservation. $a_\ell$ and ${\cal F}_\ell$ may depend on scalar products such as $\vec p^2$, $\boldsymbol{\e}_A \cdot \vec p$ and $\boldsymbol{\e}_A \cdot \boldsymbol{\e}_B$. In the following, we consider only the $a_\ell$ contribution to $\tilde a_\ell$; any corrections arising from the $\vec q$-dependent terms of \eq{eq:a ell expansion} may be included only in conjunction with similar corrections arising from the normalisation factor of \eq{eq:Norm NR} and from the off-shellness of the perturbative amplitude $\M\ann^{\rm pert}$.

We may now insert \eq{eq:M ann expansion} into \eqs{eq:Ampl ann NR} and \eqref{eq:Ampl dec NR}, and use the formula
\begin{multline}
\int d\W_{\vec p} \, P_{\ell'} (\cos\theta_{\vec p}) \int \frac{d^3 q}{(2\p)^3}  \: \tilde \f_{\vec k} (\vec q) 
\: |\vec q|^\ell P_\ell (\cos\theta_{\vec q, \vec p}) 
= \\ =
\d_{\ell\ell'} \: \frac{(2\ell+1)!!}{i^\ell \, (2\ell +1) \, \ell!}
\[\frac{d^\ell}{dr^\ell} \int d\W_{\vec r} \: 
P_\ell (\cos \theta_{\vec r}) \: \f_{\vec k} (\vec r) 
\]_{\vec r=0} \, ,
\label{eq:partial waves}
\end{multline}
and similarly for $\ps_n$. We prove \eq{eq:partial waves} in appendix~\ref{App:Partial Waves}. Keeping only the $\vec q$-independent term from the expansion of \eq{eq:a ell expansion}, we find
\begin{align}
\M_{{\rm ann}, \ell} &\simeq 
\frac{a_\ell \, |\vec p|^\ell}{(m\mu)^\ell} 
\: \frac{(2\ell+1)!!}{i^\ell (2\ell+1) \ell!} 
\: \[\frac{d^\ell}{dr^\ell} \int d\W_{\vec r} P_\ell (\cos \theta_{\vec r}) \f_{\vec k} (\vec r) \]_{\vec r=0} ,
\\
\M_{{\rm dec}, \ell}^{(n)} &\simeq 
\frac{a_\ell \, |\vec p|^\ell}{\sqrt{2\mu} \, (m\mu)^\ell} 
\: \frac{(2\ell+1)!!}{i^\ell (2\ell+1) \ell!} 
\: \[\frac{d^\ell}{dr^\ell} \int d\W_{\vec r} P_\ell (\cos \theta_{\vec r}) \ps_n (\vec r) \]_{\vec r=0} .
\label{eq:Mann J}
\end{align}
Using \eqs{eq:sigma ann 2-body} and \eqref{eq:decay rate 2-body}, we find the contribution of the $\ell$ partial wave to $\s\ann \vrel$ and $\G\dec^{(n)}$ to be
\begin{align}
(\s_{\rm ann} \, \vrel)_\ell &= \s_\ell \: S_{\ell, \rm ann} \, ,
\label{eq:sigma ann ell}
\\
\G_{\ell, \rm dec}^{(n)} &= \s_\ell \, S_{\ell, \rm dec} \, ,
\label{eq:decay rate ell}
\end{align}
where
\beq
\s_\ell =  \frac{\[\ell! / (2\ell)!!\]^2}{2\ell+1} 
\: \frac{f_s \, |a_\ell|^2}{32\p m\mu} \, ,
\label{eq:sigma ell}
\eeq
and 
\begin{align}
S_{\ell, \rm ann} &= 
\frac{\[(2\ell+1)! / (\ell!)^2 \]^2}{4^{\ell+2} \,\p^2 \, \mu^{2\ell} }
\: \left|\frac{d^\ell}{dr^\ell} \int d\W_{\vec r} P_\ell (\cos \theta_{\vec r}) 
\f_{\vec k} (\vec r) \right|^2_{\vec r=0} \, ,
\label{eq:S ann ell}
\\
S_{\ell, \rm dec}^{(n)} &= 
\frac{\[(2\ell+1)! / (\ell!)^2 \]^2}{4^{\ell+2} \,\p^2 \, \mu^{2\ell} }
\: \left|\frac{d^\ell}{dr^\ell} \int d\W_{\vec r} P_\ell (\cos \theta_{\vec r}) 
\psi_n (\vec r) \right|^2_{\vec r=0} \, .
\label{eq:S dec ell}
\end{align}
In the limit where the interaction in the two-particle state can be neglected, $\f_{\vec k} (\vec r) = e^{i \vec k \cdot \vec r}$ and $S_{\ell, \rm ann} = (|\vec k|/\mu)^{2\ell} = \vrel^{2\ell}$.

Similar analyses to the above for the non-perturbative annihilation cross-section, have been performed in Refs.~\cite{Iengo:2009ni,Cassel:2009wt}, where also the Sommerfeld enhancement factors of \eq{eq:S ann ell} have been computed for a Yukawa potential.


\section{Bound-state formation, de-excitation and decay rates for specific interactions \label{Sec:Interactions}}

We now focus on specific interactions and apply the formalism of the previous sections to calculate the BSF cross-sections, and the rates for de-excitation or decay of bound states, where relevant. We consider the interaction of two scalar particles (i) via a light scalar boson (Sec.~\ref{sec:SSS}), and (ii) via an Abelian gauge vector boson (Sec.~\ref{sec:SSV}).

In the instantaneous approximation, these interactions are described in general by a Yukawa potential.\footnote{For a classification of the low-energy effective potentials generated by long-range interactions, and a systematic renormalisation procedure of singular potentials, see Ref.~\cite{Bellazzini:2013foa}.}  
(Of course, an unbroken gauge symmetry gives rise to a Coulomb potential.)  A Yukawa potential admits bound state solutions if $m_\vf < \a \mu$, where $\a$ is the fine-structure constant of the interaction. On the other hand, the radiative formation of bound states via emission of a force mediator is kinematically possible if $m_\vf < (\a^2 +\vrel^2)\mu/2$; for $\vrel < \a$ -- which is when the Sommerfeld effect renders bound-state formation efficient -- this is a much stronger condition. Provided that this condition holds, the distortion of the wavefunctions due to the non-zero mediator mass, from their Coulomb limit, is expected to be negligible. For simplicity, we shall thus perform our computations in the Coulomb limit.

As is well known, in the presence of an attractive Coulomb potential 
\beq
V(\vec r) = - \frac{\a}{r} \, , \quad \a>0 \, ,
\label{eq:Coulomb pot}
\eeq
there is a discrete spectrum and a continuous spectrum of energy eigenstates. The continuous spectrum corresponds to the two-particle states, and is characterised by a continuous quantum number that stands for the expectation value of the momentum of the reduced system,  $\vec k = \mu \vec\vrel$, with $\vec \vrel$ being the expectation value of the relative velocity. The discrete spectrum corresponds to the bound states, and is characterised by the integer-valued quantum numbers $\{n\ell m\}$. In the discrete spectrum, the expectation value of the momentum of the reduced system is $\kappa/n$, with $\kappa \equiv \mu \a$ being the Bohr momentum. As we shall see, the parameter that essentially determines the efficiency of BSF is the ratio of the momentum expectation values of the bound and the scattering states,
\beq 
\z \equiv  \frac{\k}{k} = \frac{\a}{\vrel} \: .
\label{eq:zeta def}
\eeq

The energies of the states of the discrete and the continuous spectra are
\begin{align}
\w_{\vec P, n} = m + \frac{\vec P^2}{2m} + {\cal E}_n \, ,
\qquad
\w_{\vec K, \vec k} = m + \frac{\vec K^2}{2m} + {\cal E}_{\vec k} \: ,
\end{align}
where $\vec P$, $\vec K$ are the momenta of the CM of the bound and the two-particle states respectively, and
\beq  
{\cal E}_n = - \frac{\k^2}{2\mu} = -\frac{\mu \a^2}{2n^2} \, ,
\qquad 
{\cal E}_{\vec k} = \frac{\vec k^2}{2\mu} = \frac{\mu \vrel^2}{2} \, .
\label{eq:En,Ek}
\eeq
The wavefunctions are given in appendix~\ref{app:Schr sol}.

\subsubsection*{Useful integrals}

For the calculation of the amplitudes $\M_{\vec k \to n}$ and $\M_{n' \to n}$, we will find it useful to define the integrals 
\begin{align}
\Ks_1 (\vec q, \vec p;  K,  P) \equiv
\int \frac{dq^0}{2\p} \: S(q;K)
\int \frac{dp^0}{2\p} \tilde{S}_1(\h_1 P + p) 
\: (2\p) \d(q^0-p^0-\h_2 P_\vf^0)  
\: ,
\label{eq:Ksi 1 def}
\\
\Ks_2 (\vec q, \vec p;  K,  P) \equiv
\int \frac{dq^0}{2\p} \: S(q;K)
\int \frac{dp^0}{2\p} \tilde{S}_2(\h_2 P - p) 
\: (2\p) \d(q^0-p^0+\h_1 P_\vf^0)  
\: .
\label{eq:Ksi 2 def}
\end{align}
We evaluate $\Ks_1$ and $\Ks_2$ in appendix~\ref{App:Ksi}. We will also need the following integrals involving the initial and final state wavefunctions
\begin{align}
{\cal I}_{\vec k, n} (\vec b) 
&\equiv 
\int \frac{d^3p}{(2\p)^3} \:\tilde \psi_n^\star (\vec p) \: \tilde \f_{\vec k}  (\vec p + \vec b) \: ,
\label{eq:I cal k-n def}
\\
\boldsymbol{\cal J}_{\vec k, n} (\vec b) 
&\equiv 
\int \frac{d^3p}{(2\p)^3} \: \vec p \: \tilde \psi_n^\star (\vec p) \: \tilde \f_{\vec k}  (\vec p + \vec b) \: ,
\label{eq:J cal k-n def}
\\
{\cal K}_{\vec k, n} (\vec b) 
&\equiv 
\int \frac{d^3p}{(2\p)^3} \: \vec p^2 \: \tilde \psi_n^\star (\vec p) \: \tilde \f_{\vec k}  (\vec p + \vec b) \: ,
\label{eq:K cal k-n def}
\end{align}
and 
\begin{align}
{\cal I}_{n', n} (\vec b) 
&\equiv 
\int \frac{d^3p}{(2\p)^3} \:\tilde \psi_n^\star (\vec p) \: \tilde \psi_{n'}  (\vec p + \vec b) \: ,
\label{eq:I cal n'-n def}
\\
\boldsymbol{\cal J}_{n', n} (\vec b) 
&\equiv 
\int \frac{d^3p}{(2\p)^3} \: \vec p \: \tilde \psi_n^\star (\vec p) \: \tilde \psi_{n'}  (\vec p + \vec b) \: ,
\label{eq:J cal n'-n def}
\\
{\cal K}_{n', n} (\vec b) 
&\equiv 
\int \frac{d^3p}{(2\p)^3} \: \vec p^2 \: \tilde \psi_n^\star (\vec p) \: \tilde \psi_{n'}  (\vec p + \vec b) \: .
\label{eq:K cal n'-n def}
\end{align}
We evaluate ${\cal I}, \boldsymbol{\cal J}$ and ${\cal K}$ in appendix~\ref{App:WaveFun}, for the initial and final states of interest. We shall use the integrals \eqref{eq:Ksi 1 def} -- \eqref{eq:K cal n'-n def} in sections~\ref{sec:SSS} and \ref{sec:SSV}.

\subsection{Scalar mediator \label{sec:SSS}}

We consider the interaction Lagrangians
\begin{align}
\d{\cal L}_{S,r} 
&= \frac{1}{2} \partial_\m \x_1 \, \partial^\m \x_1
 + \frac{1}{2} \partial_\m \x_2 \, \partial^\m \x_2
 + \frac{1}{2} \partial_\m \vf  \, \partial^\m \vf
- \frac{1}{2} m_1^2 \x_1^2 - \frac{1}{2} m_2^2 \x_2^2 - \frac{1}{2} m_\vf^2 \f^2
\nn \\
&- \frac{1}{2} g_1 m_1 \vf \x_1^2 - \frac{1}{2} g_2 m_2 \vf \x_2^2
\: ,
\label{eq:Langr SSS r}
\\
\d{\cal L}_{S,c} 
&= \partial_\m \x_1^\dagger \, \partial^\m \x_1
+ \partial_\m \x_2^\dagger \, \partial^\m \x_2
+ \frac{1}{2}\partial_\m \vf \, \partial^\m \vf
- m_1^2 |\x_1|^2 - m_2^2 |\x_2|^2 - \frac{1}{2} m_\vf^2 \f^2
\nn \\
&- g_1 m_1 \vf |\x_1|^2 - g_2 m_2 \vf |\x_2|^2
\: .
\label{eq:Langr SSS c}
\end{align}
In \eq{eq:Langr SSS r}, $\x_1$ and $\x_2$ are real scalar fields, while in \eq{eq:Langr SSS c} they are complex. $\f$ is a real scalar boson, with mass $m_\vf \ll m_1, m_2$, and $g_1, g_2$ are dimensionless couplings.\footnote{In \eqs{eq:Langr SSS r}, \eqref{eq:Langr SSS c} and \eqref{eq:Langr SSV}, we omit the quartic couplings in the scalar potential, since they do not enter our calculations. \label{foot:quartic}}

\begin{figure}[t]
\centering
\begin{tikzpicture}[line width=1.5 pt, scale=1.3]
\node at (-2.5,0) {$\tilde W (p,p';Q) \ = $};
\begin{scope}
\node at (-0.8, 0.7) {$\h_1 Q + p$};
\node at (-0.8,-0.7) {$\h_2 Q - p$};
\draw[violet] (-0.5, 0.5) -- (0.5, 0.5);
\draw[purple] (-0.5,-0.5) -- (0.5,-0.5);
\node at (0.8, 0.7) {$\h_1 Q + p'$};
\node at (0.8,-0.7) {$\h_2 Q - p'$};
\draw[fill=white] (0,0) ellipse (0.3cm and 0.6cm);
\end{scope}
\node at (2,0) {$\simeq$};
\begin{scope}[shift={(3.9,0)}]
\node at (-0.8, 0.7) {$\h_1 Q + p$};
\node at (-0.8,-0.7) {$\h_2 Q - p$};
\draw[violet] (-0.5, 0.5) -- (0.5, 0.5);
\draw[purple] (-0.5,-0.5) -- (0.5,-0.5);
\node at (0.8, 0.7) {$\h_1 Q + p'$};
\node at (0.8,-0.7) {$\h_2 Q - p'$};
\draw[vector] (0,-0.5) -- (0,0.5);
\end{scope}
\end{tikzpicture}
\caption{In the non-relativistic regime, the one-boson exchange is the dominant contribution to the $\x_1 - \x_2$ interaction. Consequently, the 4-point function is approximated by an infinite ladder of one-boson exchange diagrams. The exchanged boson may be either scalar or vector.}
\label{fig:one boson}
\end{figure}

To lowest order, the interaction between $\x_1$ and $\x_2$ is mediated by one-boson exchange, as shown in Fig.~\ref{fig:one boson}. Then
\beq
\tilde W(p,p';Q)  =  -\frac{i g_1 g_2 m_1 m_2}{(p-p')^2-m_\vf^2} \: .
\label{eq:W SSS}
\eeq
In the instantaneous approximation\footnote{ \label{foot:corrections}
Note that the leading-order correction to the approximation of \eq{eq:W instant SSS}, and similarly of \eq{eq:W instant SSV}, are of order $(p^0-p'^0)^2$. Due to the linearity of the Bethe-Salpeter \eqs{eq:BSE bound}, \eqref{eq:BSE free}, the corresponding corrections to the approximations of the Bethe-Salpeter wavefunctions, $\tilde \Ps_{\vec Q, n}(p)$ and $\tilde \F_{\vec Q, \vec q}(p)$, that are introduced via \eqs{eq:Schr WF bound def} and \eqref{eq:Schr WF free def}, would also be of the same order. As discussed in earlier sections, in the regime of interest, $p^0 \sim \O(\vec p^2/\mu)$, and such corrections would then be $\O(\vec p^4/\mu^2)$, i.e. of higher order than the corrections considered in \eq{eq:Norm NR} and which appear in the $\mathsf{S}$-matrix elements \eqref{eq:Ampl BSF Inst} and \eqref{eq:Ampl LevelTrans Inst}. This confirms the consistency of the approximation.}
\beq
{\cal W}(|\vec p - \vec p'|)  = \frac{i g_1 g_2 m_1 m_2}{({\bf p-p'})^2+m_\vf^2} \, .
\label{eq:W instant SSS}
\eeq
From \eq{eq:U def}, we find the non-relativistic potential,
\beq
V(\vec r) 
= -\frac{1}{i 4 m \m}\int \frac{d^3p}{(2\pi)^3} {\cal W}({\bf p}) \, e^{i \bf p \cdot r}  
= -\frac{i\,g_1 g_2 m_1 m_2}{i 4 m \m} \int \frac{d^3p}{(2\pi)^3} \frac{e^{i \bf p \cdot r}}{\vec p^2+m_\vf^2} 
\: .
\eeq
That is,
\beq
V({\bf r}) = -\frac{\a e^{-m_\vf r}}{r} \: , \quad \text{with} \quad \a = \frac{g_1g_2}{16\p} \: .
\eeq
The interaction is attractive if $g_1 g_2 >0$, i.e. it is always attractive between particles of the same species or particles and antiparticles, but it can be either attractive or repulsive between particles of different species. As already mentioned, for our computations, we shall consider the limit $m_\vf \to 0$.

\subsubsection{Bound-state formation amplitudes \label{sec:BSF ampl SSS}}

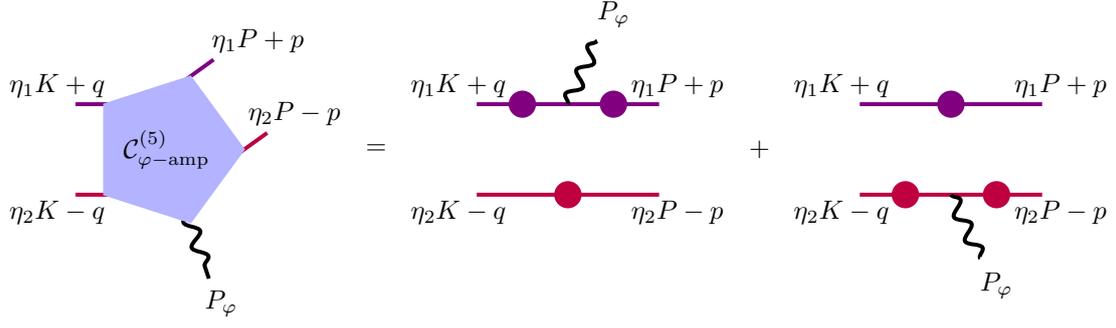
\begin{figure}[t]
\centering
\begin{tikzpicture}[line width=1.5 pt, scale=1.2]
\begin{scope}
\node at (-1.2, 0.7) {$\h_1 K + q$};
\node at (-1.2,-0.7) {$\h_2 K - q$};
\draw[violet] (-1, 0.5)--(0, 0.5);
\draw[purple] (-1,-0.5)--(0,-0.5);
\draw[rotate=36,violet] (0, 0.5)--(1, 0.5);
\draw[rotate=36,purple] (0,-0.5)--(1,-0.5);
\node at (1,  1.2) {$\h_1 P + p$};
\node at (1.4,0.4) {$\h_2 P - p$};
\draw[vector,rotate=-72] (0,0)--(1.5,0);
\node at (0.6,-1.7) {$P_\vf$};
\draw[fill=blue!30,draw=none] (0,0)
+(0:.86) -- +(72:0.86) -- +(144:0.86) -- +(216:0.86) -- +(288:0.86) -- cycle;
\node at (0,0){${\cal C}_{\vf-{\rm amp}}^{(5)}$};
\end{scope}
\node at (2.3,0) {$=$};
\begin{scope}[shift={(4.4,0)}]
\begin{scope}
\node at (-1.2, 0.7) {$\h_1 K + q$};
\node at (-1.2,-0.7) {$\h_2 K - q$};
\draw[violet] (-1, 0.5)--(1, 0.5);
\draw[purple] (-1,-0.5)--(1,-0.5);
\node at (1.2, 0.7) {$\h_1 P + p$};
\node at (1.2,-0.7) {$\h_2 P - p$};
\draw[vector] (0,.5)--(0.32,1.2);
\node at (0.5,1.5) {$P_\vf$};
\draw[fill=violet,draw=none] (-0.5, 0.5) circle (0.15);
\draw[fill=violet,draw=none] ( 0.5, 0.5) circle (0.15);
\draw[fill=purple,draw=none] (   0,-0.5) circle (0.15);
\end{scope}
\node at (2.1,0) {$+$};
\begin{scope}[shift={(4.2,0)}]
\node at (-1.2, 0.7) {$\h_1 K + q$};
\node at (-1.2,-0.7) {$\h_2 K - q$};
\draw[violet] (-1, 0.5)--(1, 0.5);
\draw[purple] (-1,-0.5)--(1,-0.5);
\node at (1.2, 0.7) {$\h_1 P + p$};
\node at (1.2,-0.7) {$\h_2 P - p$};
\draw[vector] (0,-.5)--(0.32,-1.2);
\node at (0.5,-1.5) {$P_\vf$};
\draw[fill=violet,draw=none] (0,0.5) circle (0.15);
\draw[fill=purple,draw=none] (-0.5,-0.5) circle (0.15);
\draw[fill=purple,draw=none] ( 0.5,-0.5) circle (0.15);
\end{scope}
\end{scope}
\end{tikzpicture}
\caption{The lowest order contribution to level transition amplitudes, including bound-state formation.}
\label{fig:C phi-amp}
\end{figure}

The lowest order contribution to the perturbative part of the radiative BSF amplitude arises from the diagrams shown in Fig.~\ref{fig:C phi-amp}. In our approximation, the entire BSF amplitude corresponds to the ladder diagrams of Fig.~\ref{fig:Transition}.  Recalling \eqs{eq:C 5p} and \eqref{eq:C varphi}, the diagrams of Fig.~\ref{fig:C phi-amp} evaluate to
\begin{multline}
(2\p)^4 \d^4(q_1 + q_2 -p_1 -p_2 -P_\vf) \,
i \, {\cal C}_{\vf-{\rm amp}}^{(5)} (P_\vf, p_1,p_2,q_1,q_2) =
\\ 
-i g_1 m_1 \tilde S_1(q_1) \tilde S_1(p_1)
\, (2\p)^4 \d^4 (P_\vf + p_1 - q_1) 
\: \tilde S_2(q_2) \, (2\p)^4 \d^4(p_2 - q_2)
\\
-i g_2 m_2 \tilde S_2(p_2) \tilde S_2(q_2)
\, (2\p)^4 \d^4 (P_\vf + p_2 - q_2) 
\: \tilde S_1(q_1) \, (2\p)^4 \d^4(p_1 - q_1) 
\: . \nn
\end{multline}
Equivalently
\begin{multline}
{\cal C}_{\vf-{\rm amp}}^{(5)} (P_\vf, \h_1 P+p, \h_2 P-p, \h_1 K+q, \h_2 K-q) 
= - S(q;K) \times
\\ 
\[g_1 m_1 \, S_1(\h_1 P+p) \, (2\p)^4 \d^4(q-p-\h_2 P_\vf) 
 + g_2 m_2 \, S_2(\h_2 P-p) \, (2\p)^4 \d^4(q-p+\h_1 P_\vf)\]
\, . \nn
\end{multline}
Using the above, \eq{eq:M trans} can be expressed in terms of the $\Ks_1, \Ks_2$ integrals defined in \eqs{eq:Ksi 1 def} and \eqref{eq:Ksi 2 def}, as follows
\begin{align}
&\M_{\rm trans}(\vec q;\vec p) = \nn \\
&= - \frac{\[g_1 m_1 \, \Ks_1(\vec q, \vec p; K,P) \, (2\p)^3 \d^3(\vec q - \vec p - \h_2 \vec P_\vf) 
+ g_2 m_2 \, \Ks_2(\vec q, \vec p; K,P) \, (2\p)^3 \d^3(\vec q - \vec p + \h_1 \vec P_\vf) \]}
{{\cal S}_0(\vec p;P) \, {\cal S}_0(\vec q;K)}
\nn \\
&\simeq - 2 m \mu \[1 + \frac{\vec p^2}{2\mu^2} \(1-\frac{2\mu}{m}\) \]
\[g_1 \, (2\p)^3 \d^3(\vec q - \vec p - \h_2 \vec P_\vf) 
+ g_2 \, (2\p)^3 \d^3(\vec q - \vec p + \h_1 \vec P_\vf) \] \, , 
\label{eq:M trans SSS}
\end{align}
where in the second step, we used the non-relativistic approximations of $\Ks_1, \, \Ks_2$, given in  \eqs{eq:Ksi 1 NR}, \eqref{eq:Ksi 2 NR}. Inserting this into \eq{eq:Ampl BSF Inst}, we obtain
\begin{align}
\M_{\vec k \to n} \simeq - m\sqrt{2\mu} \int \! \frac{d^3p}{(2\p)^3}   
\(1 + \frac{\vec p^2}{2m \mu} \)
\tilde \psi_n^\star (\vec p)
& \[ g_1 \: \tilde \f_{\vec k}  (\vec p + \h_2 \vec P_\vf)  +  g_2 \: \tilde \f_{\vec k}  (\vec p - \h_1 \vec P_\vf)\] .
\label{eq:M SSS 0}
\end{align}
The $\vec p^2$ term in the square brackets, will prove to be important in the cases of identical particles and particle-antiparticle pairs (see below). In terms of the integrals \eqref{eq:I cal k-n def} -- \eqref{eq:K cal k-n def}, \eq{eq:M SSS 0} becomes
\begin{multline}
\M_{\vec k \to n} \simeq - m \sqrt{2\mu} \: 
\[ g_1 {\cal I}_{\vec k, n} (\h_2 \vec P_\vf) + g_2 {\cal I}_{\vec k, n} (-\h_1 \vec P_\vf)  
+ \frac{g_1 {\cal K}_{\vec k, n} (\h_2 \vec P_\vf)+ g_2 {\cal K}_{\vec k, n} (-\h_1 \vec P_\vf)}{2m\mu}   
\]  \, .
\label{eq:M SSS}
\end{multline}

\subsubsection*{Capture in the ground state}

For capture in the $\{100\}$ state, \eq{eq:M SSS} becomes
\begin{multline}
\M_{\vec k \to \{100\}} \simeq -m \sqrt{2\mu} \: \left\{
\[ g_1 {\cal I}_{\vec k, \{100\}} (\h_2 \vec P_\vf) + g_2 {\cal I}_{\vec k, \{100\}} (-\h_1 \vec P_\vf)  \] \(1-\frac{\mu \a^2}{2m}\)
\right. \\ \left. 
+ \frac{g_1 \bar{\cal K}_{\vec k} (\h_2 \vec P_\vf) 
+ g_2 \bar{\cal K}_{\vec k} (-\h_1 \vec P_\vf)}{2m\mu} 
\right\}  ,
\nn
\end{multline}
where we used \eq{eq:K cal inter1}. We shall drop the $\a^2$ correction in the coefficient of the ${\cal I}_{\vec k,\{100\}}$ integrals. 
The mediator momentum is $|\vec P_\vf| = {\cal E}_{\vec k} - {\cal E}_{\{100\}} = (1+\z^2) k^2/(2\mu)$ (c.f.~\eqs{eq:P CM}, \eqref{eq:En,Ek}). Using \eqs{eq:I cal fin} and \eqref{eq:bar K cal fin} for the integrals ${\cal I}_{\vec k, \{100\}}$ and $\bar{\cal K}_{\vec k}$, we find
\begin{multline}
\M_{\vec k \to \{100\}} \simeq -{\cal R} (\z) \: \sqrt{\frac{2m^2}{\mu k}} \: \times 
\[ (g_1\h_2-g_2\h_1) \cos \theta   
\right. \\ \left. 
+ \frac{k}{2\mu} 
\left\{ (g_1 \h_2^2 + g_2 \h_1^2) \[(-1+i\z) + 2(2-i\z)\cos^2 \theta\]  
+\frac{\mu}{m}(g_1 + g_2) (1+i\z) \right\}
\] \, ,
\label{eq:M SSS n=1 num}
\end{multline}
where $\theta$ is the angle between $\vec k$ and $\vec P_\vf$, and ${\cal R}(\z)$ is given in \eq{eq:R cal}. (We emphasise that the above expression is not a consistent expansion in $\a$, but rather only in $\vrel^2$). We discern the following cases:
\bit
\item
For a particle-antiparticle pair, or for identical particles, $g_1=g_2=g$, $\h_1=\h_2=1/2$ and $\mu=m/4$; the first term in \eq{eq:M SSS n=1 num} vanishes, and we obtain
\begin{align}
\M_{\vec k \to \{100\}} &\simeq 
- 8  \sqrt{2\p\z} \: {\cal R}(\z) \: \vrel \:
\[i\z + (2-i\z) \cos^2\th\] \nn \\
&= -\frac{16 \sqrt{2\p\z} \: {\cal R}(\z)}{3} \: \vrel \:  
\[(1+i\z) \: P_0(\cos\th) + (2-i\z) \: P_2(\cos\th)\] \: ,
\label{eq:M SSS n=1 deg}
\end{align}
where in the second line, we decomposed the amplitude in partial waves.

\item
For non-degenerate particles, the first term in \eq{eq:M SSS n=1 num} dominates, and
\beq
\M_{\vec k \to \{100\}} \simeq -4 \sqrt{2\p\z} \, {\cal R}(\z) \(\frac{m}{\mu}\)
\[\frac{(g_1\h_2-g_2\h_1)^2}{16\p\a}\]^{1/2} \, P_1(\cos\th) \, .
\label{eq:M SSS n=1 non-deg}
\eeq
(The factor inside the square brackets becomes equal to 1 in the limit $g_1 = g_2$, $\h_1 \gg \h_2$.)
 
\eit

\subsubsection*{Capture in excited state with non-zero angular momentum}

Since for a pair of degenerate particles, the cross-section for radiative capture to the ground state is either $\vrel^2$ or $\a^2$ suppressed (as seen by comparing \eqs{eq:M SSS n=1 deg} and \eqref{eq:M SSS n=1 non-deg}), we shall now calculate the amplitude for capture in the $\{210\}$ state. In this case, $|\vec P_\vf| = {\cal E}_{\vec k} - {\cal E}_{\{210\}} = (4+\z^2) k^2/(8\mu)$. 
From \eq{eq:M SSS}, and keeping only the leading order terms of \eq{eq:I cal 210 fin} for the ${\cal I}_{\vec k, \{210\}}$ integrals, we find
\begin{align}
\M_{\vec k \to \{210\}} 
&\simeq - m \sqrt{2\mu} \: \[
g_1 {\cal I}_{\vec k, \{210\}} (\h_2 \vec P_\vf) + g_2 {\cal I}_{\vec k, \{210\}} (-\h_1 \vec P_\vf) \] 
\nn \\ 
&\simeq  
\frac{m}{\mu} \(\frac{g_1\h_2+g_2\h_1}{\sqrt{16\p\a}}\) 
\frac{2^7 \p i}{3} \frac{\z^4 (2-i\z) \, e^{\p\z/2} \, \G(1-i\z)}{(4+\z^2)^3}
\(\frac{i\z+2}{i\z-2}\)^{-i\z} 
\nn \\
&\times \[(i\z+2)P_0(\cos\theta) + 8(i\z-1) P_2(\cos\theta) \] .
\nn 
\end{align}
The $J=0$ and $J=2$ contributions to the above yield the squared amplitudes
\begin{align}
|\M_{\vec k \to \{210\}, J=0}|^2 &\simeq
\frac{m^2}{\mu^2} \[\frac{(g_1\h_2+g_2\h_1)^2}{16\p\a}\] 
\: \frac{2^{19} \p^5}{3^2} \: \frac{\z^9}{(4+\z^2)^4}  
\: \frac{e^{-4\z {\rm arccot}(\z/2)}}{1-e^{-2\p\z}} \: ,
\label{eq:M J=0 SSS n=2}
\\
|\M_{\vec k \to \{210\}, J=2}|^2 &\simeq
\frac{m^2}{\mu^2} \[\frac{(g_1\h_2+g_2\h_1)^2}{16\p\a}\] 
\: \frac{2^{25} \p^5}{3^2 \, 5^2} \: \frac{\z^9 (1+\z^2)}{(4+\z^2)^5}  
\: \frac{e^{-4\z {\rm arccot}(\z/2)}}{1-e^{-2\p\z}} \: .
\label{eq:M J=2 SSS n=2}
\end{align}
(We remind that $\M_J$ is defined in \eq{eq:MJ}, and note that the factor inside the square brackets becomes equal to 1 in the limit $g_1=g_2$, independently of $\h_1, \h_2$.)

\begin{figure}[t]
\centering
\begin{tikzpicture}[line width=1.5 pt, scale=1.3]
\begin{scope}
\draw[violet] (-2.2,1) -- (2.2,1);
\draw[purple] (-2.2,0) -- (2.2,0);
\node at (-1.9,0.5) {$\cdots$};
\draw[vector] (-1.4,0) -- (-1.4,1);
\draw[vector] (-1.0,0) -- (-1.0,1);
\draw[vector] (-0.6,0) -- (-0.6,1);
\draw[vector] (0,1) -- (0.8,1.5);
\draw[vector] (0.6,0) -- (0.6,1);
\draw[vector] (1.0,0) -- (1.0,1);
\draw[vector] (1.4,0) -- (1.4,1);
\node at (1.9,0.5) {$\cdots$};
\end{scope}
\node at (2.8,0.5) {$+$};
\begin{scope}[shift={(5.6,0)}]
\draw[violet] (-2.2,1) -- (2.2,1);
\draw[purple] (-2.2,0) -- (2.2,0);
\node at (-1.9,0.5) {$\cdots$};
\draw[vector] (-1.4,0) -- (-1.4,1);
\draw[vector] (-1.0,0) -- (-1.0,1);
\draw[vector] (-0.6,0) -- (-0.6,1);
\draw[vector] (0,0) -- (0.8,-0.5);
\draw[vector] (0.6,0) -- (0.6,1);
\draw[vector] (1.0,0) -- (1.0,1);
\draw[vector] (1.4,0) -- (1.4,1);
\node at (1.9,0.5) {$\cdots$};
\end{scope}
\end{tikzpicture}
\caption{The ladder diagrams giving the dominant contribution to bound-state formation and other level transitions. The mediator can be either a scalar or a vector boson.}
\label{fig:Transition}
\end{figure}
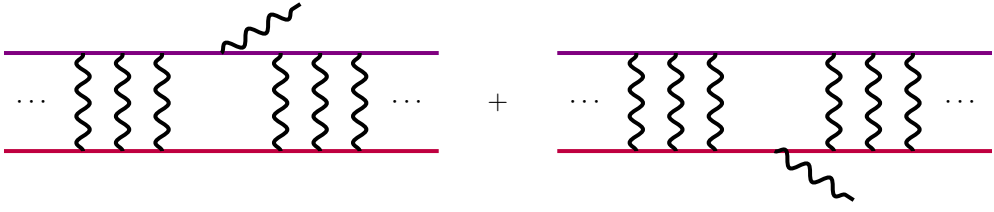

\subsubsection{Bound-state formation cross-sections and partial-wave unitarity}

Combining \eq{eq:sigmaJ} for the partial-wave cross-section and the amplitudes \eqref{eq:M SSS n=1 deg} -- \eqref{eq:M J=2 SSS n=2}, we obtain

\bit

\item
For $g_1 = g_2 = g$ and $\h_1 = \h_2 = 1/2$ (i.e.~$\mu = m/4$),
\begin{align}
\s_{_{{\rm BSF},J=0}}^{\{100\}}  &\simeq
\frac{2^7 \p^2 \, \a^3}{3^2 \mu^2} \ \frac{\z^4}{1+\z^2} 
\ \frac{e^{-4\z {\rm arccot} \z}}{1-e^{-2\p\z}} \, ,
\label{eq:sigma BSF SSS J=0 deg}
\\
\s_{_{{\rm BSF},J=2}}^{\{100\}}  &\simeq
\frac{2^7 \p^2 \, \a^3}{3^2 5\mu^2} \ \frac{\z^4 (4+\z^2)}{(1+\z^2)^2} 
\ \frac{e^{-4\z {\rm arccot} \z}}{1-e^{-2\p\z}} \, ,
\label{eq:sigma BSF SSS J=2 deg}
\\
\s_{_{\rm BSF}}^{\{100\}} \simeq 
\s_{_{{\rm BSF},J=0}}^{\{100\}}  +  \s_{_{{\rm BSF},J=2}}^{\{100\}}
&\simeq
\frac{2^7 \p^2 \, \a^3}{15\mu^2} \ \frac{\z^4 (3+2\z^2)}{(1+\z^2)^2} 
\ \frac{e^{-4\z {\rm arccot} \z}}{1-e^{-2\p\z}} \, .
\label{eq:sigma BSF SSS deg}
\end{align}

\item
For non-degenerate particles,
\beq
\s\BSFgr \ \simeq \ \s_{_{{\rm BSF},J=1}}^{\{100\}} \ \simeq \
\frac{2^7 \p^2 \, \a}{3\mu^2} \ \[\frac{(g_1\h_2-g_2\h_1)^2}{16\p\a}\]
\ \frac{\z^6}{(1+\z^2)^2} \ \frac{e^{-4\z {\rm arccot}\z}}{1-e^{-2\p\z}} \, .
\label{eq:sigma BSF SSS non-deg}
\eeq

\item
For capture to the $\{210\}$ state,
\begin{align}
\s_{_{{\rm BSF}, J=0}}^{\{210\}} &\simeq
\frac{2^8 \p^2 \, \a}{3^2 \mu^2} \[\frac{(g_1\h_2+g_2\h_1)^2}{16\p\a}\] 
\frac{\z^8}{(4+\z^2)^3} 
\, \frac{e^{-4\z {\rm arccot}(\z/2)}}{1-e^{-2\p\z}} \, ,
\label{eq:sigma BSF SSS J=0 210}
\\
\s_{_{{\rm BSF}, J=2}}^{\{210\}} &\simeq
\frac{2^{14} \p^2 \, \a}{3^2 5 \mu^2} \[\frac{(g_1\h_2+g_2\h_1)^2}{16\p\a}\]
\frac{\z^8 (1+\z^2)}{(4+\z^2)^4} 
\, \frac{e^{-4\z {\rm arccot}(\z/2)}}{1-e^{-2\p\z}} \, ,
\label{eq:sigma BSF SSS J=2 210}
\\
\s\BSF^{\{210\}}
\simeq \s_{_{{\rm BSF}, J=0}}^{\{210\}} + \s_{_{{\rm BSF}, J=2}}^{\{210\}} 
&\simeq  
\frac{2^8 \p^2 \, \a}{15 \mu^2} \[\frac{(g_1\h_2+g_2\h_1)^2}{16\p\a}\] 
\! \frac{\z^8 (28+23\z^2)}{(4+\z^2)^4} 
\, \frac{e^{-4\z {\rm arccot}(\z/2)}}{1-e^{-2\p\z}} .
\label{eq:sigma BSF SSS 210}
\end{align}

\eit

\subsubsection*{Unitarity limit \label{sec:unitarity SSS}}

For $\z \gg 1$, \eqs{eq:sigma BSF SSS J=0 deg}, \eqref{eq:sigma BSF SSS J=2 deg}, \eqref{eq:sigma BSF SSS non-deg}, \eqref{eq:sigma BSF SSS J=0 210} and \eqref{eq:sigma BSF SSS J=2 210} become
\begin{align}
\s_{_{{\rm BSF},J=0}}^{\{100\}}  & \ \to \
\frac{\p}{\mu^2 \vrel^2} \times \frac{2^7 \p \a^5}{3^2 e^4} 
\qquad\quad \text{(degenerate particles: $g_1=g_2, \h_1=\h_2=1/2$)} \: ,
\label{eq:sigma BSF SSS J=0 deg zeta>>1}
\\
\s_{_{{\rm BSF},J=2}}^{\{100\}}  & \ \to \
\frac{5\p}{\mu^2 \vrel^2} \times \frac{2^7 \p \a^5}{3^2 5^2e^4} 
\qquad\quad \text{(degenerate particles: $g_1=g_2, \h_1=\h_2=1/2$)}  \: ,
\label{eq:sigma BSF SSS J=2 deg zeta>>1}
\\
\s_{_{{\rm BSF},J=1}}^{\{100\}}  & \ \to \
\frac{3\p}{\mu^2 \vrel^2} \times \frac{2^7 \p \a^3}{3^2 e^4} 
\[\frac{(g_1\h_2-g_2\h_1)^2}{16\p\a}\]
\qquad\quad \text{(non-degenerate particles)} \: ,
\label{eq:sigma BSF SSS non-deg zeta>>1}
\\
\s_{_{{\rm BSF},J=0}}^{\{210\}}  & \ \to \
\frac{\p}{\mu^2 \vrel^2} \times \frac{2^8 \p \a^3}{3^2 e^8} 
\[\frac{(g_1\h_2+g_2\h_1)^2}{16\p\a}\] \: ,
\label{eq:sigma BSF SSS J=0 210 zeta>>1}
\\
\s_{_{{\rm BSF},J=2}}^{\{210\}}  & \ \to \
\frac{5\p}{\mu^2 \vrel^2} \times \frac{2^{14} \p \a^3}{3^2 5^2 e^8} 
\[\frac{(g_1\h_2+g_2\h_1)^2}{16\p\a}\] \: .
\label{eq:sigma BSF SSS J=2 210 zeta>>1}
\end{align}
It is interesting to note that in this low-velocity regime, the velocity dependence of all partial waves is the same.\footnote{This was also noted in Ref.~\cite{Cassel:2009wt}, for Sommerfeld-enhanced annihilation processes.}  
This is in fact expected by unitarity, since $\z\gg 1$ is both the large coupling and the low-velocity limit. Indeed, the unitarity bounds on the partial-wave inelastic cross-sections, shown in \eq{eq:unitarity}, all have the same velocity dependence. They are realised when the factors to the right of the $\times$ symbols in \eqs{eq:sigma BSF SSS J=0 deg zeta>>1} -- \eqref{eq:sigma BSF SSS J=2 210 zeta>>1} become $\approx 1$. The validity of our calculation is thus limited to at most $\a \lesssim \a\uni$, with the strongest bound, $\a\uni \approx 1$, obtained from $\s_{_{{\rm BSF},J=0}}^{\{100\}}$.

\subsubsection{De-excitation rate \label{sec:210 de-excitation}}

The radiative capture to the $\{210\}$ state is the dominant BSF process for particle-antiparticle pairs and pairs of self-conjugate particles. Moreover, for non-degenerate particles, it is slower but comparable to the capture to the ground state. Here, we shall thus compute the de-excitation rate of the $\{210\}$ state. 

The radiative de-excitation of a bound state arises from the same diagrams as the radiative capture to a bound state, albeit for different initial and final states. In our approximation, these are the ladder diagrams shown in Fig. \ref{fig:Transition}. Inserting \eq{eq:M trans SSS} into \eq{eq:Ampl LevelTrans Inst}, we find
\begin{multline}
\M_{n' \to n} \simeq -m\left\{
g_1 {\cal I}_{n', n} (\h_2 \vec P_\vf) + g_2 {\cal I}_{n', n} (-\h_1 \vec P_\vf)  
+ \frac{g_1 {\cal K}_{n', n} (\h_2 \vec P_\vf) + g_2 {\cal K}_{n', n} (-\h_1 \vec P_\vf)}{2m\mu}  
\right\}  \: .
\label{eq:M n'n SSS}
\end{multline}

In the $\{210\} \to \{100\}$ transition, the mediator is emitted with momentum $|\vec P_\vf| = {\cal E}_{\{210\}} - {\cal E}_{\{100\}} = (3/8) \mu \a^2$. Then, using \eq{eq:I cal 210-100}, we find 
\begin{align}
\M_{\{210\} \to \{100\}} 
&\simeq -m\[g_1 {\cal I}_{\{210\} , \{100\}} (\h_2 \vec P_\vf) 
+ g_2 {\cal I}_{\{210\} , \{100\}} (-\h_1 \vec P_\vf)  \]
\nn \\
&\simeq \frac{i \, 2^6 m \, \a\sqrt{2\p\a}}{3^4} 
\[\frac{g_1\h_2 + g_2 \h_1}{\sqrt{16 \p \a}}\]
\: .
\label{eq:M 210-100 SSS}
\end{align}
The de-excitation rate, given by \eq{eq:de-excitation rate}, becomes
\beq
\G_{\{210\} \to \{100\}+\vf} \simeq \frac{2^7 \a^5 \mu}{3^7}
\[\frac{(g_1 \h_2 + g_2 \h_1)^2}{16\p\a}\]  \: .
\label{eq:decay 210-100 SSS}
\eeq

\subsubsection{Annihilation vs bound-state formation for particle-antiparticle pairs}

\begin{figure}[t]
\centering
\begin{tikzpicture}[line width=1.5 pt, scale=1.5]
\begin{scope}[shift={(5.6,0)}]
\draw[violet] (-2.2,1) -- (0,1);
\draw[violet] (-2.2,0) -- (0,0);
\draw[violet] (0,0) -- (0,1);
\node at (-1.9,0.5) {$\cdots$};
\draw[vector] (-1.4,0) -- (-1.4,1);
\draw[vector] (-1.0,0) -- (-1.0,1);
\draw[vector] (-0.6,0) -- (-0.6,1);
\draw[vector] (0,1) -- (1,1);
\draw[vector] (0,0) -- (1,0);
\end{scope}
\end{tikzpicture}
\caption{Annihilation of an unbound particle-antiparticle pair or decay of a bound particle-antiparticle pair into two force mediators. The mediator may be either a scalar or a vector boson.}
\label{fig:AnnDec 2-body}
\end{figure}
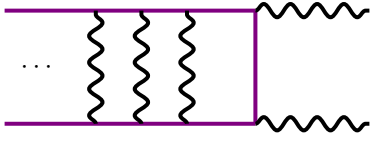

A particle-antiparticle pair or a pair of self-conjugate identical particles coupled to a light scalar, can annihilate into two mediators, $\x^\dagger\x \to \vf\vf$ or $\x\x \to \vf\vf$, as shown in Fig.~\ref{fig:AnnDec 2-body}. In both cases, the perturbative annihilation amplitude, evaluated in the CM frame, in the non-relativistic regime, is 
\beq 
\hat\M\ann^{\rm pert}(\vec q, -\vec q; \vec p, -\vec p) 
\simeq a_0 + \frac{a_1}{m\mu} (\vec q \cdot \vec p)
\label{eq:Mann SSS}
\eeq
with $a_0 \simeq a_1 \simeq 16\p\a $. Using \eqs{eq:sigma ann ell}, \eqref{eq:sigma ell} and \eqref{eq:S ann ell}, we find that the annihilation cross-section is $s$-wave dominated, $\s\ann\vrel \simeq  \s_0 \, S_{0, \rm ann}$, with $\s_0 = f_s |a_0|^2/(32 \p m\mu) \simeq \p \a^2/\mu^2$ being the perturbative $s$-wave annihilation cross-section, and $S_{0, \rm ann} = |\f_{\vec k} (0)|^2$ being the $s$-wave Sommerfeld-enhancement factor for annihilation. Using the wavefunction of \eq{eq:phi k}, we obtain
\beq
\s\ann\vrel \simeq   \frac{\p \a^2 }{\mu^2} \, \frac{2\p\z}{1-e^{-2\p\z}} \: .
\label{eq:sigma ann SSS}
\eeq
Note that here, $\mu = m_\x/2= m/4$, with $m_\x$ being the mass of the annihilating particles.

We may now compare the BSF and annihilation cross-sections,
\begin{align}
\frac{\s\BSFgr}{\s\ann} &\simeq 
\frac{2^6 \a^2 \z^2 (3+2\z^2)}{15 (1+\z^2)^2}  
\: e^{-4\z{\rm arccot \z}} \, ,
\label{eq:BSF/ann 100 SSS}
\\
\frac{\s\BSF^{\{210\}}}{\s\ann} &\simeq
\frac{2^7 \z^6 (28+23\z^2)}{15 (4+\z^2)^4} 
\: e^{-4\z{\rm arccot (\z/2)}} \: .
\label{eq:BSF/ann 210 SSS}
\end{align}
At $\z \gg 1$, $\s\BSFgr / \s\ann \simeq 0.15 \a^2$ and  $\s\BSF^{\{210\}} / \s\ann \simeq 0.066$.\footnote{Note that for a Dirac fermion-antifermion pair, the annihilation into scalars is dominantly $p$-wave. The spin-averaged annihilation cross-section times relative velocity is $\s\ann \vrel \simeq \s_1 S_{1,\rm ann}$, with $\s_1 \simeq 3\p \a^2 / (2\mu^2)$ and $S_{1,\rm ann} = \vrel^2 (1+\z^2) \, 2\p\z /(1-e^{-2\p\z})$. On the other hand, the BSF cross-sections do not depend on the spins of the interacting particles, since in the non-relativistic regime, the spin is conserved separately from the orbital angular momentum. Thus, for Dirac fermions, at $\z \gg 1$, $\s\BSFgr / \s\ann \simeq 0.1$ and  $\s\BSF^{\{210\}} / \s\ann \simeq 0.044/\a^2$. This means that for $\a\lesssim 0.2$, BSF is faster than annihilation in the regime where the Sommerfeld effect is important.} 
We compare $\s\ann$ and $\s\BSF^{\{210\}}$ in Fig.~\ref{fig:AnnVsBSF}. Since annihilation is the dominant inelastic process, it lowers the value of $\a$ at which the unitarity bound appears to be realised, to $\a\uni \approx 0.54$.

\subsubsection{Particle-antiparticle bound-state decay rates}

From \eqref{eq:decay rate ell}, \eqref{eq:sigma ell} and \eqref{eq:S dec ell}, we find that the decay of the $\{100\}$ particle-antiparticle bound state into two mediators, is dominated by the $s$-wave contribution,
\beq
\G_{\{100\} \to \vf\vf} =  \s_0 |\ps_{100} (0)|^2 
= \frac{\p \a^2}{\mu^2} \: \frac{\k^3}{\p} = \mu \a^5 \: .
\label{eq:decay 100 SSS}
\eeq

For the $\{210\}$ bound state, the $s$-wave decay mode vanishes, since $\ps_{210} (0) = 0$. However, a non-vanishing contribution arises from the $p$-wave mode. From \eq{eq:sigma ell}, we find 
$\s_1 = f_s |a_1|^2/(2^7 3 \p m\mu) = \p \a^2 /(12\mu^2)$. From \eq{eq:S dec ell} and the wavefunction \eqref{eq:psi 210}, we obtain $S_{1,\rm dec} = \k^5/(32\p \mu^2)$. Then
\beq
\G_{\{210\}\to\vf\vf} = \s_1 \, S_{1, \rm dec} =  \frac{\p \a^2 }{12\mu^2} \: \frac{\k^5}{32\p\mu^2} = \frac{\mu \a^7}{2^7 \, 3} \: .
\label{eq:decay 210 SSS}
\eeq

The decay rates into three mediators are expected to be suppressed by one additional power of $\a$ with respect to the above. Recalling \eq{eq:decay 210-100 SSS}, this suggests that for the excited state $\{210\}$, the transition to the ground state is the dominant decay mode.

\subsection{Vector mediator \label{sec:SSV}}

We now consider two scalar particles $\x_1, \, \x_2$ coupled to a gauged $U(1)$ force. The interaction Lagrangian is
\begin{align}
\d {\cal L}_V
&= (D_\m \x_1)^\dagger (D^\m \x_1)
+ (D_\m \x_2)^\dagger (D^\m \x_2)
- \frac{1}{4} F_{\m\n} F^{\m\n}
- m_1^2 |\x_1|^2 - m_2^2 |\x_2|^2
\: ,
\label{eq:Langr SSV}
\end{align}
where $\x_1, \: \x_2$ are complex scalar bosons, $F^{\m\n} = \partial^\m \vf^\n - \partial^\n \vf^\m$ and $D^\m = \partial^\m - i c_j g \vf^\m$, with $c_1, \: c_2$ being the charges of $\x_1, \: \x_2$.${}^{\ref{foot:quartic}}$

The one-boson exchange diagram gives
\begin{align}
\tilde W(p,p';Q) &= \frac{i c_1 c_2 g^2}{(p-p')^2} \ 
\[ (2\h_1 Q + p + p') \cdot (2\h_2 Q - p - p')  \] . \nn
\end{align}
In the non-relativistic regime, we shall approximate the above with 
\beq
\tilde W(p,p';Q) \simeq \frac{i 4\h_1\h_2 Q^2 \, c_1 c_2 g^2 }{(p-p')^2} \, .
\eeq
In the instantaneous approximation, and setting $Q^2 \simeq m^2$
\beq
{\cal W}(|{\bf p-p'}|) \simeq -\frac{i 4\h_1\h_2 \, c_1 c_2  g^2 \, m^2}{({\bf p-p'})^2} ,
\label{eq:W instant SSV}
\eeq
and
\beq
V(\vec r) = -\frac{1}{i 4m\m}
\int \frac{d^3p}{(2\pi)^3} {\cal W}(\vec p) \, e^{i \bf p \cdot r}  
= c_1 c_2 g^2 \int \frac{d^3p}{(2\pi)^3} \frac{e^{i \bf p \cdot r}}{{\bf p}^2} 
\: .
\eeq
That is,
\beq
V({\bf r}) = -\frac{\a}{r} \: , \quad \text{with} \quad \a = -\frac{c_1 c_2 g^2}{4\p} \: .
\eeq
The interaction is attractive if $c_1 c_2 < 0$.

\subsubsection{Bound-state formation amplitude, cross-section and partial-wave unitarity}

The perturbative part of the level-transition amplitudes is ${\cal C}_{\vf-\rm amp}^{(5)} = \e_\mu {\cal C}_{\vf-\rm amp}^\mu$, with the lowest order contribution depicted in Fig.~\ref{fig:C phi-amp},
\begin{multline}
(2\p)^4 \d^4(q_1 + q_2 -p_1 -p_2 -P_\vf) \, i \, 
{\cal C}_{\vf-\rm amp}^\mu (P_\vf, p_1,p_2,q_1,q_2) \simeq
\\
-i g  c_1 (p_1^\mu + q_1^\mu) 
\, \tilde S_1(p_1) \tilde S_1(q_1)
\, (2\p)^4 \d^4 (P_\vf + p_1 - q_1) 
\, \tilde S_2(q_2)
\, (2\p)^4 \d^4(p_2 - q_2)
\nn \\ 
-i g  c_2 (p_2^\mu + q_2^\mu)
\, \tilde S_2(p_2) \tilde S_2(q_2)
\, (2\p)^4 \d^4 (P_\vf + p_2 - q_2) 
\, \tilde S_1(q_1)
\, (2\p)^4 \d^4(p_1 - q_1) 
\: . \nn
\end{multline}
From this, we obtain
\begin{multline}
{\cal C}_{\vf-\rm amp}^\mu (P_\vf, \h_1 P +p, \h_2 P-p,  \h_1 K+q, \h_2 K-q) =
\\
= - g \: S(q;K) 
\left\{
c_1  [2\h_1 K^\mu - (\h_1-\h_2)P_\vf^\mu + 2p^\mu]
\, \tilde S_1(\h_1 P+p) \, (2\p)^4 \d^4(q-p-\h_2 P_\vf) 
\right.
\\
+ \left.
c_2  [2\h_2 K^\mu + (\h_1-\h_2) P_\vf^\mu - 2p^\mu] 
\, \tilde S_2(\h_2 P-p) \, (2\p)^4 \d^4(q-p+\h_1 P_\vf) \right\}
\: . \nn
\end{multline}
Then, $\M_{\rm trans} = \e_\mu \M_{\rm trans}^\mu$, where for $j=1,2,3$,
\begin{align}
\M_{\rm trans}^j (\vec q;\vec p)
=& - \frac{g}{ {\cal S}_0 (\vec q;K) \, {\cal S}_0 (\vec p; P)} 
\nn \\
\times& \left\{ c_1 \[2\h_1 K^j - (\h_1-\h_2)P_\vf^j + 2p^j\] \, \Ks_1(\vec q, \vec p; K, P) 
\, (2\p)^3 \d^3(\vec q - \vec p - \h_2 \vec P_\vf) \right. 
\nn \\
+ & \left. c_2 \[2\h_2 K^j + (\h_1-\h_2)P_\vf^j - 2p^j\] \, \Ks_2(\vec q, \vec p; K, P) 
\, (2\p)^3 \d^3(\vec q - \vec p + \h_1 \vec P_\vf) \right\} \, .
\end{align}
We remind that $\Ks_1, \Ks_2$ are defined in \eqs{eq:Ksi 1 def}, \eqref{eq:Ksi 2 def}. Using their non-relativistic approximations \eqref{eq:Ksi 1 NR},  \eqref{eq:Ksi 2 NR}, we find the amplitude of \eq{eq:Ampl BSF Inst} to be 
$\M_{\vec k \to n} =  \e_\mu \M_{\vec k \to n}^\mu$, where for $j=1,2,3$, 
\begin{multline}
{\cal M}_{\vec k \to n}^j = - g \sqrt{2\mu} 
\int \frac{d^3p}{(2\p)^3} \: \tilde \psi_n^\star (\vec p)
\left\{ \frac{c_1}{\h_1} \[2\h_1 K^j - (\h_1-\h_2) P_\vf^j + 2p^j\]
\ \tilde \f_{\vec k}  (\vec p + \h_2 \vec P_\vf) 
\right.
\nn \\
\left.
+  \frac{c_2}{\h_2} \[2\h_2 K^j + (\h_1 -\h_2) P_\vf^j - 2p^j\]  
\ \tilde \f_{\vec k}  (\vec p- \h_1 \vec P_\vf)  
\right\}
\, . \nn
\end{multline}
We may rewrite the above in terms of the ${\cal I, \, J}$ integrals, defined in \eqs{eq:I cal k-n def} and \eqref{eq:J cal k-n def}, as follows
\begin{multline}
\M_{\vec k \to n}^j = 
- 2g\sqrt{2\mu} \ \left\{
 \frac{c_1}{\h_1}\, {\cal J}_{\vec k, n}^j (\h_2 \vec P_\vf) 
-\frac{c_2}{\h_2}\, {\cal J}_{\vec k, n}^j (-\h_1 \vec P_\vf) 
\right. \\  \left.
+ \[c_1 \(K^j - \frac{\h_1 -\h_2}{2\h_1} P_\vf^j \) \: {\cal I}_{\vec k, n} (\h_2 \vec P_\vf)
+c_2 \(K^j + \frac{\h_1 -\h_2}{2\h_2} P_\vf^j \) \: {\cal I}_{\vec k, n} (-\h_1 \vec P_\vf) \]
\right\} \: .
\label{eq:M SSV}
\end{multline}
Because the vector boson $\vf^\mu$ is transverse, the $\mu = 0$ component and the component parallel to $\vec P_\vf$ do not contribute to the amplitude $\M_{\vec k \to n} = \e_\mu \M_{\vec k \to n}^\mu$. Dropping those components, we obtain $\M_{\vec k \to n}^j \to \tilde \M_{\vec k \to n}^j$. In the rest frame, $\vec K = 0$, and for capture in the ground state $\{100\}$, using \eq{eq:J cal fin} for the ${\cal J}_{\vec k, \{100\}}$ integrals, and keeping only the leading term, we find
\begin{align}
\tilde \M_{\vec k \to n}^j 
&= - 2g\sqrt{2\mu} \ {\cal R}(\z) 
\ \frac{k^j \sin \theta}{k^{3/2}}
\ \( \frac{c_1}{\h_1} - \frac{c_2}{\h_2} \) 
\: ,
\label{eq:tilde Mj}
\end{align}
where the $\sin \theta$ factor arises from the projection of $\vec k$ on the plane vertical to $\vec P_\vf$, and ${\cal R}(\z)$ is defined in \eq{eq:R cal}. (Recall that $c_1 c_2 < 0$ for an attractive potential.) Note that the partial wave decomposition of the $\theta$-dependent factor of \eq{eq:tilde Mj} is
\beq
\sin \theta  = \sqrt{1-\cos^2 \theta} 
= \frac{\p}{4} \, P_0 (\cos\theta) -\frac{5\p}{2^5} \, P_2(\cos\theta) - \frac{3^2\p}{2^8}\, P_4(\cos\theta) 
+ \dots \: .
\label{eq:sin theta}
\eeq

The sum over the vector-boson polarisations is
\beq
\sum_\e |\e_\mu \M_{\vec k \to n}^\mu|^2 
= \M_{\vec k \to n}^{\mu *} \M_{\vec k \to n}^\nu  \sum_\e \e_\mu^* \e_\nu  
=-\M_{\vec k \to n}^{\mu *} \M_{\vec k \to n}^\nu  g_{\mu\nu}  
= \tilde\M_{\vec k \to n}^{j*} \tilde\M_{\vec k \to n}^j
\: .
\label{eq:sum M^2 SSV}
\eeq
In \eq{eq:sum M^2 SSV}, the contribution of the $\mu=0$ and $\nu=0$ components cancels the contribution from the component of $\M$ that is parallel to $\vec P_\vf$, yielding the final result. Using \eqs{eq:tilde Mj} and \eqref{eq:R cal square}, we obtain
\beq
\sum_\e |\e_\mu \M_{\vec k \to n}^\mu|^2  =
\[\frac{(\h_2 c_1 - \h_1 c_1)^2}{-c_1c_2} \]
\frac{2^{12} \p^3}{\h_1^2 \h_2^2} 
\frac{\z^7}{(1+\z^2)^3} \frac{e^{-4\z {\rm arc cot} \z}}{1-e^{-2\p\z}} \: \sin^2 \theta\: .
\label{eq:M^2 SSV}
\eeq
Note that for $c_1 = -c_2 = 1$, which includes the case of a particle-antiparticle pair, the factor in the square brackets in the above expression becomes $[(\h_2 c_1 -\h_1 c_2)^2 / (-c_1 c_2)] = 1$.

\bigskip

Using \eqs{eq:diff sigma} and \eqref{eq:M^2 SSV}, we find the unpolarised cross-section for radiative capture to the ground state to be
\beq
\s\BSFgr \vrel = 
\[ \frac{(\h_2 c_1 -\h_1 c_2)^2}{-c_1 c_2} \]
\ \frac{2^8\p^2 \, \a^2}{3\mu^2} 
\ \frac{\z^5}{(1+\z^2)^2}
\ \frac{e^{-4\z{\rm arccot}\z}}{1-e^{-2\p\z}} \: .
\label{eq:sigma BSF SSV}
\eeq
This result agrees with Ref.~\cite{BetheSalpeter_QM} (see Eq.~(75.6)), and is smaller than that of Ref.~\cite{AkhiezerMerenkov_sigmaHydrogen} by a factor of 2.
At $\z \gg 1$, \eq{eq:sigma BSF SSV} becomes
\beq
\s\BSFgr \vrel \simeq
\[ \frac{(\h_2 c_1 -\h_1 c_2)^2}{-c_1 c_2} \]
\: \frac{2^8 \p^2 \, \a^2}{3 e^4  \, \mu^2} \, \z
\, .
\label{eq:sigma BSF SSV zeta>>1}
\eeq
The contribution from the $J=0$ partial wave is 
$\s_{_{{\rm BSF},J=0}}^{\{100\}} = (\p/4)^2 \: \s\BSFgr$ (c.f. \eq{eq:sin theta}). The unitarity limit on the $s$-wave inelastic cross-section, given in \eq{eq:unitarity}, is realised for $\a = \a\uni$, with
\beq
\a\uni \approx 0.69 \[ \frac{-c_1 c_2}{(\h_2 c_1 -\h_1 c_2)^2} \]^{1/3} 
\: .
\label{eq:unit SSV}
\eeq

\subsubsection{Annihilation vs bound-state formation for particle-antiparticle pairs}

In the non-relativistic regime, the annihilation of a particle-antiparticle pair into two vector mediators, $\x \x^\dagger \to \vf\vf$, is $s$-wave dominated, with the perturbative unpolarised cross-section times relative velocity being $\s_0 = \p\a^2/(2\mu^2)$. From \eqs{eq:sigma ann ell} and \eqref{eq:S ann ell}, we find
\beq
\s\ann \vrel  
= \s_0 \, |\f_{\vec k}(0)|^2
= \frac{\p \a^2}{2\mu^2} \, \frac{2\p\z}{1-e^{-2\p\z}} \: ,
\label{eq:sigma ann SSV}
\eeq
where  here $\mu = m_\x /2 =m/4$, with $m_\x$ being the mass of the annihilating particles.

We may now compare the BSF and annihilation cross-sections:
\beq
\frac{\s\BSFgr}{\s\ann} \simeq \frac{2^8 \z^4 \, e^{-4\z \, {\rm arccot} \z}}{3(1+\z^2)^2} \: .
\label{eq:sBSF/sann vector}
\eeq
While at $\z < 1$ BSF is negligible, at $\z\gtrsim 1.11$ the contribution of BSF to the inelastic scattering dominates over annihilation. 
At $\z \gg1$,  
$\s\BSFgr / \s\ann \simeq 1.56$.\footnote{For a fermionic particle-antiparticle pair coupled to a gauged $U(1)$, $\s_0 \simeq \p\a^2/(4\mu^2)$. Then, at $\z\gg 1$, the importance of BSF relative to annihilation is even greater, with $\s\BSFgr/\s\ann \simeq 3.1$.}  We compare $\s\ann$ and $\s\BSF^{\{100\}}$ in Fig.~\ref{fig:AnnVsBSF}.

\subsubsection{Particle-antiparticle bound-state decay rate}

From \eq{eq:decay rate ell}, we find the unpolarised decay rate of a particle-antiparticle bound state into two vector mediators to be
\begin{align}
\G_{\{100\} \to \vf\vf} &\simeq  \s_0  |\ps_{100} (0)|^2 
= \frac{\p \a^2}{2\mu^2}  \frac{\k^3}{\p}
= \frac{\mu \a^5}{2} \: .
\label{eq:decay 100 SSV}
\end{align}

\section{Discussion \label{Sec:Conc}}

The formation of bound states affects the phenomenology of dark matter in a variety of ways. Computing the rates for bound-state formation and other related processes is essential in calculating the cosmology of DM and accurately estimating the expected DM signals and detection prospects.

In the non-relativistic regime, the formation of bound states is enhanced by the Sommerfeld effect. The Sommerfeld effect has already been incorporated in computations of the DM annihilation rate, in the context of various theories, and has been shown to have important phenomenological implications. Besides enhancing the total DM annihilation rate, it may also modify -- depending on the nature of the DM interactions -- the relative strength of the various annihilation channels, thus changing the spectrum of the annihilation products~\cite{Hryczuk:2014hpa}. Our results demonstrate that, for particle-antiparticle pairs or pairs of self-conjugate particles, the radiative formation of bound states can be faster than annihilation, in the entire regime where the Sommerfeld effect is important. This suggests that bound-state formation and decay may affect the annihilation signals of symmetric thermal-relic dark matter, as well as its relic abundance, well beyond the experimental uncertainty in the DM density~\cite{vonHarling:2014kha}. Bound-state dynamics should then be incorporated in any relevant analyses.

The importance of this point is underscored by present experimental results, which strongly constrain sub-TeV DM with electroweak interactions, and thus motivate investigations in the multi-TeV regime. As is well known, for symmetric (or self-conjugate) thermal-relic DM heavier than a few TeV, including WIMP DM, the Sommerfeld effect is important, both in the determination of the relic abundance and in the estimation of the expected indirect-detection signals. For the indirect detection of hidden-sector DM, the Sommerfeld effect -- and therefore the formation of bound states -- can be important even for lower DM masses.

Asymmetric dark matter can couple even more strongly to light force mediators than symmetric DM; indeed, in the presence of a particle-antiparticle asymmetry, the very efficient annihilation which such a coupling would imply, cannot destroy the DM relic abundance. It follows that, for a much larger range of masses, asymmetric DM may efficiently form stable bound states in the early universe. This has important implications for its phenomenology. On one hand, the formation of bound states typically curtails the DM self-interactions and hastens the kinetic decoupling of DM from dark radiation in the early universe; consequently, it regulates the potential effect of the DM dynamics on the galactic structure. On the other hand, DM may participate in a variety of radiative processes inside haloes, such as excitations and de-excitations of bound states, or outright formation of bound states. In addition, the scattering of DM on nucleons may involve a variety of interactions, including both elastic and inelastic processes. This interplay between cosmology and the fundamental interactions of the DM constituents, determines all manifestations of DM today, and can be calculated only with precise knowledge of the rates governing bound-state-related processes.

In this work, we established a field-theoretic framework for computing rates for processes involving bound states. This framework can be employed in future investigations of related effects, in a variety of theories. In particular, the computation of bound-state formation rates in theories which involve non-Abelian interactions -- including the electroweak theory of the Standard Model -- necessitates adopting a field-theoretic formalism. Moreover, this framework allows for systematic expansions in the interaction strength and in the momentum exchange between the interacting degrees of freedom; these higher-order corrections are important when the leading-order terms cancel, as was explicitly shown in our computations.

The significance of long-range interactions -- and therefore, the importance of comprehending their implications -- is affirmed by unitarity. Unitarity sets an upper bound on the partial-wave inelastic cross-section, $(\s_{{\rm inel}, J})_{\rm max}$, shown in \eq{eq:unitarity}. This, in turn, yields an upper bound on the mass of thermal-relic DM~\cite{Griest:1989wd,vonHarling:2014kha}. Notably, the velocity dependence of $(\s_{{\rm inel}, J})_{\rm max}$ suggests that the unitarity bound can be realised only if the underlying interactions are long-ranged~\cite{vonHarling:2014kha}. However, in the presence of long-range interactions, the formation of bound states may be the dominant inelastic process, as shown in the present work. The realisation of the unitarity bound, and its phenomenological implications, are thus largely determined by the dynamics of bound states, which should be fully incorporated in any related study. For example, the DM self-destruction in the early universe via bound-state formation and decay involves an interplay between capture, disassociation and decay processes that is absent in the case of direct annihilation into radiation~\cite{vonHarling:2014kha}. 

Moreover, our results show that in the large-coupling (or low-velocity) regime, the dominant inelastic channel often belongs to a higher partial wave than usually assumed. Since the unitarity bound on higher partial waves is more relaxed, this implies that thermal-relic DM may be significantly heavier than previously estimated. 

Lastly, using the computed bound-state-formation cross-sections, we may estimate the interaction strength for which the unitarity bound is seemingly realised. Our leading-order computations show that this is at $\a \sim 0.5$, i.e. well below what is often considered to be the perturbativity limit, $\a\sim \pi$ or $4\pi$.

\section*{Acknowledgments}

We thank Paul Hoyer for very helpful discussions. This work was supported by the Netherlands Foundation for Fundamental Research of Matter (FOM) and the Netherlands Organisation for Scientific Research (NWO).


\clearpage
\appendix

\section*{\LARGE Appendices}

\section{Bethe-Salpeter wavefunctions: Summary of definitions \label{App:BSW def}}

We summarise the definitions of the Bethe-Salpeter wavefunctions and their Fourier transforms.
Following Sec.~\ref{sec:BSW}, we find
\begin{align}
\Psi_{\vec Q, n} (x_1, x_2) &= e^{-i Q X} \Psi_{\vec Q, n} (x) \: ,
\label{eq:Psi hat}
\\
\Psi_{\vec Q, n}^\star (x_1, x_2) &= e^{i Q X} \Psi_{\vec Q, n}^\star (x) \: ,
\label{eq:Psi* hat}
\\
\F_{\vec Q, \vec q} (x_1, x_2) &= e^{-i Q X} \F_{\vec Q, \vec q} (x) \: ,
\label{eq:Phi hat}
\\
\F_{\vec Q, \vec q}^\star (x_1, x_2) &= e^{i Q X} \F_{\vec Q, \vec q}^\star (x) \: ,
\label{eq:Phi* hat}
\end{align}
with $Q^0 = \w_{n}(\vec Q)$ and $Q^0 = \w_{\vec Q, \vec q}$ being the energies of $|{\cal B}_{\vec Q, n}\>$ and $|{\cal U}_{\vec Q, \vec q}\>$, and where we defined
\begin{align}
\Psi_{\vec Q, n} (x) &\equiv \< \W | T \x_1 (\h_2 x) \x_2 (-\h_1 x) | {\cal B}_{\vec Q, n} \> \: , \label{eq:Psi}\\
\Psi_{\vec Q, n}^\star (x) &\equiv \< {\cal B}_{\vec Q, n} | T \x_1^\dagger (\h_2 x) \x_2^\dagger (-\h_1 x) | \W \> \: , \label{eq:Psi star} \\
\F_{\vec Q, \vec q} (x) &\equiv \< \W | T \x_1 (\h_2 x) \x_2 (-\h_1 x) | {\cal U}_{\vec Q, \vec q} \> \: , \label{eq:Phi} \\
\F_{\vec Q, \vec q}^\star (x) &\equiv \< {\cal U}_{\vec Q, \vec q} | T \x_1^\dagger (\h_2 x) \x_2^\dagger (-\h_1 x) | \W \>  \: .  \label{eq:Phi star}
\end{align}
The Fourier transforms are
\begin{align}
\Psi_{\vec Q,n} (x) &\equiv \int \frac{d^4 p}{(2\p)^4} \, \tilde \Psi_{\vec Q,n} (p) \, e^{- i p x} 
\:, \qquad
&\tilde \Psi_{\vec Q,n} (p) &\equiv \int d^4 x \, \Psi_{\vec Q,n} (x) \, e^{i p x} \ .
\\
\Psi_{\vec Q, n}^\star (x) &\equiv \int \frac{d^4 p}{(2\p)^4} \, \tilde \Psi_{\vec Q, n}^\star (p) \, e^{i p x}
\:, \qquad
&\tilde \Psi_{\vec Q, n}^\star (p) &\equiv \int d^4 x \, \Psi_{\vec Q, n}^\star (x) \, e^{-i p x} \: ,
\\
\F_{\vec Q, \vec q} (x) &\equiv \int \frac{d^4 p}{(2\p)^4} \, \tilde \F_{\vec Q, \vec q} (p) \, e^{-i p x}
\:, \qquad
&\tilde \F_{\vec Q, \vec q} (p) &\equiv \int d^4 x \, \F_{\vec Q, \vec q} (x) \, e^{i p x} \: , 
\\
\F_{\vec Q, \vec q}^\star (x) &\equiv \int \frac{d^4 p}{(2\p)^4} \, \tilde \F_{\vec Q, \vec q} (p) \, e^{i p x}
\:, \qquad
&\tilde \F_{\vec Q, \vec q}^\star (p) &\equiv \int d^4 x \, \F_{\vec Q, \vec q}^\star (x) \, e^{-i p x} \: .
\end{align}


\section{Calculation of $h(x^0)$  \label{App:h}}

As in~\eq{eq:x,X}, we define
\begin{align}
x_1 &= X + \h_2 x \: , \qquad  x_2 = X - \h_1 x \: , \\
y_1 &= Y + \h_2 y \: , \qquad  y_2 = Y - \h_1 y \: .
\end{align}
Then
\begin{align}
&\min(x_1^0,x_2^0) - \max (y_1^0, y_2^0) =
\nn \\
&= \min (X^0 + \h_2 x^0, X^0 - \h_1 x^0) 
- \max (Y^0 + \h_2 y^0, Y^0 - \h_1 y^0) 
\nn \\
&= X^0 + \min (\h_2 x^0, - \h_1 x^0) 
 - Y^0 - \max (\h_2 y^0, - \h_1 y^0) 
\nn \\
&= X^0 - Y^0 + \left\{
\bal{6}
(-&\h_1 x^0 - \h_2 y^0) \: ,& \quad &\text{ if } x^0>0, y^0 >0& \\
(-&\h_1 x^0 + \h_1 y^0) \: ,& \quad &\text{ if } x^0>0, y^0 <0& \\
( &\h_2 x^0 - \h_2 y^0) \: ,& \quad &\text{ if } x^0<0, y^0 >0& \\
( &\h_2 x^0 + \h_1 y^0) \: ,& \quad &\text{ if } x^0<0, y^0 <0&
\eal
\right.
\nn \\
&= X^0 - Y^0 
- \frac{1}{2}\[(\h_1-\h_2) x^0 + (\h_1+\h_2) |x^0|\]
+ \frac{1}{2}\[(\h_1-\h_2) y^0 - (\h_1+\h_2) |y^0|\]
\nn \\
&= X^0 - Y^0 + h(x^0) - h(y^0) \: ,
\end{align}
where, as defined in \eq{eq:h},
\beq 
h_\pm(x^0) \equiv \frac{1}{2} (\h_2 - \h_1) x^0 \pm \frac{1}{2}|x^0| \: .  
\eeq


\section{Calculation of ${\cal S}_0(\vec p; P)$, ${\cal S}(t; \vec p; P)$ and ${\cal S}_{1,2}(t; \vec p; P)$  \label{App:S cal}}

Here we calculate the integrals
\begin{align}
{\cal S}_0 (\vec p;P) 
&\equiv
\int \frac{dp^0}{2\p} \: \tilde S_1(\h_1 P + p) \tilde S_2(\h_2 P - p) \: ,
\label{eq:S0 cal def re}
\\
{\cal S} (t;\vec p;P) 
&\equiv
\int \frac{dp^0}{2\p} \: \tilde S_1(\h_1 P + p) \tilde S_2(\h_2 P - p) e^{i p^0 t} \: ,
\label{eq:S cal def}
\\
{\cal S}_1 (t;\vec p;P) 
&\equiv
\int \frac{dp^0}{2\p} \: \tilde S_1(\h_1 P + p) e^{-i p^0 t}  \: ,
\label{eq:S1 cal def}
\\
{\cal S}_2 (t;\vec p;P) 
&\equiv
\int \frac{dp^0}{2\p} \: \tilde S_2(\h_2 P - p) e^{-i p^0 t}  \: .
\label{eq:S2 cal def}
\end{align}
(Note the different sign in the exponential between the definitions \eqref{eq:S cal def} and \eqref{eq:S1 cal def}, \eqref{eq:S2 cal def}, chosen so for later convenience.)

We shall use the perturbative propagators
\beq
\tilde S_1(p_1) = \frac{i}{p_1^2 - m_1^2 + i \e} \, , \quad 
\tilde S_2(p_2) = \frac{i}{p_2^2 - m_2^2 + i \e} \, ,
\label{eq:prop}
\eeq
and we define $E_1(\vec p; \vec P) = \sqrt{(\h_1 \vec P + \vec p)^2+m_1^2}$ and $E_2(\vec p; \vec P) = \sqrt{(\h_2 \vec P - \vec p)^2 + m_2^2}$. For convenience, we also define
\begin{align}
\rho_1 (\vec p ; P)
&\equiv \h_1 P^0 - E_1(\vec p; \vec P)
\: ,
\label{eq:rho 1}
\\
\rho_2 (\vec p ; P)
&\equiv  \h_2 P^0 - E_2(\vec p; \vec P)
\: ,
\label{eq:rho 2}
\\
\s_1 (\vec p ; P)
&\equiv \h_1 P^0 + E_1(\vec p; \vec P)
\: ,
\label{eq:sigma 1}
\\
\s_2 (\vec p ; P)
&\equiv  \h_2 P^0 + E_2(\vec p; \vec P)
\: .
\label{eq:sigma 2}
\end{align}

We first consider the integral \eqref{eq:S cal def},
\begin{align}
{\cal S} (t;\vec p;P) 
&= \int \frac{dp^0}{2\p} \:
\frac{i}{[\h_1 P^0 + p^0 - E_1(\vec p; \vec P) +i\e] \: [\h_1 P^0 + p^0 + E_1(\vec p; \vec P)-i\e]}
\nn \\ 
&\times
\frac{i}{[\h_2 P^0 - p^0 + E_2(\vec p; \vec P)-i\e] \: [\h_2 P^0 - p^0 - E_2(\vec p; \vec P)+i\e]} 
 \: e^{i p^0 t}
\: .
\end{align}
We may evaluate ${\cal S} (t;\vec p; P)$ by closing the $p^0$ contour above and below the real axis for $t>0$ and $t<0$, respectively.  The poles that contribute in each case are
\begin{align}
t>0 : & \qquad   p^0 = -\h_1 P^0 - E_1(\vec p; \vec P) + i\e \: , 
\qquad  p^0 =  \h_2 P^0 - E_2(\vec p; \vec P) +i\e \: ,
\label{eq:poles t>0}
\\
t<0 : & \qquad   p^0 = -\h_1 P^0 + E_1(\vec p; \vec P) - i\e \: , 
\qquad  p^0 =  \h_2 P^0 + E_2(\vec p; \vec P) -i\e \: .
\label{eq:poles t<0}
\end{align}
Then 
\begin{align}
{\cal S} (t; \vec p ;P)  = \left\{
\bal{6}
&i \[
\frac{ e^{-i\r_1 (\vec p ; P) t} }  {2E_1 \[(P^0 - E_1)^2 -E_2^2 \]} +
\frac{ e^{i\s_2 (\vec p ; P) t} }  {2E_2 \[(P^0 + E_2)^2 -E_1^2 \]} \] \, ,&
\qquad
&\text{for } t < 0 \, ,& 
\\
&i \[
\frac{ e^{-i\s_1 (\vec p ; P) t} }  {2E_1 \[(P^0 + E_1)^2 -E_2^2 \]} +
\frac{ e^{i\r_2 (\vec p ; P) t} }  {2E_2 \[(P^0 - E_2)^2 -E_1^2 \]} \] \, ,&
\qquad
&\text{for } t > 0 \, , & 
\eal
\right.
\label{eq:S cal}
\end{align}
and 
\beq
{\cal S}_0 (\vec p; P) = {\cal S} (t = 0;\vec p; P) = 
\frac{i(E_1 +E_2)}{2E_1 E_2 \[(P^0)^2 - (E_1 + E_2)^2\]}
\: .
\label{eq:S0 cal}
\eeq
It will be useful to rewrite \eq{eq:S cal}, using \eq{eq:S0 cal}, as follows
\begin{multline}
{\cal S} (t; \vec p ;P)  = \frac{{\cal S}_0 (\vec p;P)}{E_1+E_2} \times 
\\
\times \left\{
\bal{6}
&\[
\(\frac{P^0 +E_1+E_2}{P^0-E_1+E_2}\) E_2 \, e^{-i\r_1 (\vec p ; P) t} +
\(\frac{P^0-E_1-E_2}{P^0-E_1+E_2}\)  E_1 \, e^{i\s_2 (\vec p ; P) t} \]
\, ,&
\quad
&\text{for } t < 0 \, ,& 
\\
&\[
\(\frac{P^0-E_1-E_2}{P^0 + E_1 -E_2}\)  E_2\, e^{-i\s_1 (\vec p ; P) t}+
\(\frac{P^0+E_1+E_2}{P^0 + E_1 -E_2}\)  E_1\, e^{ i\r_2 (\vec p ; P) t} \]
\, ,&
\quad
&\text{for } t > 0 \, . & 
\eal
\right.
\label{eq:S cal mod}
\end{multline}
(Note that in \eqref{eq:S cal} -- \eqref{eq:S cal mod}, it is implied that $E_1$ and $E_2$ come with the arguments $(\vec p; \vec P)$.)

Similarly to the above, we find
\beq
{\cal S}_1(t; \vec p;P) = 
\left\{
\bal{6}
&\frac{e^{i \s_1(\vec p;P) t}}{2E_1(\vec p; \vec P)} \, , 
&  \quad \text{for } \ t<0 \, ,
\\
&\frac{e^{i \rho_1(\vec p;P) t}}{2E_1(\vec p; \vec P)} \, , 
&  \quad \text{for } \ t>0 \, ,
\eal
\right.
\label{eq:S1 cal}
\eeq
and
\beq
{\cal S}_2(t; \vec p;P) =
\left\{
\bal{6}
&\frac{e^{-i \rho_2(\vec p;P) t}}{2E_2(\vec p; \vec P)}  \, ,
&  \quad \text{for } \ t<0 \, ,
\\
&\frac{e^{-i \s_2(\vec p;P) t}}{2E_2(\vec p; \vec P)}  \, , 
&  \quad \text{for } \ t>0 \, .
\eal
\right.
\label{eq:S2 cal}
\eeq

\subsection*{Non-relativistic approximation}

In the non-relativistic regime, $\vec P, \vec p \ll P^0, m_1, m_2$. Then 
\begin{align}
E_1(\vec p; \vec P) &\simeq \h_1 \(m + \frac{\vec P^2}{2 m}\)  + \frac{\vec P \cdot \vec p}{m} + \frac{\vec p^2}{2 m_1} \label{eq:E1 NR} \\
E_2(\vec p; \vec P) &\simeq \h_2 \(m + \frac{\vec P^2}{2 m}\)  - \frac{\vec P \cdot \vec p}{m} + \frac{\vec p^2}{2 m_2} \label{eq:E2 NR} \\
E_1(\vec p; \vec P) + E_2(\vec p; \vec P) & \simeq m + \frac{\vec P^2}{2 m} + \frac{\vec p^2}{2 \m} \label{eq:E1+E2 NR}
\end{align}
Note that in the last expression, the cancellation of the mixed terms proportional to ${\vec p \cdot \vec P}$, can be traced to \eq{eq:eta}. This reflects the fact that in the non-relativistic regime, the relative motion can be separated from the motion of the CM. For convenience, we set
\beq
P^0 = m + \frac{\vec P^2}{2 m} + {\cal E} \: .
\label{eq:P0}
\eeq
With these approximations, \eq{eq:S0 cal} becomes
\beq
{\cal S}_0 (\vec p; P) 
\simeq  -\frac{1}{i 4 m\mu \(P^0 -m -\frac{\vec P^2}{2m} - \frac{\vec p^2}{2\mu}\)} 
=-\frac{1}{i 4 m\mu \({\cal E} - \frac{\vec p^2}{2\mu}\)} 
\: .
\label{eq:S0 cal NR}
\eeq
%

\section{Partial-wave analysis for (co-)annihilation and decay processes \label{App:Partial Waves}}

We will prove the following two relations
\begin{align}
\int \frac{d^3 q}{(2\p)^3}  \: \tilde \f_{\vec k} (\vec q) 
\: |\vec q|^\ell P_\ell (\cos\theta_{\vec q, \vec p}) 
=
\frac{(2\ell+1)!!}{4\p i^\ell \, \ell!}
\[\frac{d^\ell}{dr^\ell} \int d\W_{\vec r} \: 
P_\ell (\cos \theta_{\vec p,\vec r}) \: \f_{\vec k} (\vec r) 
\]_{\vec r=0} \, ,
\label{eq:partial waves 1}
\end{align}
and
\begin{multline}
\int d\W_{\vec p} \, P_{\ell'} (\cos\theta_{\vec p}) \int \frac{d^3 q}{(2\p)^3}  \: \tilde \f_{\vec k} (\vec q) 
\: |\vec q|^\ell P_\ell (\cos\theta_{\vec q, \vec p}) 
= \\ =
\d_{\ell\ell'} \: \frac{(2\ell+1)!!}{i^\ell \, (2\ell +1) \, \ell!}
\[\frac{d^\ell}{dr^\ell} \int d\W_{\vec r} \: 
P_\ell (\cos \theta_{\vec r}) \: \f_{\vec k} (\vec r) 
\]_{\vec r=0} \, .
\label{eq:partial waves 2}
\end{multline}
We shall use the addition theorem of spherical harmonics,
\beq
P_\ell (\vec{\hat x} \cdot \vec{\hat y}) = \frac{4\p}{2\ell + 1}
\: \sum_{m = -\ell}^\ell Y_{\ell m}^* (\W_{\vec x}) Y_{\ell m} (\W_{\vec y}) \, ,
\label{eq:addition theorem}
\eeq
where $\vec{\hat x}, \, \vec{\hat y}$ are unit vectors, and $Y_{\ell m}$ are the spherical harmonics. From \eq{eq:addition theorem} and the orthonormality of $Y_{\ell m}$, it follows that
\beq
\int d\W_{\vec x} \, Y_{\ell m} (\W_{\vec x}) P_{\ell'} (\cos\theta_{\vec x, \vec y}) 
= \frac{4\p}{2\ell + 1} \,  Y_{\ell m} (\W_{\vec y}) \, \d_{\ell \ell'} \, .
\label{eq:Y and P}
\eeq
We will also need the expansion
\beq 
e^{i \vec q \cdot \vec r} = \sum_{\ell = 0}^\infty (2\ell +1) i^\ell \,
j_\ell (qr) \, P_\ell (\cos\theta_{\vec q, \vec r}) \, ,
\label{eq:exp expansion}
\eeq
where $j_\ell$ is the spherical Bessel function, which satisfies
\beq
\left.\frac{d^\ell j_\ell(x)}{dx^\ell} \right|_{x = 0} 
=  \frac{\ell!}{(2\ell +1)!!} \, .
\label{eq:Bessel}
\eeq

\bigskip
\bigskip

We begin with the right side of \eq{eq:partial waves 1}. Using \eqs{eq:addition theorem} -- \eqref{eq:Bessel} and the Fourier transform of $\tilde\f_{\vec k}(\vec q)$, we find
\begin{align}
&\[\frac{d^\ell}{dr^\ell} \int d\W_{\vec r} \: P_\ell (\cos \theta_{\vec p,\vec r}) 
\: \f_{\vec k} (\vec r) \]_{\vec r=0} =
\nn \\
&= \frac{4\p}{2\ell + 1} \sum_{m = -\ell}^\ell Y_{\ell m}^* (\W_{\vec p})
\int \frac{d^3 q}{(2\p)^3} \, \tilde{\f}_{\vec k} (\vec q)
\[ \frac{d^\ell}{d r^\ell} \int d\W_{\vec r} \, Y_{\ell m} (\W_{\vec r}) e^{i \vec q \cdot \vec r}
\]_{\vec r=0}
\nn \\
&= \frac{4\p}{2\ell + 1} \sum_{m = -\ell}^\ell Y_{\ell m}^* (\W_{\vec p})
\int \frac{d^3 q}{(2\p)^3} \, \tilde{\f}_{\vec k} (\vec q)
\sum_{\ell'=0}^\infty (2\ell'+1) \, i^{\ell'}
\[ \frac{d^\ell}{d r^\ell} j_{\ell'} (qr)\]_{\vec r=0}
\int \!d\W_{\vec r} \, Y_{\ell m} (\W_{\vec r}) P_{\ell'} (\cos\theta_{\vec q, \vec r})
\nn \\
&= \(\frac{4\p}{2\ell + 1}\)^2 \sum_{m = -\ell}^\ell Y_{\ell m}^* (\W_{\vec p})
\int \frac{d^3 q}{(2\p)^3} \, \tilde{\f}_{\vec k} (\vec q) \, |\vec q|^\ell  
\, (2\ell+1) \, i^{\ell} \[ \frac{d^\ell j_{\ell} (x)}{d x^\ell} \]_{x=0} 
\,  Y_{\ell m}(\W_{\vec q})
\nn \\
&= \frac{4\p \, i^{\ell} \, \ell!}{(2\ell+1)!!}
\int \frac{d^3 q}{(2\p)^3} \, \tilde{\f}_{\vec k} (\vec q) \, |\vec q|^\ell  
\,  P_{\ell }(\cos\theta_{\vec p, \vec q}) \, .
\end{align}
This proves \eq{eq:partial waves 1}.

Acting on the left side of \eq{eq:partial waves 1} with $\int d\W_{\vec p} \, P_{\ell'} (\cos\theta_{\vec p})$, we find
\begin{align}
\int d\W_{\vec p}
& \, P_{\ell'} (\cos\theta_{\vec p}) 
\int \frac{d^3 q}{(2\p)^3} \: \tilde \f_{\vec k} (\vec q) 
\: |\vec q|^\ell P_\ell (\cos\theta_{\vec q, \vec p}) = 
\nn \\
&=\frac{(2\ell+1)!!}{4\p i^\ell \, \ell!} \sqrt{\frac{4\pi}{2\ell+1}}
\int d\W_{\vec p} \, Y_{\ell',0} (\W_{\vec p}) 
\[\frac{d^\ell}{dr^\ell} \int d\W_{\vec r} \: 
P_\ell (\cos \theta_{\vec p,\vec r}) \: \f_{\vec k} (\vec r) 
\]_{\vec r=0}
\nn \\
&=\frac{(2\ell+1)!!}{4\p i^\ell \, \ell!} \(\frac{4\pi}{2\ell+1}\)^{3/2} \, \d_{\ell\ell'}
\[\frac{d^\ell}{dr^\ell} \int d\W_{\vec r} \: 
Y_{\ell,0} (\W_{\vec r}) \: \f_{\vec k} (\vec r) 
\]_{\vec r=0}
\nn \\
&=
\d_{\ell\ell'} \: \frac{(2\ell+1)!!}{i^\ell \, (2\ell +1) \, \ell!}
\[\frac{d^\ell}{dr^\ell} \int d\W_{\vec r} \: 
P_\ell (\cos \theta_{\vec r}) \: \f_{\vec k} (\vec r) 
\]_{\vec r=0} \, .
\end{align}
This proves \eq{eq:partial waves 2}.

\section{Integrals for the non-relativistic reduction of transition amplitudes  \label{App:Ksi}}

The transition amplitudes of Sec.~\ref{Sec:Interactions} contain integrals of the following forms
\begin{align}
\Ks_1 (\vec q, \vec p;  K,  P) \equiv
\int \frac{dq^0}{2\p} \: S(q;K)
\int \frac{dp^0}{2\p} \tilde{S}_1(\h_1 P + p) 
\: (2\p) \d(q^0-p^0-\h_2 P_\vf^0)  
\: ,
\label{eq:Ksi 1 def re}
\\
\Ks_2 (\vec q, \vec p;  K,  P) \equiv
\int \frac{dq^0}{2\p} \: S(q;K)
\int \frac{dp^0}{2\p} \tilde{S}_2(\h_2 P - p) 
\: (2\p) \d(q^0-p^0+\h_1 P_\vf^0)  
\: .
\label{eq:Ksi 2 def re}
\end{align}
We are interested in evaluating $\Ks_1 (\vec q, \vec p; K,P)$ at $\vec q - \vec p - \h_2 \vec P_\vf=0$, and $\Ks_2 (\vec q, \vec p; K,P)$ at $\vec q - \vec p + \h_1 \vec P_\vf=0$.  To evaluate $\Ks_1$ and $\Ks_2$, we Fourier-transform the $\d$-function, and use \eqs{eq:S cal mod}, \eqref{eq:S1 cal} and \eqref{eq:S2 cal}. 

\medskip

For $\Ks_1$, we obtain:
\begin{align}
\Ks_1 (\vec q, \vec p;  K,  P)
&= \int_{-\infty}^{\infty} dt \: e^{-i \h_2 P_\vf^0 t} 
\int \frac{dq^0}{2\p} \: S(q;K) \: e^{i q^0 t}
\int \frac{dp^0}{2\p} \tilde{S}_1(\h_1 P + p) \: e^{-i p^0 t}
\nn \\
&= \int_{-\infty}^0 dt \: e^{-i \h_2 P_\vf^0 t} \ 
{\cal S}(t;\vec q;K) \ {\cal S}_1 (t;\vec p;P)
+ \int_0^{\infty} dt \: e^{-i \h_2 P_\vf^0 t} \ 
{\cal S}(t;\vec q;K) \ {\cal S}_1 (t;\vec p;P)
\nn \\
&= \frac{{\cal S}_0(\vec q;K)}{\[E_1(\vec q;\vec K)+E_2(\vec q;\vec K)\] \, 2E_1(\vec p;\vec P)} \times 
\[\Ks_1^- (\vec q, \vec p;  K,  P) + \Ks_1^+ (\vec q, \vec p;  K,  P)\] \, ,
\label{eq:Ksi1 decomp}
\end{align}
where
\begin{align}
\Ks_1^- (\vec q, \vec p;  K,  P) &= 
\frac{K^0+E_1(\vec q;\vec K)+E_2(\vec q;\vec K)}{K^0 - E_1(\vec q;\vec K) + E_2(\vec q;\vec K)} E_2(\vec q;\vec K) 
\int_{-\infty}^0 dt \: e^{i [-\h_2 P_\vf^0 - \r_1(\vec q;K)  + \s_1(\vec p;P)]t} 
\nn \\
& + \frac{K^0-E_1(\vec q;\vec K)-E_2(\vec q;\vec K)}{K^0 - E_1(\vec q;\vec K) + E_2(\vec q;\vec K)} E_1(\vec q;\vec K)
\int_{-\infty}^0 dt \: e^{i [-\h_2 P_\vf^0 + \s_2(\vec q;K)  + \s_1(\vec p;P)]t} \, ,
\label{eq:Ksi1- def}
\end{align}
and
\begin{align}
\Ks_1^+ (\vec q, \vec p;  K,  P) &= 
\frac{K^0-E_1(\vec q;\vec K)-E_2(\vec q;\vec K)}{K^0 + E_1(\vec q;\vec K) - E_2(\vec q;\vec K)}
E_2(\vec q;\vec K)
\int_0^{\infty} dt \: e^{i [-\h_2 P_\vf^0 - \s_1(\vec q;K)  + \r_1(\vec p;P)]t} 
\nn \\
&+ \frac{K^0+E_1(\vec q;\vec K)+E_2(\vec q;\vec K)}{K^0 + E_1(\vec q;\vec K) - E_2(\vec q;\vec K)}
E_1(\vec q;\vec K)
\int_0^{\infty} dt \: e^{i [-\h_2 P_\vf^0 + \r_2(\vec q;K)  + \r_1(\vec p;P)]t} \, . 
\label{eq:Ksi1+ def}
\end{align}
Using the definitions \eqref{eq:rho 1} -- \eqref{eq:sigma 2} and the overall energy-momentum conservation, $K = P + P_\vf$, we may simplify the phases appearing in the above integrals, as follows
\begin{align}
-\h_2 P_\vf^0 -\rho_1 (\vec q; K) + \s_1(\vec p;P) 
&\ = \ P^0-K^0 + E_1(\vec q;\vec K) +E_1(\vec p;\vec P) \, , 
\\
-\h_2 P_\vf^0 +\s_2(\vec q;K) + \s_1(\vec p;P) 
&\ = \ P^0 + E_2(\vec q;\vec K) + E_1(\vec p;\vec P) \, ,
\\
-\h_2 P_\vf^0 - \s_1(\vec q; K) +\rho_1 (\vec p;P)
&\ = \ P^0-K^0 -E_1(\vec q;\vec K) -E_1(\vec p;\vec P) \, ,
\\
-\h_2 P_\vf^0 + \rho_2(\vec q; K) +\rho_1 (\vec p;P)
&\ = \ P^0 -E_2(\vec q;\vec K) -E_1(\vec p;\vec P) \, .
\end{align}
Adding a small imaginary parts to the integration variable $t$, such that the integrals in \eqref{eq:Ksi1- def}, \eqref{eq:Ksi1+ def} converge, we find
\begin{align}
&\Ks_1^- (\vec q, \vec p;  K,  P) = -\frac{i}{K^0 - E_1(\vec q;\vec K) + E_2(\vec q;\vec K)}
\nn \\
&\times \(\frac{\[K^0+E_1(\vec q;\vec K)+E_2(\vec q;\vec K)\] E_2(\vec q;\vec K)}{P^0-K^0 + E_1(\vec q;\vec K) + E_1(\vec p;\vec P)} 
+ \frac{\[K^0-E_1(\vec q;\vec K)-E_2(\vec q;\vec K)\] E_1(\vec q;\vec K)}{P^0 + E_2(\vec q;\vec K) + E_1(\vec p;\vec P)} \) 
\nn \\
&= - i \ 
\frac{\[P^0+E_1(\vec p;\vec P)+E_1(\vec q;\vec K)+E_2(\vec q;\vec K)\]\[E_1(\vec q;\vec K) + E_2(\vec q;\vec K)\] -K^0 E_1(\vec q;K)}
{\[P^0-K^0 + E_1(\vec q;\vec K)  + E_1(\vec p;\vec P)\] \, \[P^0 + E_2(\vec q;\vec K) + E_1(\vec p;\vec P)\]}
\, ,
\label{eq:Ksi1-}
\\
&\Ks_1^+ (\vec q, \vec p;  K,  P) = \frac{i}{K^0 + E_1(\vec q;\vec K) - E_2(\vec q;\vec K)}
\nn \\
&\times \( \frac{\[K^0-E_1(\vec q;\vec K)-E_2(\vec q;\vec K)\] E_2(\vec q;\vec K)}{P^0 - K^0 - E_1(\vec q;\vec K) - E_1(\vec p;\vec P)}
+ \frac{\[K^0+E_1(\vec q;\vec K)+E_2(\vec q;\vec K)\] E_1(\vec q;\vec K)}{P^0 - E_2(\vec q;\vec K) - E_1(\vec p;\vec P)} \) 
\nn \\
&=i \ 
\frac{\[P^0-E_1(\vec p;\vec P)-E_1(\vec q;\vec K)-E_2(\vec q;\vec K)\]\[E_1(\vec q;\vec K) + E_2(\vec q;\vec K)\] -K^0 E_1(\vec q;K)}
{\[P^0-K^0 - E_1(\vec q;\vec K)  - E_1(\vec p;\vec P)\]  \, \[P^0 - E_2(\vec q;\vec K) - E_1(\vec p;\vec P)\]}
\, . 
\label{eq:Ksi1+}
\end{align}

\medskip

For $\Ks_2$, we obtain:
\begin{align}
\Ks_2 (\vec q, \vec p;  K,  P)
&= \int_{-\infty}^{\infty} dt \: e^{i \h_1 P_\vf^0 t} 
\int \frac{dq^0}{2\p} \: S(q;K) \: e^{i q^0 t}
\int \frac{dp^0}{2\p} \tilde{S}_2(\h_1 P - p) \: e^{-i p^0 t}\: f(p)
\nn \\
&= \int_{-\infty}^0 dt \: e^{i \h_1 P_\vf^0 t} \ 
{\cal S}(t;\vec q;K) \ {\cal S}_2 (t;\vec p;P)
+ \int_0^{\infty} dt \: e^{i \h_1 P_\vf^0 t} \ 
{\cal S}(t;\vec q;K) \ {\cal S}_2 (t;\vec p;P)
\nn \\
&= \frac{{\cal S}_0(\vec q;K)}{\[E_1(\vec q;\vec K)+E_2(\vec q;\vec K)\] \, 2E_2(\vec p;\vec P)} \times 
\[\Ks_2^- (\vec q, \vec p;  K,  P) + \Ks_2^+ (\vec q, \vec p;  K,  P)\] \, ,
\label{eq:Ksi2 decomp}
\end{align}
where
\begin{align}
\Ks_2^- (\vec q, \vec p;  K,  P) &= 
\frac{K^0+E_1(\vec q;\vec K)+E_2(\vec q;\vec K)}{K^0 - E_1(\vec q;\vec K) + E_2(\vec q;\vec K)} 
E_2(\vec q;\vec K) 
\int_{-\infty}^0 dt \: e^{i [\h_1 P_\vf^0 - \r_1(\vec q;K)  - \r_2(\vec p;P)]t} 
\nn \\
& + \frac{K^0-E_1(\vec q;\vec K)-E_2(\vec q;\vec K)}{K^0 - E_1(\vec q;\vec K) + E_2(\vec q;\vec K)} 
E_1(\vec q;\vec K)
\int_{-\infty}^0 dt \: e^{i [\h_1 P_\vf^0 + \s_2(\vec q;K)  - \r_2(\vec p;P)]t} \, ,
\label{eq:Ksi2- def}
\end{align}
and
\begin{align}
\Ks_2^+ (\vec q, \vec p;  K,  P) &= 
\frac{K^0-E_1(\vec q;\vec K)-E_2(\vec q;\vec K)}{K^0 + E_1(\vec q;\vec K) - E_2(\vec q;\vec K)}
E_2(\vec q;\vec K)
\int_0^{\infty} dt \: e^{i [\h_1 P_\vf^0 - \s_1(\vec q;K)  - \s_2(\vec p;P)]t} 
\nn \\
&+ \frac{K^0+E_1(\vec q;\vec K)+E_2(\vec q;\vec K)}{K^0 + E_1(\vec q;\vec K) - E_2(\vec q;\vec K)}
E_1(\vec q;\vec K)
\int_0^{\infty} dt \: e^{i [\h_1 P_\vf^0 + \r_2(\vec q;K)  - \s_2(\vec p;P)]t} \, . 
\label{eq:Ksi2+ def}
\end{align}
Using the definitions \eqref{eq:rho 1} -- \eqref{eq:sigma 2} and the overall energy-momentum conservation, $K = P + P_\vf$, we simplify the phases appearing in the above integrals, as follows
\begin{align}
\h_1 P_\vf^0 -\rho_1 (\vec q; K) - \rho_2(\vec p;P) 
&\ = \ -P^0 + E_1(\vec q;\vec K) +E_2(\vec p;\vec P) \, , 
\\
\h_1 P_\vf^0 +\s_2(\vec q;K) - \rho_2(\vec p;P) 
&\ = \ - P^0 + K^0 + E_2(\vec q;\vec K) + E_2(\vec p;\vec P) \, ,
\\
\h_1 P_\vf^0 - \s_1(\vec q; K) - \s_2 (\vec p;P)
&\ = \ -P^0 -E_1(\vec q;\vec K) -E_2(\vec p;\vec P) \, ,
\\
\h_1 P_\vf^0 + \rho_2(\vec q; K) - \s_2 (\vec p;P)
&\ = \ - P^0 + K^0 -E_2(\vec q;\vec K) -E_2(\vec p;\vec P) \, .
\end{align}
Adding a small imaginary parts to the integration variable $t$, as before, we find
\begin{align}
&\Ks_2^- (\vec q, \vec p;  K,  P) = -\frac{i}{K^0 - E_1(\vec q;\vec K) + E_2(\vec q;\vec K)}
\nn \\
&\times \(\frac{\[K^0+E_1(\vec q;\vec K)+E_2(\vec q;\vec K)\] E_2(\vec q;\vec K)}{-P^0 + E_1(\vec q;\vec K) + E_2(\vec p;\vec P)} 
+ \frac{\[K^0-E_1(\vec q;\vec K)-E_2(\vec q;\vec K)\] E_1(\vec q;\vec K)}{-P^0 +K^0 + E_2(\vec q;\vec K) + E_2(\vec p;\vec P)} \) 
\nn \\
&= i \ 
\frac{\[P^0 - E_2(\vec p;\vec P) - E_1(\vec q;\vec K) - E_2(\vec q;\vec K)\]\[E_1(\vec q;\vec K) + E_2(\vec q;\vec K)\] - K^0 E_2(\vec q; \vec K)}
{\[P^0-K^0 - E_2(\vec q;\vec K) - E_2(\vec p;\vec P)\] \, \[P^0 - E_1(\vec q;\vec K) - E_2(\vec p;\vec P)\]}
\, ,
\label{eq:Ksi2-}
\\
&\Ks_2^+ (\vec q, \vec p;  K,  P) = \frac{i}{K^0 + E_1(\vec q;\vec K) - E_2(\vec q;\vec K)}
\nn \\
&\times \( \frac{\[K^0-E_1(\vec q;\vec K)-E_2(\vec q;\vec K)\] E_2(\vec q;\vec K)}{-P^0 - E_1(\vec q;\vec K) - E_2(\vec p;\vec P)}
+ \frac{\[K^0+E_1(\vec q;\vec K)+E_2(\vec q;\vec K)\] E_1(\vec q;\vec K)}{-P^0 +K^0 - E_2(\vec q;\vec K) - E_2(\vec p;\vec P)} \) 
\nn \\
&= - i \ 
\frac{\[P^0 + E_2(\vec p;\vec P) + E_1(\vec q;\vec K) + E_2(\vec q;\vec K)\]\[E_1(\vec q;\vec K) + E_2(\vec q;\vec K)\] -K^0 E_2(\vec q; \vec K)}
{\[P^0-K^0 + E_2(\vec q;\vec K) + E_2(\vec p;\vec P)\]  \, \[P^0 + E_1(\vec q;\vec K) + E_2(\vec p;\vec P)\]}
\, . 
\label{eq:Ksi2+}
\end{align}

\subsection*{Non-relativistic approximation}

In the following, we consider the CM frame, $\vec K=0$, as in Sec.~\ref{Sec:Interactions}. We shall evaluate $\Ks_1(\vec q, \vec p;K,P)$ and $\Ks_2(\vec q, \vec p;K,P)$ at next-to-leading order in the momenta $\vec q, \vec p$, applying the non-relativistic approximations of \eqs{eq:E1 NR} -- \eqref{eq:E1+E2 NR}, and setting, according to \eq{eq:P0},
\begin{align}
K^0 &= m + \frac{\vec K^2}{2m} + {\cal E}_{\vec k} \, , \\
P^0 &= m + \frac{\vec P^2}{2m} + {\cal E}_n \, ,
\end{align}
where ${\cal E}_n$ and ${\cal E}_{\vec k}$ are given by \eqs{eq:En} and \eqref{eq:Ek}. 
The next-to-leading order corrections in $\vec q, \vec p$ become important when the leading term in a $\vrel^2$ expansion cancels, as is the case for the interaction of two degenerate scalar particles via a scalar mediator (c.f. Sec.~\ref{sec:SSS}). 
Note though that we drop subleading terms in the couplings; such corrections do not change the structure of the wavefunction convolution integrals (c.f.~\eqs{eq:I cal k-n def}, \eqref{eq:J cal k-n def}) that enter into the transition amplitudes of Sec.~\ref{Sec:Interactions}, and thus do not avert the cancellation of the leading term in the $\vrel^2$ expansion.  The same holds for $\vec P^2$ corrections.  In addition, as seen from \eq{eq:P CM}, $\vec P^2$ corrections are of order $\vrel^4, \a^4$ and $\a^2\vrel^2$, while $\vec p^2$ corrections are only of order $\vrel^2$. Similarly, $\vec P \cdot \vec p$ corrections are suppressed with respect to $\vec p^2$.  As mentioned, we are interested, in particular, in evaluating $\Ks_1(\vec q, \vec p;K,P)$ at $\vec q - \vec p - \h_2 \vec P_\vf = 0$ and $\Ks_2(\vec q, \vec p;K,P)$ at $\vec q - \vec p + \h_1 \vec P_\vf = 0$. Then, according to the above, we shall keep only the $\vec p^2$ corrections.

At $\vec q - \vec p - \h_2 \vec P_\vf = 0$, we find
\beq
\Ks_1 (\vec q, \vec p; K,P) \simeq 2 m_2 \, {\cal S}_0(\vec q;K) \, {\cal S}_0(\vec p;P)
\[1 + \frac{\vec p^2}{2\mu^2} \(1 - \frac{2\mu}{m}\) 
\] \, ,
\label{eq:Ksi 1 NR}
\eeq
and at $\vec q - \vec p + \h_1 \vec P_\vf = 0$, we find 
\beq
\Ks_2 (\vec q, \vec p; K,P) \simeq 2 m_1 \, {\cal S}_0(\vec q;K) \, {\cal S}_0(\vec p;P)
\[1 + \frac{\vec p^2}{2\mu^2} \(1 - \frac{2\mu}{m}\) 
\] \, .
\label{eq:Ksi 2 NR}
\eeq


\addtocontents{toc}{\protect\enlargethispage{50pt}}
\section{Schr\"{o}dinger wavefunctions and convolution integrals \label{App:WaveFun}}

In the following, we shall consider the attractive Coulomb potential
\beq
V(\vec r) = -\frac{\a}{r} \: .
\label{eq:Coulomb}
\eeq

\subsection{Solutions of the Schr\"{o}dinger equation \label{app:Schr sol}}

The discrete spectrum of solutions to the Schr\"{o}dinger equation \eqref{eq:SE coord bound}, 
\beq
\[-\frac{\nabla^2}{2\m} + V({\bf r})\]\ps_n ({\bf r}) = {\cal E}_n \ps_n ({\bf r}) \: ,
\eeq
for the Coulomb potential of \eq{eq:Coulomb}, are (see e.g.~\cite{Messiah:1962})
\begin{align}
\psi_{n \ell m} (\vec r) &= R_{n\ell}(r) \, Y_{\ell m}(\W) \: , 
\label{eq:psi nlm}
\\
R_{n\ell}(r) &= (2\k/n)^{3/2} \[\frac{(n-\ell-1)!}{2n (n+\ell)!}\]^{1/2}
\ e^{-\k r/n}  \: \(2 \k r / n\)^\ell \: L_{n-\ell-1}^{(2\ell+1)} \(2 \k r/n\) \: ,
\label{eq:R}
\end{align}
where $Y_{\ell m}(\W)$ are the spherical harmonics and $L_{n-\ell-1}$ are the generalized Laguerre polynomials of degree $n-\ell-1$.\footnote{We assume the following normalisation for the Laguerre polynomials
\beq
\int_0^\infty x^a  e^{-x} \, L_n^{(a)} (x) \, L_m^{(a)}(x) \, dx = \frac{\Gamma(n+a+1)}{n!} \: \d_{n,m}
\eeq
} 
In the above, 
\beq \k \equiv \mu \a \eeq 
is the Bohr momentum, and the energy eigenvalues are
\beq
{\cal E}_n = -\frac{\mu \a^2}{2n^2} \: .
\label{eq:En}
\eeq

In the following, we shall consider transitions to the ground state $\{100\}$. For the case of a scalar mediator, we shall also consider radiative capture to the first excited state with non-zero angular momentum $\{210\}$. The corresponding wavefuctions are
\begin{align}
\psi_{100} (\vec r) &= \frac{\kappa^{3/2}}{\sqrt{\p}} \: e^{-\k r} \: , \label{eq:psi 100} \\
\psi_{210} (\vec r) &= \frac{\kappa^{3/2}}{4 \sqrt{2\p}} \: \k r \: e^{-\k r/2} \: \cos \theta_{\vec r} \: , \label{eq:psi 210}
\end{align}
where $\theta_{\vec r}$ is the polar angle of the position vector $\vec r$.

\bigskip

The continuous spectrum of solutions to the Schr\"{o}dinger equation \eqref{eq:SE coord free}
\beq
\[-\frac{\nabla^2}{2\m} + V({\bf r})\]\f_{\vec k} ({\bf r}) = {\cal E}_{\vec k} \f_{\vec k} ({\bf r}) \: ,
\eeq
is characterized by the quantum number $\vec k$, which is the  expectation value of the momentum of the reduced system, with $\vec k = \mu \vec \vrel$ and $\vec \vrel$ being the expectation value of the relative velocity. The solutions are (see e.g.~\cite{Messiah:1962})
\beq
\f_{\vec k} (\vec r) = e^{\pi \z/2} \: \G(1-i \z) 
\: F \[i\z, 1, i (kr - \vec k \cdot \vec r)\] 
\: e^{i \vec k \cdot \vec r} \: ,
\label{eq:phi k}
\eeq
where $k = |\vec k|$, and
\beq 
\z \equiv \kappa / k \: . 
\label{eq:zeta}
\eeq
The energy eigenvalues are
\beq
{\cal E}_{\vec k} = \frac{\vec k^2}{2\mu} = \frac{1}{2} \, \mu \vrel^2 \: .
\label{eq:Ek}
\eeq
Thus, $\z = \a/\vrel$.

\subsection{Convolution of the wavefunctions \label{app:WF convol}}

We now want to calculate the integrals \eqref{eq:I cal k-n def} -- \eqref{eq:K cal k-n def}, appearing in the amplitudes of Sec.~\ref{Sec:Interactions},
\begin{align}
{\cal I}_{\vec k, n} (\vec b) 
&\equiv 
\int \frac{d^3p}{(2\p)^3} \:\tilde \psi_n^\star (\vec p) \: \tilde \f_{\vec k}  (\vec p + \vec b)
= \int d^3 r \: \psi_n^\star (\vec r) \:  \f_{\vec k}  (\vec r) e^{-i \vec b \cdot \vec r} \: ,
\label{eq:I cal k-n def re}
\\
\boldsymbol{\cal J}_{\vec k, n} (\vec b) 
&\equiv 
\int \frac{d^3p}{(2\p)^3} \: \vec p \: \tilde \psi_n^\star (\vec p) \: \tilde \f_{\vec k}  (\vec p +\vec b) 
= i \int d^3 r \: [\nabla \psi_n^\star (\vec r)] \:  \f_{\vec k}  (\vec r) e^{-i \vec b \cdot \vec r}
\: ,
\label{eq:J cal k-n def re}
\\
{\cal K}_{\vec k, n} (\vec b) 
&\equiv 
\int \frac{d^3p}{(2\p)^3} \: \vec p^2 \: \tilde \psi_n^\star (\vec p) \: \tilde \f_{\vec k}  (\vec p + \vec b) 
= - \int d^3 r \: [\nabla^2 \psi_n^\star (\vec r)] \:  \f_{\vec k}  (\vec r) e^{-i \vec b \cdot \vec r}
\: ,
\label{eq:K cal k-n def re}
\end{align}
and
\begin{align}
{\cal I}_{n', n} (\vec b) 
&\equiv 
\int \frac{d^3p}{(2\p)^3} \:\tilde \psi_n^\star (\vec p) \: \tilde \psi_{n'}  (\vec p + \vec b) 
= \int d^3 r \: \psi_n^\star (\vec r) \:  \psi_{n'}  (\vec r) e^{-i \vec b \cdot \vec r} \: ,
\label{eq:I cal n'-n def re}
\\
\boldsymbol{\cal J}_{n', n} (\vec b) 
&\equiv 
\int \frac{d^3p}{(2\p)^3} \: \vec p \: \tilde \psi_n^\star (\vec p) \: \tilde \psi_{n'}  (\vec p + \vec b) 
= i \int d^3 r \: [\nabla \psi_n^\star (\vec r)] \:  \psi_{n'}  (\vec r) e^{-i \vec b \cdot \vec r} \: ,
\label{eq:J cal n'-n def re}
\\
{\cal K}_{n', n} (\vec b) 
&\equiv 
\int \frac{d^3p}{(2\p)^3} \: \vec p^2 \: \tilde \psi_n^\star (\vec p) \: \tilde \psi_{n'}  (\vec p + \vec b) 
= - \int d^3 r \: [\nabla^2 \psi_n^\star (\vec r)] \:  \psi_{n'}  (\vec r) e^{-i \vec b \cdot \vec r} \: ,
\label{eq:K cal n'-n def re}
\end{align}
where we transformed into the coordinate space using \eqs{eq:psi FT} and \eqref{eq:phi FT}.

\bigskip

For the $\{100\}$ state, the integrals \eqref{eq:I cal k-n def re} -- \eqref{eq:K cal k-n def re} become
\begin{align}
{\cal I}_{\vec k, \{100\}} (\vec b)
&= \frac{\k^{3/2} \, e^{\p\z/2} \: \G(1-i \z) }{\sqrt{\p}}
\int d^3 r \ e^{i(\vec k - \vec b) \cdot \vec r -\kappa r}
\ F\[i\z,1, i (kr - \vec k\cdot \vec r)\]  \: ,
\label{eq:I cal inter1}
\\
\boldsymbol{\cal J}_{\vec k, \{100\}} (\vec b)
&= - \frac{i\, \k^{5/2} \, e^{\p\z/2} \: \G(1-i \z)}{\sqrt{\p}}
\int d^3 r \ \hat{\vec{r}}\ e^{i(\vec k - \vec b) \cdot \vec r -\kappa r}
\ F\[i\z,1, i (kr - \vec k\cdot \vec r)\]  \: ,
\label{eq:J cal inter1}
\\
{\cal K}_{\vec k, \{100\}} (\vec b)
&= -\k^2 {\cal I}_{\vec k, n=1} (\vec b) + \bar{\cal K}_{\vec k} (\vec b) \: ,
\label{eq:K cal inter1}
\end{align}
where $\hat{\vec{r}} = \vec r/r$ and 
\beq
\bar{\cal K}_{\vec k} (\vec b) \equiv 
\frac{2\kappa^{5/2} \, e^{\p\z/2} \: \G(1-i \z) }{\sqrt{\p}}
\int \frac{d^3 r}{r} \ e^{i(\vec k - \vec b) \cdot \vec r -\kappa r}
\ F\[i\z,1, i (kr - \vec k\cdot \vec r)\]  \: .
\label{eq:bar K cal def}
\eeq
For the $\{210\}$ state, we will need only the integral \eqref{eq:I cal k-n def re},
\begin{align}
{\cal I}_{\vec k, \{210\}} (\vec b)
&= \frac{\k^{5/2} \, e^{\p\z/2} \: \G(1-i \z) }{4\sqrt{2\p}}
\int d^3 r \ r \: \cos \theta_{\vec r} \: e^{i(\vec k - \vec b) \cdot \vec r -\kappa r/2}
\ F\[i\z,1, i (kr - \vec k\cdot \vec r)\]  \: .
\label{eq:I cal 210 inter1}
\end{align}

To evaluate these expressions, we make use of the identity~\cite{AkhiezerMerenkov_sigmaHydrogen}
\beq
\int d^3 r \ \frac{e^{i(\vec k - \vec b) \cdot \vec r -\kappa r}}{4\p r}
\ F\[i\z,1, i (kr - \vec k\cdot \vec r)\]  =
\frac{\[\vec b^2 + (\kappa - i k)^2\]^{-i\z}}{\[(\vec k - \vec b)^2 + \kappa^2 \]^{1-i\z}} 
\ \equiv \ f_{\vec k, \vec b} (\kappa)  \: .
\label{eq:AM identity}
\eeq
Differentiating \eq{eq:AM identity} with respect to $\k$~\cite{Pearce:2013ola}, with respect to $b^j$~\cite{AkhiezerMerenkov_sigmaHydrogen}, and with respect to $b$, we obtain the following expressions
\begin{align}
{\cal I}_{\vec k, \{100\}} (\vec b)
&=- 4\sqrt{\p} \: e^{\p\z/2} \: \G(1-i \z) \, \k^{3/2}
\ \frac{\partial f_{\vec k, \vec b} (\kappa)}{\partial \kappa}
\: ,
\label{eq:I cal inter2}
\\
{\cal J}_{\vec k, \{100\}}^j (\vec b)
&= 4\sqrt{\p} \: e^{\p\z/2} \: \G(1-i \z) \, \kappa^{5/2} 
\ \frac{\partial f_{\vec k, \vec b} (\kappa)}{\partial  b^j} 
\: ,
\label{eq:J cal inter2}
\\
\bar{\cal K}_{\vec k} (\vec b)
&= 8\sqrt{\p} \: e^{\p\z/2} \: \G(1-i \z) \, \kappa^{5/2}
\ f_{\vec k, \vec b} (\kappa)
\: .
\label{eq:bar K cal inter2}
\\
{\cal I}_{\vec k, \{210\}} (\vec b)
&= -i \sqrt{\frac{\p}{2}} \: e^{\p\z/2} \: \G(1-i \z)  \, \kappa^{5/2}
\[ \frac{\partial^2 f_{\vec k, \vec b} (\kappa')}{\partial b \, \partial \kappa'} \]_{\kappa' = \kappa/2}
\: .
\label{eq:I cal 210 inter2}
\end{align}

For the cases of interest, $\vec b =  \h_2 \vec P_\vf$ or $- \h_1 \vec P_\vf$; evidently $b=|\vec b| < |\vec P_\vf|$. Moreover, $|\vec P_\vf|$ is determined by energy-momentum conservation (c.f. \eq{eq:P CM}). In the CM frame ($\vec K = 0$), and in the non-relativistic regime, $|\vec P_\vf| \simeq \mu (\a^2 + \vrel^2)/ 2$ for capture in the ground state, and  $|\vec P_\vf| \simeq  \mu (\a^2/4 + \vrel^2)/ 2$ for capture in the $\{210\}$ state. As long as $\a, \vrel < 1$, then $b \ll \kappa$ if $\a>\vrel$, or $b \ll k$, if $\a< \vrel$.   In evaluating \eqs{eq:I cal inter2} -- \eqref{eq:I cal 210 inter2}, we may thus expand in $b$ and keep only the leading orders. In particular, to estimate $\partial f_{\vec k, \vec b} (\kappa) / \partial \kappa$,  $\partial f_{\vec k, \vec b} (\kappa) / \partial b^j$ and  $\partial^2 f_{\vec k, \vec b} (\kappa') / \partial b \, \partial \kappa'$, we first differentiate $f$, then use $\k = \z k$ (c.f. \eq{eq:zeta}) and expand up to order $b^2$ around $b=0$. (Note, though, that in most applications, we will not need all of the terms included in the expansions below.) Setting 
\beq 
\cos \tilde \theta \equiv \frac{\vec k \cdot \vec b }{k \, b} \: , 
\label{eq:k,b angle}
\eeq 
we find
\beq
f_{\vec k, \vec b} (\kappa) \simeq
\frac{1}{k^2 (1+\z^2)^2}  \(\frac{i\z+1}{i\z-1}\)^{-i\z}
\[1+ \frac{2 b \cos \tilde{\theta}}{k (1+ i\z)}
+\frac{b^2 \[-(1-i\z)+2(2-i\z)\cos^2\tilde{\theta}\]}{k^2 (1+ \z^2) (1+i \z)} \]
\, , 
\label{eq:f expansion}
\eeq
\beq
\frac{\partial f_{\vec k, \vec b} (\kappa)}{\partial \kappa} \simeq
-\frac{4\z \, b}{k^4 (1+ \z^2)^2 (1+i\z)} \(\frac{i\z+1}{i\z-1}\)^{-i\z}
\!\[\cos \tilde{\theta} + \frac{b \[-(1-i\z) + 2(2-i\z)\cos^2\tilde{\theta}\]}{k (1+\z^2)}\]
\, , 
\label{eq:df/dkappa}
\eeq
\begin{multline}
\frac{\partial f_{\vec k, \vec b} (\kappa)}{\partial b^j} \simeq
\frac{2}{k^4 (1+\z^2) (1+i\z)}  \(\frac{i\z+1}{i\z-1}\)^{-i\z} 
\left\{- \frac{b^j}{1+i\z} \[1 
+ \frac{4b \cos \tilde{\theta}}{k (1+\z^2)} \] \right. 
\\
+ \left. k^j
\[1+ \frac{2 (2-i\z) b  \cos\tilde{\theta} }{k(1+\z^2)}
- \frac{2b^2 \[1-i\z-(6-\z^2-i5\z)\cos^2\theta\]}{k^2 (1+\z^2)^2} \] 
\right\} \, ,
\label{eq:df/dbj}
\end{multline}
and
\begin{multline}
\[\frac{\partial^2 f_{\vec k, \vec b} (\kappa')}{\partial b \, \partial \kappa'}\]_{\kappa' = \kappa/2} 
\simeq
\frac{2^8 b \, \z (i\z-2)}{k^5 \, (4+\z^2)^4} 
\: \(\frac{i\z+2}{i\z-2}\)^{-i\z} \times
\\
\[ i\z-2 -4 (i\z-1)\cos^2\tilde{\theta} 
- \frac{8 b (i\z-1) \[3(i\z-2) -4 (i\z-3)\cos^2\tilde{\theta}\] \cos\tilde{\theta}}{k(4+\z^2) } 
\] \, .
\label{eq:df/dbdkappa}
\end{multline}

For convenience, we define
\beq
{\cal R}(\z)
\equiv 
\frac{8\sqrt{\p} \: \z^{5/2} \: e^{\p\z/2}}{1+\z^2} 
\ \frac{\G(1-i \z)}{1+i\z} \, \(\frac{i\z+1}{i\z-1}\)^{-i\z} \, .
\label{eq:R cal}
\eeq
In evaluating the cross-sections of interest, we shall need
\beq
|{\cal R}(\z)|^2 = \frac{2^7 \p^2 \z^6}{(1+\z^2)^3} \:
\frac{e^{-4\z {\rm arccot}\, \z}}{1-e^{-2\p\z}} \, ,
\label{eq:R cal square}
\eeq
where we used the identities
\beq 
|\G(1-i\z)|^2 = 2\p\z \: \frac{e^{-\p \z} }{1- e^{-2\p\z}} \, , 
\label{eq:identity1}
\eeq
\beq 
\abs{\(\frac{i\z+1}{i\z-1}\)^{-i\z}}^2 = e^{-4\z \, {\rm arccot}\,\z} \, . 
\label{eq:identity2}
\eeq

Collecting the above, we obtain
\beq
{\cal I}_{\vec k, \{100\}} (\vec b) \simeq 
\frac{2{\cal R}(\z)}{1+\z^2} \: \frac{b}{k^{5/2}}
\(\cos\tilde{\theta} + \frac{b \[-1+i\z + 2(2-i\z)\cos^2\tilde{\theta}\]}{k (1+\z^2)}\)
\, ,
\label{eq:I cal fin}
\eeq
\begin{multline}
\boldsymbol{\cal J}_{\vec k, \{100\}} (\vec b) \simeq
\frac{{\cal R}(\z)}{k^{3/2}}
\left\{ - \frac{\vec b}{1+i\z} \[1 + \frac{4 b \cos \tilde{\theta}}{k (1+\z^2)}\] 
\right.  \\ \left. 
+ \vec k 
\[1+ \frac{2 (2-i\z) b  \cos\tilde{\theta} }{k(1+\z^2)}
+ \frac{2b^2 \[-1+i\z+(6-\z^2-i5\z)\cos^2\tilde\theta\]}{k^2 (1+\z^2)^2} \] 
\right\}
\, ,
\label{eq:J cal fin}
\end{multline}
\beq
\bar{\cal K}_{\vec k} (\vec b) \simeq  {\cal R}(\z) \, k^{1/2}
\, \[1+i\z + \frac{2b \cos \tilde \theta}{k} 
+ \frac{b^2 \[-1+i\z + 2(2-i\z) \cos^2\tilde\theta\]}{k^2 (1+\z^2)}\]
\, ,
\label{eq:bar K cal fin}
\eeq
and
\begin{multline}
{\cal I}_{\vec k, \{210\}} (\vec b) \simeq 
-\frac{i 2^7 \sqrt{2\p} \, b}{k^{5/2}}
\: \frac{\z^{7/2} \, (i\z-2) \, e^{\p\z/2} \, \G(1-i\z)}{(4+\z^2)^4} 
\: \(\frac{i\z+2}{i\z-2}\)^{-i\z}  
\\ \times \[ i\z-2 -4 (i\z-1)\cos^2\tilde{\theta} 
- \frac{8 b (i\z-1) \[3(i\z-2) -4 (i\z-3)\cos^2\tilde{\theta}\] \cos\tilde{\theta}}{k(4+\z^2) } 
\]
\, .
\label{eq:I cal 210 fin}
\end{multline}

\bigskip

Note that in the above, the angle $\tilde{\theta}$ is related to the angle $\theta$ between $\vec k$ and $\vec P_\vf$, defined by
\beq
\cos \theta = \frac{\vec k \cdot \vec P_\vf}{k |\vec P_\vf|} \: ,
\label{eq:theta}
\eeq
as follows
\beq
\tilde{\theta} = \left\{
\bal{6}
&\theta,& \quad &\text{for } \vec b = \h_2 \vec P_\vf \, ,& \\
&\p + \theta,& \quad &\text{for } \vec b = -\h_1 \vec P_\vf \, .& 
\eal
\right.
\label{eq:theta - theta tilde}
\eeq

\bigskip
\bigskip

For the computation of the $\{210\}$ state de-excitation rate (Sec.~\ref{sec:210 de-excitation}), we will need the integral ${\cal I}_{\{210\}, \{100\}} (\vec b)$. Starting from \eq{eq:I cal n'-n def re}, and using the wavefunctions \eqref{eq:psi 100} and \eqref{eq:psi 210}, it is easy to show that 
\beq
{\cal I}_{\{210\}, \{100\}} (\vec b)
= -\frac{i 2^6 \sqrt{2}}{3^4} \: \frac{(2b)/(3\kappa)}{\[1+ (2b)/(3\kappa)\]^3}
\simeq  -\frac{i\, 2^7 \sqrt{2} \: b}{3^5 \, \kappa} \: ,
\label{eq:I cal 210-100}
\eeq
where we took into account that in the $\{210\} \to \{100\}$ radiative transition, the mediator is emitted with momentum $|\vec P_\vf| = (3/8)\mu \a^2$; then, for $b = \h_{1,2} |\vec P_\vf|$, 
$2b/(3\kappa) = \h_{1,2} \, \a / 4\ll 1$.

\clearpage
\bibliography{BoundStates_arXiv_v2}
\end{document}

%% file: TIKZ_prelim.tex
\usepackage{tikz}
\usetikzlibrary{arrows,shapes}
\usetikzlibrary{trees,patterns}
\usetikzlibrary{matrix,arrows} 				
\usetikzlibrary{positioning}				  
\usetikzlibrary{calc,through}				  
\usetikzlibrary{decorations.pathreplacing}  
\usepackage{pgffor}							

\usetikzlibrary{decorations.pathmorphing}	
\usetikzlibrary{decorations.markings}
\tikzset{
	>=stealth', 
    vector/.style={decorate, decoration={snake}, draw},
	provector/.style={decorate, decoration={snake,amplitude=2.5pt}, draw},
	antivector/.style={decorate, decoration={snake,amplitude=-2.5pt}, draw},
	bigvector/.style={decorate, decoration={snake,amplitude=4pt}, draw},
    fermion/.style={draw=black, postaction={decorate},
        decoration={markings,mark=at position .55 with {\arrow[draw=black]{>}}}},
    fermionbar/.style={draw=black, postaction={decorate},
        decoration={markings,mark=at position .55 with {\arrow[draw=black]{<}}}},
    fermionnoarrow/.style={draw=black},
    gluon/.style={decorate, draw=black,
        decoration={coil,amplitude=4pt, segment length=5pt}},
    scalar/.style={dashed,draw=black, postaction={decorate},
        decoration={markings,mark=at position .55 with {\arrow[draw=black]{>}}}},
    scalarbar/.style={dashed,draw=black, postaction={decorate},
        decoration={markings,mark=at position .55 with {\arrow[draw=black]{<}}}},
    scalarnoarrow/.style={dashed,draw=black},
    momentum/.style={draw=black, postaction={decorate},
        decoration={markings,mark=at position 1 with {\arrow[draw=black]{>}}}},
    antimomentum/.style={draw=black, postaction={decorate},
        decoration={markings,mark=at position 0.1 with {\arrow[draw=black]{<}}}}
}

\tikzstyle{block} = [draw, rectangle, 
    minimum height=3em, minimum width=6em]

%% file: mymacros.tex
\newcommand{\nc}{\newcommand}

\nc{\eq}[1]{Eq.~(\ref{#1})}
\nc{\eqs}[1]{Eqs.~(\ref{#1})}
\nc{\papref}[1]{[\ref{#1}]}

\nc{\pd}{\partial}
\nc{\bea}{\begin{eqnarray}}
\nc{\eea}{\end{eqnarray}}
\nc{\bal}{\begin{alignedat}}
\nc{\eal}{\end{alignedat}}
\nc{\beq}{\begin{equation}}
\nc{\eeq}{\end{equation}}
\nc{\bit}{\begin{itemize}}
\nc{\eit}{\end{itemize}}
\nc{\benu}{\begin{enumerate}}
\nc{\eenu}{\end{enumerate}}
\nc{\bdes}{\begin{description}}
\nc{\edes}{\end{description}}

\nc{\nn}{\nonumber}

\nc{\hc}{\rm{h.c.}}
\nc{\cc}{\rm{c.c.}}

\nc{\sub}[1]{_{\rm{#1}}}
\nc{\ssub}[1]{_{_\rm{#1}}}
\nc{\super}[1]{^{\rm{#1}}}
\nc{\ssuper}[1]{^{^\rm{#1}}}

\nc{\slashed}[1]{{#1}\hspace{-2mm}/}

\nc{\pare}[1]{\left( #1 \right)}
\nc{\sqpare}[1]{\left[ #1 \right]}
\nc{\ang}[1]{\left\langle #1 \right\rangle}
\nc{\abs}[1]{\left| #1 \right|}

\def\g5{\gamma_{5}}

\def\a{\alpha}

\def\g{\gamma}
\def\d{\delta}
\def\e{\epsilon}
\def\z{\zeta}
\def\h{\eta}
\def\th{\theta}

\def\k{\kappa}

\def\m{\mu}
\def\n{\nu}

\def\p{\pi}
\def\r{\rho}
\def\s{\sigma}

\def\f{\phi}
\def\x{\chi}
\def\ps{\psi}
\def\w{\omega}

\def\ve{\varepsilon}

\def\vf{\varphi}

\def\G{\Gamma}

\def\Ks{\Xi}

\def\F{\Phi}
\def\Ps{\Psi}
\def\W{\Omega}


%% file: BoundStates_arXiv_v2.bbl
\providecommand{\href}[2]{#2}\begingroup\raggedright\begin{thebibliography}{10}

\bibitem{Spergel:1999mh}
D.~N. Spergel and P.~J. Steinhardt, {\it {Observational evidence for
  selfinteracting cold dark matter}},  {\em Phys.Rev.Lett.} {\bf 84} (2000)
  3760--3763, [\href{http://arxiv.org/abs/astro-ph/9909386}{{\tt
  astro-ph/9909386}}].

\bibitem{Kusenko:2001vu}
A.~Kusenko and P.~J. Steinhardt, {\it {Q ball candidates for selfinteracting
  dark matter}},  {\em Phys.Rev.Lett.} {\bf 87} (2001) 141301,
  [\href{http://arxiv.org/abs/astro-ph/0106008}{{\tt astro-ph/0106008}}].

\bibitem{Feng:2008mu}
J.~L. Feng, H.~Tu, and H.-B. Yu, {\it {Thermal Relics in Hidden Sectors}},
  {\em JCAP} {\bf 0810} (2008) 043, [\href{http://arxiv.org/abs/0808.2318}{{\tt
  arXiv:0808.2318}}].

\bibitem{Loeb:2010gj}
A.~Loeb and N.~Weiner, {\it {Cores in Dwarf Galaxies from Dark Matter with a
  Yukawa Potential}},  {\em Phys.Rev.Lett.} {\bf 106} (2011) 171302,
  [\href{http://arxiv.org/abs/1011.6374}{{\tt arXiv:1011.6374}}].

\bibitem{Weinberg:2013aya}
D.~H. Weinberg, J.~S. Bullock, F.~Governato, R.~K. de~Naray, and A.~H.~G.
  Peter, {\it {Cold dark matter: controversies on small scales}},
  \href{http://arxiv.org/abs/1306.0913}{{\tt arXiv:1306.0913}}.

\bibitem{Peter:2012jh}
A.~H. Peter, M.~Rocha, J.~S. Bullock, and M.~Kaplinghat, {\it {Cosmological
  Simulations with Self-Interacting Dark Matter II: Halo Shapes vs.
  Observations}},  {\em Mon.Not.Roy.Astron.Soc.} {\bf 430} (2013) 105,
  [\href{http://arxiv.org/abs/1208.3026}{{\tt arXiv:1208.3026}}].

\bibitem{Rocha:2012jg}
M.~Rocha, A.~H. Peter, J.~S. Bullock, M.~Kaplinghat, S.~Garrison-Kimmel,
  et~al., {\it {Cosmological Simulations with Self-Interacting Dark Matter I:
  Constant Density Cores and Substructure}},  {\em Mon.Not.Roy.Astron.Soc.}
  {\bf 430} (2013) 81--104, [\href{http://arxiv.org/abs/1208.3025}{{\tt
  arXiv:1208.3025}}].

\bibitem{Vogelsberger:2014pda}
M.~Vogelsberger, J.~Zavala, C.~Simpson, and A.~Jenkins, {\it {Dwarf galaxies in
  CDM and SIDM with baryons: observational probes of the nature of dark
  matter}},  {\em Mon.Not.Roy.Astron.Soc.} {\bf 444} (2014) 3684,
  [\href{http://arxiv.org/abs/1405.5216}{{\tt arXiv:1405.5216}}].

\bibitem{Zavala:2012us}
J.~Zavala, M.~Vogelsberger, and M.~G. Walker, {\it {Constraining
  Self-Interacting Dark Matter with the Milky Way's dwarf spheroidals}},  {\em
  Monthly Notices of the Royal Astronomical Society: Letters} {\bf 431} (2013)
  L20--L24, [\href{http://arxiv.org/abs/1211.6426}{{\tt arXiv:1211.6426}}].

\bibitem{Vogelsberger:2012ku}
M.~Vogelsberger, J.~Zavala, and A.~Loeb, {\it {Subhaloes in Self-Interacting
  Galactic Dark Matter Haloes}},  {\em Mon.Not.Roy.Astron.Soc.} {\bf 423}
  (2012) 3740, [\href{http://arxiv.org/abs/1201.5892}{{\tt arXiv:1201.5892}}].

\bibitem{Davoudiasl:2012uw}
H.~Davoudiasl and R.~N. Mohapatra, {\it {On Relating the Genesis of Cosmic
  Baryons and Dark Matter}},  {\em New J.Phys.} {\bf 14} (2012) 095011,
  [\href{http://arxiv.org/abs/1203.1247}{{\tt arXiv:1203.1247}}].

\bibitem{Petraki:2013wwa}
K.~Petraki and R.~R. Volkas, {\it {Review of asymmetric dark matter}},  {\em
  Int.J.Mod.Phys.} {\bf A28} (2013) 1330028,
  [\href{http://arxiv.org/abs/1305.4939}{{\tt arXiv:1305.4939}}].

\bibitem{Zurek:2013wia}
K.~M. Zurek, {\it {Asymmetric Dark Matter: Theories, Signatures, and
  Constraints}},  {\em Phys.Rept.} {\bf 537} (2014) 91--121,
  [\href{http://arxiv.org/abs/1308.0338}{{\tt arXiv:1308.0338}}].

\bibitem{Boucenna:2013wba}
S.~Boucenna and S.~Morisi, {\it {Theories relating baryon asymmetry and dark
  matter: A mini review}},  {\em Front.Phys.} {\bf 1} (2014) 33,
  [\href{http://arxiv.org/abs/1310.1904}{{\tt arXiv:1310.1904}}].

\bibitem{Foot:2003jt}
R.~Foot and R.~Volkas, {\it {Was ordinary matter synthesized from mirror
  matter? An Attempt to explain why Omega(Baryon) approximately equal to 0.2
  Omega(Dark)}},  {\em Phys.Rev.} {\bf D68} (2003) 021304,
  [\href{http://arxiv.org/abs/hep-ph/0304261}{{\tt hep-ph/0304261}}].

\bibitem{Kaplan:2009de}
D.~E. Kaplan, G.~Z. Krnjaic, K.~R. Rehermann, and C.~M. Wells, {\it {Atomic
  Dark Matter}},  {\em JCAP} {\bf 1005} (2010) 021,
  [\href{http://arxiv.org/abs/0909.0753}{{\tt arXiv:0909.0753}}].

\bibitem{Petraki:2011mv}
K.~Petraki, M.~Trodden, and R.~R. Volkas, {\it {Visible and dark matter from a
  first-order phase transition in a baryon-symmetric universe}},  {\em JCAP}
  {\bf 1202} (2012) 044, [\href{http://arxiv.org/abs/1111.4786}{{\tt
  arXiv:1111.4786}}].

\bibitem{vonHarling:2012yn}
B.~{von Harling}, K.~Petraki, and R.~R. Volkas, {\it {Affleck-Dine dynamics and
  the dark sector of pangenesis}},  {\em JCAP} {\bf 1205} (2012) 021,
  [\href{http://arxiv.org/abs/1201.2200}{{\tt arXiv:1201.2200}}].

\bibitem{Cline:2012is}
J.~M. Cline, Z.~Liu, and W.~Xue, {\it {Millicharged Atomic Dark Matter}},  {\em
  Phys.Rev.} {\bf D85} (2012) 101302,
  [\href{http://arxiv.org/abs/1201.4858}{{\tt arXiv:1201.4858}}].

\bibitem{Fargion:2005ep}
D.~Fargion, M.~Khlopov, and C.~A. Stephan, {\it {Cold dark matter by heavy
  double charged leptons?}},  {\em Class.Quant.Grav.} {\bf 23} (2006)
  7305--7354, [\href{http://arxiv.org/abs/astro-ph/0511789}{{\tt
  astro-ph/0511789}}].

\bibitem{Foot:2013nea}
R.~Foot and Z.~Silagadze, {\it {Thin disk of co-rotating dwarfs: A fingerprint
  of dissipative (mirror) dark matter?}},  {\em Phys.Dark Univ.} {\bf 2} (2013)
  163--165, [\href{http://arxiv.org/abs/1306.1305}{{\tt arXiv:1306.1305}}].

\bibitem{Foot:2013lxa}
R.~Foot, {\it {Tully-Fisher relation, galactic rotation curves and dissipative
  mirror dark matter}},  {\em JCAP} {\bf 1412} (2014) 047,
  [\href{http://arxiv.org/abs/1307.1755}{{\tt arXiv:1307.1755}}].

\bibitem{Foot:2013uxa}
R.~Foot, {\it {A dark matter scaling relation from mirror dark matter}},  {\em
  Phys.Dark Univ.} {\bf 5-6} (2014) 236--239,
  [\href{http://arxiv.org/abs/1303.1727}{{\tt arXiv:1303.1727}}].

\bibitem{Foot:2014uba}
R.~Foot and S.~Vagnozzi, {\it {Dissipative hidden sector dark matter}},  {\em
  Phys.Rev.} {\bf D91} (2015) 023512,
  [\href{http://arxiv.org/abs/1409.7174}{{\tt arXiv:1409.7174}}].

\bibitem{Foot:2015sia}
R.~Foot, {\it {Dissipative dark matter explains rotation curves}},
  \href{http://arxiv.org/abs/1502.07817}{{\tt arXiv:1502.07817}}.

\bibitem{Fan:2013tia}
J.~Fan, A.~Katz, L.~Randall, and M.~Reece, {\it {Dark-Disk Universe}},  {\em
  Phys.Rev.Lett.} {\bf 110} (2013), no.~21 211302,
  [\href{http://arxiv.org/abs/1303.3271}{{\tt arXiv:1303.3271}}].

\bibitem{ArkaniHamed:2008qn}
N.~Arkani-Hamed, D.~P. Finkbeiner, T.~R. Slatyer, and N.~Weiner, {\it {A Theory
  of Dark Matter}},  {\em Phys.Rev.} {\bf D79} (2009) 015014,
  [\href{http://arxiv.org/abs/0810.0713}{{\tt arXiv:0810.0713}}].

\bibitem{Pospelov:2008jd}
M.~Pospelov and A.~Ritz, {\it {Astrophysical Signatures of Secluded Dark
  Matter}},  {\em Phys.Lett.} {\bf B671} (2009) 391--397,
  [\href{http://arxiv.org/abs/0810.1502}{{\tt arXiv:0810.1502}}].

\bibitem{Pearce:2013ola}
L.~Pearce and A.~Kusenko, {\it {Indirect Detection of Self-Interacting
  Asymmetric Dark Matter}},  {\em Phys.Rev.} {\bf D87} (2013) 123531,
  [\href{http://arxiv.org/abs/1303.7294}{{\tt arXiv:1303.7294}}].

\bibitem{Pearce:2015zca}
L.~Pearce, K.~Petraki, and A.~Kusenko, {\it {Signals from dark atom formation
  in halos}},  \href{http://arxiv.org/abs/1502.01755}{{\tt arXiv:1502.01755}}.

\bibitem{Frandsen:2014lfa}
M.~T. Frandsen, F.~Sannino, I.~M. Shoemaker, and O.~Svendsen, {\it {X-ray Lines
  from Dark Matter: The Good, The Bad, and The Unlikely}},  {\em JCAP} {\bf
  1405} (2014) 033, [\href{http://arxiv.org/abs/1403.1570}{{\tt
  arXiv:1403.1570}}].

\bibitem{Boddy:2014qxa}
K.~K. Boddy, J.~L. Feng, M.~Kaplinghat, Y.~Shadmi, and T.~M.~P. Tait, {\it
  {Strongly interacting dark matter: Self-interactions and keV lines}},  {\em
  Phys.Rev.} {\bf D90} (2014), no.~9 095016,
  [\href{http://arxiv.org/abs/1408.6532}{{\tt arXiv:1408.6532}}].

\bibitem{Cline:2014eaa}
J.~M. Cline, Y.~Farzan, Z.~Liu, G.~D. Moore, and W.~Xue, {\it {3.5 keV x rays
  as the “21 cm line” of dark atoms, and a link to light sterile
  neutrinos}},  {\em Phys.Rev.} {\bf D89} (2014), no.~12 121302,
  [\href{http://arxiv.org/abs/1404.3729}{{\tt arXiv:1404.3729}}].

\bibitem{Detmold:2014qqa}
W.~Detmold, M.~McCullough, and A.~Pochinsky, {\it {Dark Nuclei I: Cosmology and
  Indirect Detection}},  {\em Phys.Rev.} {\bf D90} (2014) 115013,
  [\href{http://arxiv.org/abs/1406.2276}{{\tt arXiv:1406.2276}}].

\bibitem{MarchRussell:2008tu}
J.~D. March-Russell and S.~M. West, {\it {WIMPonium and Boost Factors for
  Indirect Dark Matter Detection}},  {\em Phys.Lett.} {\bf B676} (2009)
  133--139, [\href{http://arxiv.org/abs/0812.0559}{{\tt arXiv:0812.0559}}].

\bibitem{Foot:2003iv}
R.~Foot, {\it {Implications of the DAMA and CRESST experiments for mirror
  matter type dark matter}},  {\em Phys.Rev.} {\bf D69} (2004) 036001,
  [\href{http://arxiv.org/abs/hep-ph/0308254}{{\tt hep-ph/0308254}}].

\bibitem{Fornengo:2011sz}
N.~Fornengo, P.~Panci, and M.~Regis, {\it {Long-Range Forces in Direct Dark
  Matter Searches}},  {\em Phys.Rev.} {\bf D84} (2011) 115002,
  [\href{http://arxiv.org/abs/1108.4661}{{\tt arXiv:1108.4661}}].

\bibitem{Foot:2012cs}
R.~Foot, {\it {Hidden sector dark matter explains the DAMA, CoGeNT, CRESST-II
  and CDMS/Si experiments}},  {\em Phys.Rev.} {\bf D88} (2013) 025032,
  [\href{http://arxiv.org/abs/1209.5602}{{\tt arXiv:1209.5602}}].

\bibitem{Foot:2013msa}
R.~Foot, {\it {Direct detection experiments explained with mirror dark
  matter}},  {\em Phys.Lett.} {\bf B728} (2014) 45--50,
  [\href{http://arxiv.org/abs/1305.4316}{{\tt arXiv:1305.4316}}].

\bibitem{Hisano:2002fk}
J.~Hisano, S.~Matsumoto, and M.~M. Nojiri, {\it {Unitarity and higher order
  corrections in neutralino dark matter annihilation into two photons}},  {\em
  Phys.Rev.} {\bf D67} (2003) 075014,
  [\href{http://arxiv.org/abs/hep-ph/0212022}{{\tt hep-ph/0212022}}].

\bibitem{Hisano:2003ec}
J.~Hisano, S.~Matsumoto, and M.~M. Nojiri, {\it {Explosive dark matter
  annihilation}},  {\em Phys.Rev.Lett.} {\bf 92} (2004) 031303,
  [\href{http://arxiv.org/abs/hep-ph/0307216}{{\tt hep-ph/0307216}}].

\bibitem{Cirelli:2007xd}
M.~Cirelli, A.~Strumia, and M.~Tamburini, {\it {Cosmology and Astrophysics of
  Minimal Dark Matter}},  {\em Nucl.Phys.} {\bf B787} (2007) 152--175,
  [\href{http://arxiv.org/abs/0706.4071}{{\tt arXiv:0706.4071}}].

\bibitem{vonHarling:2014kha}
B.~von Harling and K.~Petraki, {\it {Bound-state formation for thermal relic
  dark matter and unitarity}},  {\em JCAP} {\bf 12} (2014) 033,
  [\href{http://arxiv.org/abs/1407.7874}{{\tt arXiv:1407.7874}}].

\bibitem{Kaplan:2011yj}
D.~E. Kaplan, G.~Z. Krnjaic, K.~R. Rehermann, and C.~M. Wells, {\it {Dark
  Atoms: Asymmetry and Direct Detection}},  {\em JCAP} {\bf 1110} (2011) 011,
  [\href{http://arxiv.org/abs/1105.2073}{{\tt arXiv:1105.2073}}].

\bibitem{Behbahani:2010xa}
S.~R. Behbahani, M.~Jankowiak, T.~Rube, and J.~G. Wacker, {\it {Nearly
  Supersymmetric Dark Atoms}},  {\em Adv.High Energy Phys.} {\bf 2011} (2011)
  709492, [\href{http://arxiv.org/abs/1009.3523}{{\tt arXiv:1009.3523}}].

\bibitem{CyrRacine:2012fz}
F.-Y. Cyr-Racine and K.~Sigurdson, {\it {The Cosmology of Atomic Dark Matter}},
   {\em Phys.Rev.} {\bf D87} (2013) 103515,
  [\href{http://arxiv.org/abs/1209.5752}{{\tt arXiv:1209.5752}}].

\bibitem{Cline:2013pca}
J.~M. Cline, Z.~Liu, G.~Moore, and W.~Xue, {\it {Scattering properties of dark
  atoms and molecules}},  {\em Phys.Rev.} {\bf D89} (2014) 043514,
  [\href{http://arxiv.org/abs/1311.6468}{{\tt arXiv:1311.6468}}].

\bibitem{Petraki:2014uza}
K.~Petraki, L.~Pearce, and A.~Kusenko, {\it {Self-interacting asymmetric dark
  matter coupled to a light massive dark photon}},  {\em JCAP} {\bf 1407}
  (2014) 039, [\href{http://arxiv.org/abs/1403.1077}{{\tt arXiv:1403.1077}}].

\bibitem{Boddy:2014yra}
K.~K. Boddy, J.~L. Feng, M.~Kaplinghat, and T.~M.~P. Tait, {\it
  {Self-Interacting Dark Matter from a Non-Abelian Hidden Sector}},  {\em
  Phys.Rev.} {\bf D89} (2014), no.~11 115017,
  [\href{http://arxiv.org/abs/1402.3629}{{\tt arXiv:1402.3629}}].

\bibitem{Krnjaic:2014xza}
G.~Krnjaic and K.~Sigurdson, {\it {Big Bang Darkleosynthesis}},
  \href{http://arxiv.org/abs/1406.1171}{{\tt arXiv:1406.1171}}.

\bibitem{Wise:2014jva}
M.~B. Wise and Y.~Zhang, {\it {Stable Bound States of Asymmetric Dark Matter}},
   {\em Phys.Rev.} {\bf D90} (2014) 055030,
  [\href{http://arxiv.org/abs/1407.4121}{{\tt arXiv:1407.4121}}].

\bibitem{Wise:2014ola}
M.~B. Wise and Y.~Zhang, {\it {Yukawa Bound States of a Large Number of
  Fermions}},  \href{http://arxiv.org/abs/1411.1772}{{\tt arXiv:1411.1772}}.

\bibitem{Laha:2013gva}
R.~Laha and E.~Braaten, {\it {Direct detection of dark matter in universal
  bound states}},  {\em Phys.Rev.} {\bf D89} (2014) 103510,
  [\href{http://arxiv.org/abs/1311.6386}{{\tt arXiv:1311.6386}}].

\bibitem{Cyr-Racine:2013fsa}
F.-Y. Cyr-Racine, R.~{de Putter}, A.~Raccanelli, and K.~Sigurdson, {\it
  {Constraints on Large-Scale Dark Acoustic Oscillations from Cosmology}},
  {\em Phys.Rev.} {\bf D89} (2014) 063517,
  [\href{http://arxiv.org/abs/1310.3278}{{\tt arXiv:1310.3278}}].

\bibitem{Shepherd:2009sa}
W.~Shepherd, T.~M. Tait, and G.~Zaharijas, {\it {Bound states of weakly
  interacting dark matter}},  {\em Phys.Rev.} {\bf D79} (2009) 055022,
  [\href{http://arxiv.org/abs/0901.2125}{{\tt arXiv:0901.2125}}].

\bibitem{Kusenko:1997zq}
A.~Kusenko, {\it {Solitons in the supersymmetric extensions of the standard
  model}},  {\em Phys.Lett.} {\bf B405} (1997) 108,
  [\href{http://arxiv.org/abs/hep-ph/9704273}{{\tt hep-ph/9704273}}].

\bibitem{Kusenko:1997si}
A.~Kusenko and M.~E. Shaposhnikov, {\it {Supersymmetric Q balls as dark
  matter}},  {\em Phys.Lett.} {\bf B418} (1998) 46--54,
  [\href{http://arxiv.org/abs/hep-ph/9709492}{{\tt hep-ph/9709492}}].

\bibitem{Kasuya:1999wu}
S.~Kasuya and M.~Kawasaki, {\it {Q ball formation through Affleck-Dine
  mechanism}},  {\em Phys.Rev.} {\bf D61} (2000) 041301,
  [\href{http://arxiv.org/abs/hep-ph/9909509}{{\tt hep-ph/9909509}}].

\bibitem{Postma:2001ea}
M.~Postma, {\it {Solitosynthesis of Q balls}},  {\em Phys.Rev.} {\bf D65}
  (2002) 085035, [\href{http://arxiv.org/abs/hep-ph/0110199}{{\tt
  hep-ph/0110199}}].

\bibitem{Kusenko:2004yw}
A.~Kusenko, L.~Loveridge, and M.~Shaposhnikov, {\it {Supersymmetric dark matter
  Q-balls and their interactions in matter}},  {\em Phys.Rev.} {\bf D72} (2005)
  025015, [\href{http://arxiv.org/abs/hep-ph/0405044}{{\tt hep-ph/0405044}}].

\bibitem{Brodsky:2009gx}
S.~J. Brodsky and R.~F. Lebed, {\it {Production of the Smallest QED Atom: True
  Muonium (mu+ mu-)}},  {\em Phys.Rev.Lett.} {\bf 102} (2009) 213401,
  [\href{http://arxiv.org/abs/0904.2225}{{\tt arXiv:0904.2225}}].

\bibitem{Sommerfeld:1931}
A.~Sommerfeld, {\it {{\"U}ber die Beugung und Bremsung der Elektronen}},  {\em
  Ann. Phys.} {\bf 403} (1931), no.~3 257--330.

\bibitem{BetheSalpeter_QM}
H.~A. {Bethe} and E.~E. {Salpeter}, {\em {Quantum Mechanics of One- and
  Two-Electron Atoms}}.
\newblock 1957.

\bibitem{AkhiezerMerenkov_sigmaHydrogen}
A.~I. {Akhiezer} and N.~P. {Merenkov}, {\it {The theory of lepton bound-state
  production}},  {\em Journal of Physics B Atomic Molecular Physics} {\bf 29}
  (May, 1996) 2135--2140.

\bibitem{Iengo:2009ni}
R.~Iengo, {\it {Sommerfeld enhancement: General results from field theory
  diagrams}},  {\em JHEP} {\bf 0905} (2009) 024,
  [\href{http://arxiv.org/abs/0902.0688}{{\tt arXiv:0902.0688}}].

\bibitem{Cassel:2009wt}
S.~Cassel, {\it {Sommerfeld factor for arbitrary partial wave processes}},
  {\em J.Phys.} {\bf G37} (2010) 105009,
  [\href{http://arxiv.org/abs/0903.5307}{{\tt arXiv:0903.5307}}].

\bibitem{Itzykson:1980rh}
C.~Itzykson and J.~Zuber, {\em {Quantum field theory}}.
\newblock 1980.

\bibitem{Silagadze:1998ri}
Z.~Silagadze, {\it {Wick-Cutkosky model: An Introduction}},
  \href{http://arxiv.org/abs/hep-ph/9803307}{{\tt hep-ph/9803307}}.

\bibitem{Hoyer:2014gna}
P.~Hoyer, {\it {Bound states -- from QED to QCD}},
  \href{http://arxiv.org/abs/1402.5005}{{\tt arXiv:1402.5005}}.

\bibitem{PeskinSchroeder}
M.~E. {Peskin} and D.~V. {Schroeder}, {\em {An Introduction to Quantum Field
  Theory}}.
\newblock Westview Press, 1995.

\bibitem{Griest:1989wd}
K.~Griest and M.~Kamionkowski, {\it {Unitarity Limits on the Mass and Radius of
  Dark Matter Particles}},  {\em Phys.Rev.Lett.} {\bf 64} (1990) 615.

\bibitem{Hryczuk:2011tq}
A.~Hryczuk, {\it {The Sommerfeld enhancement for scalar particles and
  application to sfermion co-annihilation regions}},  {\em Phys.Lett.} {\bf
  B699} (2011) 271--275, [\href{http://arxiv.org/abs/1102.4295}{{\tt
  arXiv:1102.4295}}].

\bibitem{Hryczuk:2014hpa}
A.~Hryczuk, I.~Cholis, R.~Iengo, M.~Tavakoli, and P.~Ullio, {\it {Indirect
  Detection Analysis: Wino Dark Matter Case Study}},  {\em JCAP} {\bf 1407}
  (2014) 031, [\href{http://arxiv.org/abs/1401.6212}{{\tt arXiv:1401.6212}}].

\bibitem{Beneke:2014hja}
M.~Beneke, C.~Hellmann, and P.~Ruiz-Femenia, {\it {Heavy neutralino relic
  abundance with Sommerfeld enhancements - a study of pMSSM scenarios}},
  \href{http://arxiv.org/abs/1411.6930}{{\tt arXiv:1411.6930}}.

\bibitem{Bellazzini:2013foa}
B.~Bellazzini, M.~Cliche, and P.~Tanedo, {\it {The Effective Theory of
  Self-Interacting Dark Matter}},  {\em Phys.Rev.} {\bf D88} (2013) 083506,
  [\href{http://arxiv.org/abs/1307.1129}{{\tt arXiv:1307.1129}}].

\bibitem{Messiah:1962}
A.~{Messiah}, {\em {Quantum mechanics}}.
\newblock 1962.

\end{thebibliography}\endgroup
